\documentclass[usenatbib]{mnras}

\usepackage{amsmath}	% Advanced maths commands
\usepackage[T1]{fontenc}
\usepackage{ae,aecompl}

\usepackage{graphicx}	% Including figure files

\usepackage{amsfonts}
\usepackage{amssymb}	% Extra maths symbols
\usepackage{revsymb}
\usepackage{latexsym}
\usepackage{bm}
\usepackage{slashed}

\newcommand{\GN}{G}

\title[MDM and Galaxy Clusters]{Testing Modified Dark Matter with Galaxy Clusters\\
{\LARGE Does Dark Matter know about the Cosmological Constant?}}
\author[D. Edmonds et al.]{%
Doug Edmonds,$^{1,2}$\thanks{e-mails: dedmonds@ehc.edu, farrah@vt.edu, cmho@msu.edu, dminic@vt.edu, yjng@physics.unc.edu, takeuchi@vt.edu} 
Duncan Farrah,$^{2}$\footnotemark[1]
Chiu Man Ho,$^{3}$\footnotemark[1]
\newauthor
Djordje Minic,$^{2,4}$\footnotemark[1]
Y. Jack Ng,$^{5}$\footnotemark[1] and
Tatsu Takeuchi\phantom{,}$^{2,4}$\footnotemark[1]
\\
\\
$^{1}$ Department of Physics, Emory \& Henry College, Emory, VA 24327, USA \\
$^{2}$ Department of Physics, Virginia Tech, Blacksburg, VA 24061, USA \\
$^{3}$ Department of Physics and Astronomy, Michigan State University, East Lansing, MI 48824, USA  \\
$^{4}$ Center for Neutrino Physics, Department of Physics, Virginia Tech, Blacksburg, VA 24061, USA \\ 
$^{5}$ Institute of Field Physics, Department of Physics and Astronomy, University of North Carolina, Chapel Hill, NC 27599, USA 
}

\date{Accepted XXX. Received YYY; in original form ZZZ}

\pubyear{2015}

\begin{document}
\label{firstpage}
\pagerange{\pageref{firstpage}--\pageref{lastpage}}
\maketitle

\begin{abstract}

We discuss the possibility that the cold dark matter mass profiles contain information on the cosmological constant $\Lambda$, and that such information constrains the nature of cold dark matter (CDM). We call this approach Modified Dark Matter (MDM).
In particular, we examine the ability of MDM
to explain the observed mass profiles of 13 galaxy clusters.
Using general arguments from gravitational thermodynamics, we provide a theoretical justification for our MDM mass profile and successfully compare it to the NFW mass profiles both on cluster and galactic scales.
Our results suggest that indeed the CDM mass profiles contain information about the cosmological
constant in a non-trivial way.

%We examine the ability of Modified Dark Matter (MDM)
%to explain observed-mass discrepancies on sub-homogeneity length scales
%by comparing MDM mass profile fits to those from both cold dark matter (CDM) and modified Newtonian dynamics %(MOND) for 13 galaxy clusters.
%We also compare dynamical and observed masses in a sample of 93 different galaxy clusters.
%For both samples, MDM performs as well as CDM, and better by XX than MOND. 
%Moreover, using arguments from gravitational thermodynamics, we provide a theoretical justification for our MDM mass %profile.
\end{abstract}


%%%%%%%%%%%%%%%%%%%%%%%%%%%%%%%%%%%%%%%%%%%%%%%%%%
\section{Introduction}\label{sec:intro}

\subsection{Evidence for Dark Matter}

Observational evidence for the presence of dark matter \citep{Zwicky:1933,Zwicky:1937}
exists at a variety of length scales.
It is most obvious at galactic scales ($\sim\!10\,\mathrm{kpc}$) 
where the rotation curves of spiral galaxies
have been found to be asymptotically flat, way beyond the radii of their visible disks \citep{Rubin:1970,Rubin:1978,Rubin:1980,vanAlbada:1985}.
Measurements of rotation curves have also been performed for elliptical \citep{Ciardullo:1993,Norris:2012}
and low-surface-brightness (LSB) galaxies \citep{deBlok:2002} with
similar discrepancies.
The dark matter to baryonic matter ratio inferred from the rotation curves differs from galaxy to galaxy.
Extreme cases such as 
%galaxies dominated by dark matter without any visible stars (dark galaxies) \citep{Minchin:2005}
%{\color{green}{\bf this might not be real! but LSB are dark matter dominated}}, 
%as well as 
galaxies whose rotation curves do not require dark matter \citep{Ciardullo:1993,Romanowsky:2003} have also been discovered.\footnote{This does not mean that dark matter halos are ruled out in these galaxies. For example, in \citet{DeLorenzi:2008}, it is shown that a dark matter halo is consistent with observations of NGC~4697 though a halo is not required to fit the data.}$^,$
\footnote{%
In the case of Globular clusters, which have radii on the scale of $\sim 10\,\mathrm{pc}$, there seems to be some controversy on whether they have a significant dark
matter component \citep{Mashchenko:2005a,Mashchenko:2005b,Conroy:2011}.
A recent paper claims to have discovered globular clusters dominated by dark matter \citep{Taylor:2015}.
}

At the scale of galaxy clusters ($\sim\!1\,\mathrm{Mpc}$), 
the virial theorem is used (assuming virialization of the galaxies and gas within the cluster)
to infer the mass distribution within the cluster from the distribution of the radial velocities of member galaxies \citep{Zwicky:1933,Zwicky:1937},
and the temperature and density distributions of the hot gases (the intracluster medium (ICM)) 
measured by X-ray
satellite observatories \citep{Sarazin:1988,Vikhlinin:2005,Vikhlinin:2006,Moretti:2011}.
Strong and weak gravitational lensing also provide independent measurements of the
cluster's mass \citep{Clowe:2004}.
These determinations of the dynamical masses of the galaxy clusters disagree with
their visible masses, giving rise to the so-called virial discrepancy.

At the cosmological scale ($\sim\!10\,\mathrm{Gpc}$), 
the amount of dark matter in the universe
can be inferred from the positions and heights of the acoustic peaks
in the Cosmic Microwave Background (CMB) anisotropy data \citep{Planck:2015}.

%%%%%%%%%%%%%%%%%%%%%%%%%%%%%%
\subsection{CDM \& MOND}

The most popular model which is consistent with
various cosmological structure formation constraints is
the one in which a non-zero cosmological constant $\Lambda$ (dark energy) 
is assumed in addition to the existence of collisionless Cold Dark Matter (CDM)
\citep{dark,Frenk:2012}
However, the $\Lambda$CDM paradigm is not without its problems
at and below the galactic scale.
In particular, $N$-body simulations of CDM evolution 
\citep{Dubinski:1991,Navarro:1996,Navarro:1997,Moore:1999,Klypin:2001,Taylor:2001,Colin:2004,Diemand:2005}
predict a `cusp' in the dark matter distribution toward the galactic center 
whereas observations indicate the presence of a `core,' i.e.
the `core/cusp problem' \citep{deBlok:2010}.

An alternative to the introduction of dark matter to explain the
discrepancy between the visible and inferred masses would be to modify 
the laws of gravity.
The most prominent of such approaches is
Modified Newtonian Dynamics (MOND) of \citet{Milgrom:1983a,Milgrom:1983b,Milgrom:1983c}.\footnote{%
Other approaches to modifying gravity instead of introducing dark matter
include the relativistic version of MOND by 
\citet{Bekenstein:2004}, and 
%\citet{Blanchet:2008,Blanchet:2009,Blanchet:2011,Brownstein:2006a,Brownstein:2006b,Brownstein:2006c,Brownstein:2007,Khoury2015,Mannheim:2013}.
\citet{Blanchet:2008,Blanchet:2009,Blanchet:2011,Khoury2015,Mannheim:2013,Moffat:2014}.
}
In MOND, the equation of motion is modified from $F=ma$ to
\begin{equation}
F \;=\; \begin{cases}
ma & (a\gg a_c) \\
m a^2/a_c & (a\ll a_c) 
\end{cases}
\end{equation}
where $a_c$ is the critical acceleration\footnote{%
In the MOND literature, the critical acceleration is usually denoted $a_0$.
}
which separates the two regions of behavior. 
The two regions are connected by an interpolating function
\begin{equation}
F \;=\; ma\,\mu(a/a_c)\;,
\label{MOND-EQM}
\end{equation}
where
\begin{equation}
\mu(x)\;=\;
\begin{cases}
1 & (x\gg 1) \\
x & (x\ll 1)
\end{cases}
\end{equation}
The choice of interpolating functions $\mu(x)$ is arbitrary, and an often used 
functional form is
\begin{equation}
\mu(x) \;=\; \dfrac{x}{(1+x^n)^{1/n}}\;,\qquad n\in\mathbb{N}\;.
\label{muN}
\end{equation}
For a given (baryonic) source mass $M$, its gravitational attraction on a test mass $m$ is
$F=m(\GN M/r^2)\equiv ma_N$, where $a_N=\GN M/r^2$ is the usual Newtonian acceleration without dark matter.
Then the above modification to the equation of motion implies
\begin{equation}
a \;=\; 
\begin{cases}
a_N & (a\gg a_c) \\
\sqrt{a_c a_N}  & (a\ll a_c)
\end{cases}
\end{equation}
On the outskirts of galaxies, this implies
\begin{equation}
v^2 \;=\; ra \;\xrightarrow{r\rightarrow\infty}\; r\sqrt{a_c a_N}
\;=\; \sqrt{a_c \GN M} 
\;\equiv\;v_\infty^2
\;,
\end{equation}
leading to flat rotation curves\footnote{%
In reality, rotation curves are not all flat, they display a variety of properties.
See, e.g. \citet{Persic:1991,Persic:1996,Catinella:2006}.
}
as well as the Baryonic Tully-Fisher relation \citep{Tully:1977,Steinmetz:1999,McGaugh:2000,Torres-Flores:2011,McGaugh:2012}
\begin{equation}
M\;\propto\; v^4\;.
\end{equation}
The most amazing thing about the MOND approach is that it succeeds in fitting the rotation 
curves of a large set of galaxies with the single universal parameter $a_c$ which is consistently 
found to be \citep{Begeman:1991} 
\begin{equation}
a_c \;\approx\; 10^{-8}\,\mathrm{cm/s^2}\;.
\end{equation}
Other studies of MOND in the context of rotation curves include \citet{Sanders:1996,Sanders:1998,vandenBosch:2000,Swaters:2010}.
Given that the Hubble parameter is
\begin{eqnarray}
H_0 
& = & (67.74\pm 0.46)\,\mathrm{km/s/Mpc} 
\cr
& = & \left[(6.581\pm 0.045)\times 10^{-8}\mathrm{cm/s^2}\right]/c
\;,
\end{eqnarray}
(see Table 4 of \citealt{Planck:2015}) it has been noted that 
\begin{equation}
a_c \;\approx\; \dfrac{cH_0}{2\pi}\;,
\end{equation}
suggesting that MOND may have cosmological origins \citep{Milgrom:1999}.

The application of MOND to galaxy clusters has been less successful.
\citet{Sanders:1999,Sanders:2003} shows that assuming the value of $a_c$ found
from galactic rotation curves, fitting MOND to galaxy clusters does
reduce the virial discrepancy, but not sufficiently so, mainly due to
large portions of the clusters remaining in the $a\agt a_c$ regime.
MOND also has difficulty explaining the weak gravitational lensing
of the bullet cluster \citep{Clowe:2004}.
Thus, even with MOND, the introduction of dark matter may be unavoidable at cluster scales, which negates the original motivation for MOND.

%%%%%%%%%%%%%%%%%%%%%%%%%%%%%%
\subsection{Modified Dark Matter?}

If one is to stay within the dark matter paradigm, the uncanny success of MOND at 
galactic scales, where collisionless CDM has problems, makes one ponder whether 
it would be possible to combine the salient features of MOND and CDM into a new 
framework of dark matter.

From the relativistic point of view, 
the choice between the introduction of dark matter and the modification of gravity 
amounts to which side of Einstein's equation,
\begin{equation}
G_{\alpha\beta} \;=\; (8\pi \GN)\,T_{\alpha\beta}\;\;,
\label{EinsteinEq}
\end{equation}
one chooses to modify.  The same terms could be interpreted
differently depending on which side of the equation one places it,
e.g. vacuum energy vs. cosmological constant.
Thus if a particular modification works on the left-hand-side (a relativistic extension of MOND, for example),
one could move it to the right-hand-side and reinterpret it as due to 
a new type of dark matter.
This is also evident in the non-relativistic MOND equation, Eq.~(\ref{MOND-EQM}), when rewritten as 
\begin{equation}
\dfrac{1}{\mu(a/a_c)}\dfrac{\GN M}{r^2}
%\;=\; \dfrac{a_N}{\mu(a/a_c)}
\;=\; a\;.
\label{MOND-EQM2}
\end{equation}
Interpreting $M$ as the mass of baryonic matter enclosed within a sphere of
radius $r$, one could write the left-hand-side as
\begin{equation}
\dfrac{1}{\mu(a/a_c)}\dfrac{\GN M}{r^2}
\;=\; \dfrac{\GN(M+M')}{r^2}
\;,
\end{equation}
and interpret
\begin{equation}
M' 
\;=\; M\left[\dfrac{1}{\mu(a/a_c)}-1\,\right] 
\;\equiv\; M f_\mathrm{MOND}(a/a_c)
\;,
\label{fdef}
\end{equation}
as the mass of non-baryonic dark matter enclosed within the same sphere.
Then Eq.~(\ref{MOND-EQM2}) can be written as
\begin{equation}
a_N\Bigl[\,1 + f_\mathrm{MOND}(a/a_c) \,\Bigr] \;=\; a\;.
\label{MOND-EQM3}
\end{equation}
Solving this equation for the acceleration $a$ will also determine $M'=M f_\mathrm{MOND}(a/a_c)$.
The dark matter distribution determined in this fashion
would precisely reproduce the results of MOND without modifying inertia or the law of gravity.

Note, however, that this type of dark matter can be expected to be 
quite different from any other type of dark matter heretofore considered, 
be it cold, warm, hot, or mixed.
First, $M'$ is proportional to $M$, so the dark matter must
track the baryonic matter.
%a feature we may not ultimately wish to retain if we
%hope to explain the bullet cluster.
Second, $M'$ is dependent on $a_c$.
This can be interpreted as due to the dark matter being knowledgable about the
Hubble parameter $H_0 \approx 2\pi a_c/c$.
Third, the dependence on $a$, which must be introduced to cancel the dimensions of $a_c$ by 
taking the ratio $a/a_c$,
but is a parameter that must be determined by solving the equation of motion,
means that the dark matter distribution is closely linked to the
consistency of the dynamics.

As far as the knowledge of $a_c \sim c H_0$ is concerned, \citet{Kaplinghat:2002} have 
argued that the acceleration scale $c H_0$ may arise naturally within CDM models 
if one considers the energy dissipation (cooling) and collapse of baryonic matter into the central regions 
within galactic CDM haloes (which do not collapse).
Because of the baryonic collapse, gravity in the central regions of galaxies is dominated by
baryonic gravity, while that in the outskirts is dominated by CDM gravity.
Due to the scaling properties of this process,
$c H_0$ is argued to set a universal scale at which this transition occurs.
For clusters, on the other hand, gravity is everywhere CDM dominated and this transition does not occur, explaining the failure of MOND.
In the \citet{Kaplinghat:2002} approach, therefore, the CDM obtains information on $c H_0$
from the simple fact that it is evolving in a universe expanding at the rate of $H_0$,
whereas the coincidence $a_c \sim O(1) c H_0$ is more of a numerical accident.
And even with this knowledge, CDM has problems at the galactic scale as mentioned above.

If a dark-matter model is to reproduce the success of MOND, two questions must be answered:
\begin{enumerate}
\item What should the function $f_\mathrm{MOND}(x)$ appearing in Eq.~(\ref{MOND-EQM3}) be?
\item What dynamics must the dark-matter have to reproduce such a mass distribution?
\end{enumerate}
Beginning with the first question,
the asymptotics of the MOND interpolating function $\mu(x)$ only demand
\begin{equation}
f_\mathrm{MOND}(x) \;=\;
\begin{cases}
0 & (x\gg 1) \\
x^{-1} & (x\ll 1)
\end{cases}
\label{fMONDasymp}
\end{equation}
which is not much of a constraint.
Also, though we are trying to reverse engineer MOND, 
one eventually wishes to predict MONDian behavior from a dark-matter model
based on fundamental principles.

First and foremost, the dark matter must know about $cH_0$ 
via some fundamental principle and not by accident.
To this end, \citet{Ho:2010,Ho:2011,Ho:2012} used standard gravitational thermodynamics, 
and Verlinde's idea of entropic gravity \citep{Verlinde:2011},
to argue that in a de Sitter space with cosmological
constant $\Lambda$, the equation of motion should read
\begin{equation}
a_N \Big[\,1+f_\mathrm{MDM}(a/a_0)\,\Bigr]
\;=\; \sqrt{a^2 + a_0^2} - a_0^{\phantom{1}} \;,
\label{MDM-EQM}
\end{equation}
where 
\begin{equation}
\dfrac{M'}{M} \;=\; f_\mathrm{MDM}(a/a_0)\;,
\end{equation}
and $a_0$ is given by the cosmological constant $\Lambda$ as
\begin{equation}
a_0 \;=\; c^2 \sqrt{\dfrac{\Lambda}{3}} \;=\; c H_0 \;=\; 2\pi a_c\;.
\end{equation}
Despite its appearance, Eq.~(\ref{MDM-EQM}) does not modify General Relativity (GR)
since the derivation of the right-hand-side is based on GR.
The information on $cH_0 = c^2\sqrt{\Lambda/3}$ is conveyed to the dark-matter via the temperature
of the de Sitter horizon.

\citet{Ho:2010,Ho:2011,Ho:2012} dubbed the hypothetical dark matter which
satisfies Eq.~(\ref{MDM-EQM}) `MONDian Dark Matter' (MDM),
but the equation of motion is distinct from Eq.~(\ref{MOND-EQM3}).\footnote{%
This is the reason why it is inappropriate to translate the MDM distribution function
into a MONDian interpolation function \citep{Edmonds:2014}.}
Unfortunately the name conjured among the community the idea that MDM was a
hybrid theory, which it is not.
We therefore choose to rename the model `Modified Dark Matter' (MDM, same acronym)
to clarify that it is a dark-matter model, albeit an exotic one.

As the MDM distribution function, 
\citet{Ho:2010,Ho:2011,Ho:2012} chose
\begin{equation}
f_\mathrm{MDM}(a/a_0) \;=\; \dfrac{1}{\pi}\left(\dfrac{a_0}{a}\right)^2\;, 
\label{fMDM1}
\end{equation}
which goes to zero in the limit $a\gg a_0$, and reproduces the flat galactic rotation curves. Note that  $f_\mathrm{MDM}\sim x^{-2}$ and not $x^{-1}$ as in Eq.~(\ref{fMONDasymp}). As we will show in section 2, this form is suggested by the heuristic arguments based on gravitational thermodynamics presented in section 2.
This choice is, of course, arbitrary and 
a case of reverse engineering.
\citet{Ho:2012} proposed that
quanta which obey infinite statistics \citep{Doplicher:1971,Doplicher:1974,Govorkov:1983,Greenberg:1990,Shevchenko:2008}
may exhibit the necessary dynamics to predict such behavior, 
but the details have yet to be worked out.
We note in passing, however, that quanta of infinite statistics
cannot be expressed as excitations of a local quantum field theory (QFT)
and are intrinsically `non-local' in nature, indicating that they have at least one feature that may be required of MDM.

Despite lacking a dynamical theory, which precludes performing numerical simulations,
\citet{Edmonds:2014} confronted the proposed MDM distribution with 
galactic rotation curves to see if it has merit from the phenomenological 
point of view.
It was found that the MDM distribution does indeed provide a good fit to the data, and with a single fitting parameter (the mass to luminosity ratio of the galaxy) as opposed to three (two of which are correlated) for the well-known dark matter mass profile of \citet{Navarro:1996} (NFW). This could indicate some hidden structure connected to the cosmological constant within the CDM mass profiles.

In this paper, we seek to extend our investigation of MDM to galaxy clusters.
There, Eq.~(\ref{fMDM1}) is not expected to work, given that it 
closely resembles MOND which does not work either.
We therefore begin by seeking a generalization of Eq.~(\ref{fMDM1}) which
also places the expression onto a firmer theoretical footing.
We then confront the galaxy cluster data to demonstrate the merit of the
improved/generalized MDM distribution functions.

This paper is organized as follows.
In section~\ref{MDMprofile} we review our derivation of Eq.~(\ref{MDM-EQM}),
based on gravitational thermodynamics,
and provide a theoretical justification for the 
MDM profile proposed by \citet{Ho:2010,Ho:2011,Ho:2012} and 
studied by \citet{Edmonds:2014}, albeit with a slight rescaling.
In section~\ref{GalaxyClusterFits}, we generalize
the MDM mass profile to account for galaxy cluster data as well as galactic rotation curves
and confront the new distribution with the
virial mass profiles of 13 low-redshift, relaxed galaxy clusters. Section 4 is devoted to summary and discussion. In the appendix we present the successful fit
of the generalized MDM mass profile to galactic rotation curves.
We find that the MDM mass profile is sensitive to the cosmological constant on both the cluster and galactic
scales. This in turn may constrain the nature of CDM.

%%%%%%%%%%%%%%%%%%%%%%%%%%%%%%%%%%%%%%%%%%%%%%%%%%
\section{The Modified Dark Matter Mass Profile}\label{MDMprofile}

\subsection{Background}\label{mdmback}

As mentioned above, the missing mass problem in the context of Einstein's equation  
$G_{\alpha\beta} = (8\pi\GN) T_{\alpha\beta}$ 
can be solved in two ways: change the source $T_{\alpha\beta}$ by adding a new energy-momentum tensor, 
or change the Einstein tensor $G_{\alpha\beta}$. 
CDM is an example of the first option, one in which the extra energy momentum tensor is independent of the original baryonic tensor. %\footnote{mention an example of a model where it is dependent?}. 
The second option is a modification of gravity.
Modifications to inertia, such as in Milgrom's scaling, should naturally emerge from this second option 
(e.g. \citealt{Bekenstein:2004}). However, one could also recast Einstein's equation such that the energy-momentum part contains the cosmological constant term, $\Lambda g_{\alpha \beta}$, and the question arises whether the CDM mass profiles could know about the cosmological constant. This is the idea of modified dark matter (MDM).

Given the nature of the cosmological constant problem, the MDM approach should, at heart, be
based on quantum gravity. However, quantum gravity is a very difficult problem; So our proposal is to look at the thermodynamic reformulation of Einstein's theory and search for a commensurate modification of the energy momentum tensor that accounts for the cosmological constant. The reason for this is that gravitational thermodynamics (the prototype of which is black hole thermodynamics) is the only place where quantum theory and physics in accelerating frames are (currently) precisely related.

%%%%%%%%%%%%%%%%%%%%%%%%%%%%%%%%%%%%%%%%%%%%%%%%%%
\subsection{Gravitational Thermodynamics}

In presenting this idea we follow the approach of \citet{Jacobson:1995}. We consider a local observer with acceleration $a$ in a spatially flat de Sitter space (i.e. one dominated by a positive cosmological constant $\Lambda$, in which $a_0=c^2\sqrt{\Lambda/3}$ \, and $\Lambda = 3 \,H_0^2/c^2$). In such a space, the thermodynamic relation
\begin{equation} \label{firstlaw}
dE \;=\; T dS
\end{equation}
has $T$ as the Unruh temperature associated with the local accelerating (Rindler) observer \citep{Davies:1975,Unruh:1976}
\begin{equation}
T \;=\; \frac{\hbar a}{2\pi c k_B}\;.
\end{equation}
The acceleration $a$ can be interpreted as surface gravity on the associated (Rindler) horizon. The entropy $S$ is then associated with the area of this horizon\footnote{This scaling of entropy with area is sometimes called ``holography'' \citep{holography,Susskind:1995}.}
\begin{equation}\label{hbek}
S \;=\; \frac{c^3 A}{4 \GN \hbar}\;,
\end{equation}
and the energy $E$ is the integral of the energy momentum tensor
\begin{equation}\label{intgr}
E \;=\; \int T_{\alpha\beta} k^{\alpha} k^{\beta}
\end{equation}
where $k^{\alpha}$ are appropriate unit vectors.

The link between this thermodynamic equation and Einstein's equations is given by the Raychaudhuri equation.\footnote{Used by Penrose and others to discuss gravitational focusing in their seminal proofs of the singularity theorems \citep{Penrose:1965}.} For an instantaneously stationary local Rindler horizon the shear and vorticity terms can be neglected, effectively leaving:
\begin{equation}
\dfrac{\delta A}{\delta \lambda} \;=\;  R_{\alpha\beta} k^{\alpha} k^{\beta} + \cdots
\end{equation}
where $\lambda$ is the appropriate affine parameter and $k^\alpha$ are the unit tangent vectors to the appropriate geodesic whose cross sectional area $A$ is being focused.

Taking the above expressions, along with the expressions for the Hawking-Unruh temperature $T$ and the Bekenstein-Hawking entropy $S$, it then follows that
\begin{equation}
8 \pi \GN \int T_{\alpha\beta} k^{\alpha} k^{\beta} \;=\; \int R_{\alpha\beta} k^{\alpha} k^{\beta} 
\end{equation}
and thus that $T_{\alpha\beta}\propto R_{\alpha\beta} - fg_{\alpha\beta}$. The local divergenceless of $T_{\alpha\beta}$ then sets $f = -R/2 + \Lambda$ (via the Bianchi identity), and we recover Einstein's equations. Even the correct factor of $8 \pi \GN$ comes from the factor of $2 \pi$, the periodicity of the euclidean time in the temperature formula, and the $4 \GN$ from the entropy formula. The local acceleration, $a$, in this context just sets the correct units between the entropy and energy. 

Our modification to Jacobson's argument is to {\itshape introduce a fundamental acceleration that is related to the cosmological constant}. Since we wish to preserve the holographic scaling of the area (to remain consistent with Einstein's theory) {\itshape and} still have a standard energy-momentum tensor, we are compelled to change something in the thermodynamic expression that can be interpreted both from the point of view of inertia and from the point of view of a mass source. We must therefore preserve the entropy but change the temperature such that the additional part of the energy momentum tensor ``knows'' about the inertial properties that temperature knows about. This corresponds to a change in Equation~(\ref{firstlaw}) such that the change in temperature amounts to a particular change in energy that leaves the entropy unchanged. The constancy of the entropy $dS$ is a form of infinitesimal adiabaticity.

To change the temperature we turn to our local observer with acceleration $a$ in de Sitter space, and thus with total acceleration $a_0 + a$ (where $a_0 = c^2 \sqrt{\Lambda/3}$). The Unruh temperature experienced by this observer is \citep{Deser:1997,Jacobson:1998}
\begin{equation}
T_{a_0 +a} \;=\; \dfrac{\hbar}{2\pi c k_B} \sqrt{\,a^2+a_0^2\,}\;.
\end{equation}
However, since de Sitter space has a cosmological horizon, it has a horizon temperature associated with $a_0$. We thus define the following {\itshape effective} temperature, so that for zero acceleration we get zero temperature
\begin{eqnarray}\label{net}
\tilde{T} 
\;\equiv\; T_{a_0+a} - T_{a_0} 
\;=\; \frac{\hbar}{2\pi c k_B} \left(\,\sqrt{a^2+a_0^2} - a_0 \,\right)
\;.
\end{eqnarray}
This reduces to the usual temperature for Rindler observers by neglecting $a_0$. 
Note that
\begin{equation}
\sqrt{a^2+a_0^2} - a_0
\;\approx\; 
\begin{cases}
a & (a\gg a_0)\;, \\
\dfrac{a^2}{2a_0^2} & (a\ll a_0)\;.
\end{cases}
\end{equation}
Our model is thus:
\begin{equation}\label{deftilde}
d\tilde {E} \;=\; \tilde{T} dS\;,
\end{equation}
in which $dS$ remains unchanged and where we define, in analogy with the normalized temperature $\tilde{T}$, the normalized energy 
\begin{equation}
\tilde{E} \;=\; E_{a_0 +a} - E_{a_0}\;.
\end{equation}
Therefore, energy is not changed in an arbitrary way, but instead in accordance with the change in temperature that should be fixed by the background.

If we now require that $a=a_N$, so that we have {\itshape locally} what Einstein's theory would give in the Newtonian limit, then we have, at least from the point of view of the temperature, Milgrom's scaling. There is, however, a key advance in our argument: Milgrom's scaling emerges as the difference between the temperature of an accelerated observer in de Sitter space and the background temperature of the cosmological horizon, in the limit of accelerations small compared to the acceleration associated with the surface gravity of the cosmological horizon. Milgrom's scaling (MOND) ($a = a_N, a\gg a_c$ and $a= \sqrt{a_N a_c}, a\ll a_c$) sets the value of $a_c$ from the Hubble scale; $a_0$ is related to Milgrom's critical acceleration $a_c$ as $a_c = a_0/(2\pi)$ \citep{Milgrom:1983a,Milgrom:1983b,Milgrom:1983c}.

%%%%%%%%%%%%%%%%%%%%%%%%%%%%%%%%%%%%%%%%%%%%%%%%%%
\subsection{The MDM Mass Profile}

How does this argument relate to the idea of `missing' mass? It follows from Equation~(\ref{deftilde}) that 
\begin{equation}
d E_{a_0+a} \;=\; T_{a_0+a} d S\;,
\end{equation}
\noindent where the entropy $S$ is still given by Equation~(\ref{hbek}), so that the Einstein tensor is untouched. However, due to the change in temperature the energy is changed. If we rewrite the temperature as $T_{a_0 +a} = T + T'$ where the $T$ part corresponds to the Unruh temperature of the observer moving with the Newtonian acceleration $a_N$ (in the correspondence limit $a_0 \ll a=a_N$), then we can also write
\begin{equation}
d E +d E' \;=\; T d S + T' d S\;.
\end{equation}
If we interpret the original $dE = T dS$ as corresponding to baryonic matter, then
\begin{equation}
d E' \;=\; T'  dS \;=\; \frac{T'}{T} dE\;.
\end{equation}
Thus the energy momentum tensor of the extra sources (which we will identify with missing mass) has to be related to the energy momentum tensor of ordinary visible matter (via equation~(\ref{intgr})).

Finally, by expanding the formula for the de Sitter temperature we have a relation between the energy momentum tensors of the dark (prime) and visible (unprime) matter
\begin{equation}
T'_{\alpha\beta} \;=\; \dfrac{a_0^2}{2a^2} \,T_{\alpha\beta}\;,
\end{equation}
which for energy density (i.e. the $00$ components) is
\begin{equation}
M' \;=\;  \dfrac{a_0^2}{2a^2} \,M\;.
\label{fMDM2}
\end{equation}
The $M'$ in the above expression is what we call `dark matter.' 
This dark matter profile is similar to the one in our original papers on MDM \citep{Ho:2010,Ho:2011,Ho:2012}, but with the appearance of the factor of $1/2$ as opposed to the original $1/\pi$ (which was motivated by the Tully-Fisher relation). The exact forms, without any expansion, are
\begin{equation}
T_{\alpha\beta}' \;=\;  \left(\sqrt{1 +\dfrac{a_0^2}{a^2}} - 1\,\right)\, T_{\alpha\beta}\;\;,
\end{equation}
and 
\begin{equation}
M' \;=\;  \left(\sqrt{1 +\frac{a_0^2}{a^2}} - 1\,\right)\, M\;,
\end{equation}
respectively. This gives rise to the idea of `entropic force', where:
\begin{eqnarray}\label{force1}
F_\mathrm{entropic} \;=\; m\, \left(\sqrt{a^2+a_0^2}-a_0\right)\;,
\end{eqnarray}
and 
\begin{eqnarray}
\label{force2}
\sqrt{a^2+a_0^2}-a_0 
\;=\; 
\frac{\GN(M+M')}{r^2}
\;,
\end{eqnarray}
respectively.

Our dark matter mass profile thus ``knows'' about the visible matter as well as the cosmological background, and it also knows about the inertial properties of the moving masses in that background, since it is directly tied to the visible matter, the cosmological constant $\Lambda \equiv 3 a_0^2/c^2$, and the inertial properties (acceleration $a$). In essence, we have a vacuum origin of the fundamental acceleration and also a quantum origin of the mass profile that could lead to the observed galactic rotation curves. From our perspective, flat galactic rotation curves are a quantum gravity effect at large scales.

%%%%%%%%%%%%%%%%%%%%%%%%%%%%%%%%%%%%%%%%%%%%%%%%%%
\section{Galaxy Clusters}\label{GalaxyClusterFits}

%%%%%%%%%%%%%%%%%%%%%%%%%%%%%%%%%%%%%%%%%%%%%%%%%%
\subsection{MDM Distribution Function for Clusters}

In principle, the above mass profile, Eq.~(\ref{fMDM2}), 
fixed by the ratio of the corresponding Unruh-Hawking temperatures (in the limit of small
$a_0/a$) can be modified due to some well-known physical effects associated with a change of scale.
For example, the temperature can be changed via the Tolman-Ehrenfest formula \citep{Tolman:1930a,Tolman:1930b}
\begin{equation}
T \sqrt{g_{00}} \;=\; 2 \tilde{\alpha}\;, 
\end{equation}
where the $g_{00}$ is essentially
determined by the gravitational potential $\Phi$, $g_{00} = 1+2\,\Phi$ (in the units $c=1$), and the dimensionful factor $\tilde{\alpha}$ is determined by the
boundary conditions of the problem.

The questions are what gravitational potential should be used in our case and what sets the value of $\tilde{\alpha}$. 
For example, for a constant background gravitational field, i.e. the linear potential, we are led to consider the following modification
of the mass profile:
\begin{equation}
\label{clustermassp}
f_{\mathrm{MDM}}(a/a_0)
\;=\; \dfrac{M'}{M}
\;=\; \dfrac{\alpha}{\left[\,1+ \left(r/r_\textrm{MDM}\right)\,\right]}\left(\dfrac{a_0}{a}\right)^2 \;,
\end{equation}
where the dimensionless factor $\alpha$ is really determined by the ratio of dimensionful $\tilde{\alpha}$ at different scales (in our case, the cluster and galactic scales).
It is interesting that the same prefactor can be obtained in the context of conformal gravity by rewriting the FRW cosmological line element in the Schwarschild coordinate system
(the linear potential also being the direct analogue of the Newtonian potential for conformal gravity; see \citealt{Mannheim:2013}).
Note that the prefactor $\alpha/\left[1+ (r/r_\textrm{MDM})\right]$ is only the leading term in a more general expression that involves higher order terms in $r$.

Regarding the value of $\alpha$ we note that the boundary value of the temperature is set by the ratio 
$a_0^2/2a^2$,
which scales as distance squared, because $a \sim 1/r$. Thus, in principle, $\alpha$ can increase with distance.
However, our argument is too heuristic to determine the form of this scaling.

In what follows we show that the mass profile given in Equation~\ref{clustermassp} very nicely fits the cluster data and reproduces the relevant virial mass profile, with $\alpha \sim 100$ and $r_\textrm{MDM} \sim 10$~Kpc. However, on galactic scales, given our previous successful fits of the
galactic rotation curves, $\alpha \sim 1$ and also $r$ is much smaller than $r_\textrm{MDM}$. Thus, in going from the galactic to the cluster scales, $\alpha$ needs to change by two orders of magnitude. This is not completely unrealistic, given our heuristic thermodynamic reasoning, but we have to admit of being
ignorant of the underlying reason for this.
%{\bf Can we do better here and determine the ratio of dimensionless $\tilde{\alpha}$ at two different scales from the %Tolman-Ehrenfest formula, by plugging for the vacuum temperature $T$ determined
%in terms of the acceleration and by plugging for the Newtonian potential at two different scales?}

%We consider 13 low-redshift , relaxed galaxy clusters \cite{Vikhlinin:2006} and compare our results to the CDM fits as well as MOND.

%We are now competitive with CDM - in some cases our fit is better than CDM, and in the remaining cases it is worse. All our fits are much better than MOND.  

%% Start section explaining data fits to Vikhlinin et al. (2006) sample

%%%%%%%%%%%%%%%%%%%%%%%%%%%%%%%%%%%%%%%%%%%%%%%%%%
\subsection{Comparison to Galaxy Cluster Data}

We compare MDM mass profiles with the observed (virial) mass profiles in a sample of 13 relaxed galaxy clusters given in \citet{Vikhlinin:2006}. This sample was chosen for several reasons. Firstly, they analyzed all available \textit{Chandra} data which were of sufficient quality to determine mass profiles robustly out to large radii ($\sim 0.75 r_{500}$ and extended past $r_{500}$ in five clusters), spanning a temperature range of 0.7 -- 9~keV. This provides us with a significant sample size with data covering radii from the inner, cooling region out to the virial radius. Secondly, their fitting models were given enough degrees of freedom to get very accurate data fits, which allows us to compare MDM mass profiles to these very accurate models.  Thirdly, whenever possible, data from two different X-ray observatories, \textit{Chandra} and ROSAT, were compared and combined in order to ensure that derived cluster brightness profiles agree in the overlapping regions. At large radii, the \textit{Chandra} field of view limits the statistical accuracy of surface brightness profiles. To overcome this problem, \citet{Vikhlinin:2006} combined ROSAT-PSPC and $Chandra$ data in cases where the PSPC observations were sufficiently deep (see Table~\ref{tab:clusters}, adapted from \citealt{Vikhlinin:2006}). The main goals of the analysis were to extract accurate temperature profiles and surface brightness profiles out to a large fraction of the virial radius. Spectral analysis to obtain temperature profiles is presented in \citet{Vikhlinin:2005}, and details of the X-ray surface brightness profiles is presented in \citet{Vikhlinin:2006}.

The observed mass profiles are inferred from temperature ($T$) and gas density ($\rho_g$) measurements assuming spherical symmetry and hydrostatic equilibrium \citep{Sarazin:1988}:
\begin{equation}
M(r) \;=\; 
-\dfrac{k_B T(r)r}{\mu m_p \GN}\left(\dfrac{d \ln \rho_g(r)}{d \ln r} + \dfrac{d \ln T(r)}{d \ln r}\right),
\end{equation}
where $k_B$ is Boltzmann's constant, $m_p$ is the mass of a proton, and $\mu\approx 0.6$ is the mean molecular weight for a plasma with primordial abundances.

The vast majority of the baryons in galaxy clusters are contained in a hot, diffuse gas ($T>10^6$K and densities $\sim 0.3$~particles per cubic centimeter). In such conditions, the primary emission process is due to thermal \textit{bremsstrahlung}. Recombination of electrons with heavier elements also contributes to the X-ray spectrum allowing for meaurements of the ionization and chemcial abundances of the plasma. The traditional model for the gas density in galaxy clusters is given by \citep{Cavaliere:1978}
\begin{equation}
n_e n_p \;=\; \dfrac{n_0^2}{\left[\, 1+\left(r/r_c\right)^2 \,\right]^{3 \beta}}\;.
\label{betamodel}
\end{equation}
To better fit the observed surface brightnesses of galaxy clusters in our sample, the gas density is modified to be
\begin{eqnarray}
n_e n_p 
& = & 
\dfrac{n_0^2 \left(r/r_c \right)^{-\alpha}}
{\bigl[\,1+\left(r/r_c\right)^2\,\bigr]^{3 \beta - \alpha /2} 
 \bigl[\,1+\left(r/r_s\right)^\gamma\,\bigr]^{\varepsilon / \gamma}
} 
\cr
& & 
\;+\; \dfrac{n_{02}^2}{\bigl[\,{1+\left(r/r_{c2}\right)^2}\,\bigr]^{3 \beta_2}}
\;.
\label{nenp9param}
\end{eqnarray}
These modifications to the traditional model provide a cusp in the center of the galaxy cluster ($r<r_c$), a steeper X-ray brightness profile at large radii ($r>r_s>r_c$), and extra modeling freedom near the centers of clusters ($r<r_{c2}<r_c$) \citep{Vikhlinin:2006}. The gas mass density is then given by
\begin{equation}
\rho_g \;=\; 1.24\, m_p \sqrt{n_e n_p}\;,
\end{equation}
for the plasma with primordial helium abundance and a metalicity of $0.2 Z_\odot$ for heavier elements.

Temperature profiles in the inner regions of galaxy clusters differ from those in the outer regions, and these regions must be modeled separately. Toward the cluster center, temperatures decline. 
Temperatures in the cooling region are described by \citep{Allen:2001,Sanders:2002,Burns:2010,Moretti:2011}
\begin{equation}
t_\mathrm{cool}(r) \;=\; 
\dfrac{\left(r/r_\mathrm{cool}\right)^{a_\mathrm{cool}} + T_{\min}/T_0}
      {1+\left(r/r_\mathrm{cool}\right)^{a_\mathrm{cool}}}
\;.
\end{equation}
Outside of the cooling region, the temperature profile can be modeled as a broken power law,
\begin{equation}
t(r) \;=\; \dfrac{(r/r_t)^{-a}}{\bigl[\, 1+ (r/r_t)^b \,\bigr]^{c/b}}\;.
\end{equation}
The three-dimensional temperature profile is described by the product of these two models \citep{Vikhlinin:2006},
\begin{equation}
T_\mathrm{3D}(r) \;=\; T_0\, t_\mathrm{cool}(r)\, t(r)\;.
\end{equation}

%%%%%%%%%%%%%%%%%%%%%%%%%%%%%%%%%%%%%%%%%%%%%%%%%%
\subsection{Data fits of mass profiles}

The dark matter mass profile predicted by MDM is given by Equation~(\ref{clustermassp}). The total mass, the sum of dark matter and baryonic matter, is compared with the virial mass for each galaxy cluster. The data fits are depicted in Figure~\ref{fig:plots1} with results provided in Table~\ref{tab:clusters}. 
Based on the analysis of \citet{Vikhlinin:2006}, we have assigned a 1-$\sigma$ error of 10\% on the virial mass determinations indicated by the shaded regions in the figure. While function parameter errors were not reported, inspection of their figures 3 -- 15 indicate that 10\% is a reasonable estimate. They also mention in the text that Monte Carlo simulations suggest statistical errors in gas density determinations of $\lesssim 9\%$.

For comparison, we include dynamical masses predicted by CDM and MOND. For CDM, we use the 
\citet{Navarro:1996} (NFW) density profile,
\begin{equation}
\rho_\mathrm{NFW}(r) \;=\; \dfrac{\rho_0}{\dfrac{r}{r_\textrm{CDM}}\left(1+\dfrac{r}{r_\textrm{CDM}}\right)^2}
\;,
\end{equation}
to determine the mass predicted by CDM, where
\begin{equation}
r_\textrm{CDM} \;=\; \dfrac{r_{200}}{c}\;,
\label{rsr200}
\end{equation}
and $r_{200}$ designates the edge of the halo, within which objects are assumed to be virialized, usually taken to be the boundary at which the halo density exceeds 200 times that of the background. The parameter $c$  (not to be confused with the speed of light) is a dimensionless number that indicates how centrally concentrated the halo is.

For MOND, we have from Eq.~(\ref{fdef})
\begin{equation}
\dfrac{M'(r)}{M(r)} \;=\; f_\mathrm{MOND}(a(r)/a_c) \;=\; \dfrac{1}{\mu(a(r)/a_c)}-1\;,
\end{equation}
where $M(r)$ and $M'(r)$ are respectively the total baryonic and dark matter mass enclosed within
a sphere of radius $r$ around the center of the cluster.
Assuming a spherically symmetric distribution, the dark matter density profile is then
\begin{equation}
\rho_\mathrm{MOND}(r) \;=\; \dfrac{1}{4\pi r^2}\dfrac{d}{dr}M'(r)\;.
\end{equation}
This dark matter profile $\rho_\mathrm{MOND}(r)$ will reproduce the 
acceleration profile $a(r)$ for a given MOND interpolating function $\mu(x)$.
We use the interpolating function given in Eq.~(\ref{muN}) with $n=1$ \citep{Famaey:2012},
instead of the $n=2$ case orignially proposed by 
\citet{Milgrom:1983a,Milgrom:1983b,Milgrom:1983c}
and used by \citet{Sanders:1999,Sanders:2003} to fit cluster data,
since it seems to provide better fits.

We see in Figure~\ref{fig:plots1} that the MDM mass profiles fit the virial mass data well. The fits for MDM mass profiles are as good as those for NFW. The MOND (effective) mass profiles fail to reproduce the virial mass profiles both in magnitude {\em and in shape}; The mass discrepancy in the MOND model is more significant in the inner regions of the cluster than in the outer regions. 

%%%%%%%%%%%%%%%%%%%%%%%%%%%%%%%%%%%%%%%%%%%%%%%%%%
\begin{table*}
\caption{%
Sample of galaxy clusters with redshifts as given in \citet{Vikhlinin:2006}. For each cluster, the baryonic, MDM (using $\alpha = 100$) and CDM masses contained within the observed radii are given in columns 3 -- 5, respectively. The scale radius $r_\textrm{MDM}$ for MDM is given in column 6 and columns 7 and 8 contain the fitting parameters (concentration and scale radius) for the NFW mass profile. We note that MOND has no mass term other than the baryonic mass and uses no free fitting parameters.}
\begin{tabular}{lccccccc}
\hline
Name
& z
& $M_\mathrm{B}$
& $M_\mathrm{MDM}$
& $M_\mathrm{CDM}$
& $r_\textrm{MDM}$
& $c$
& $r_\textrm{CDM}$ \\
&
&
($10^{12} M_\odot$) 
& ($10^{12} M_\odot$) 
& ($10^{12} M_\odot$)
& (kpc)
&
& (kpc) \\
\hline
 A133                   & 0.0569 & \phantom{0}5.5\phantom{0} & \phantom{0}12.9  & \phantom{0}17.1  & \phantom{0}0.818 & 3.75 & 200 \\
 A262                   & 0.0162 & \phantom{0}3.2\phantom{0} & \phantom{0}38.6  & \phantom{0}34.9  & 10.3\phantom{00} & 6.00 & 167 \\  
 A383                   & 0.1883 & 11\phantom{.00} & \phantom{0}87.5  & \phantom{0}93.5  & 17.9\phantom{00} & 6.81 & 241 \\
 A478                   & 0.0881 & 18\phantom{.00} & 106\phantom{.0}  & 110\phantom{.0}  & 16.3\phantom{00} & 6.82 & 263 \\
 A907                   & 0.1603 & 11\phantom{.00} & 106\phantom{.0}  & 103\phantom{.0}  & 25.5\phantom{00} &7.48  & 225 \\
 A1413                 & 0.1429 & 13\phantom{.00} & 126\phantom{.0}  & 116\phantom{.0}  & 31.7\phantom{00} & 7.85 & 226 \\ 
 A1795                 & 0.0622 & 12\phantom{.00} & \phantom{0}85.4  & \phantom{0}98.9  & 15.5\phantom{00} & 6.14 & 284 \\
 A1991                 & 0.0592 & \phantom{0}4.3\phantom{0}  & \phantom{0}54.4  & \phantom{0}54.1  & 15.6\phantom{00} & 6.86 & 178 \\
 A2029                 & 0.0779 & 16\phantom{.00} & 147\phantom{.0}  & 143\phantom{.0}  & 39.7\phantom{00} & 8.39 & 231 \\
 A2390                 & 0.2302 & 19\phantom{.00} & 116\phantom{.0}  & 111\phantom{.0}  & 18.5\phantom{00} & 8.02 & 214 \\
 RX J1159+5531  & 0.0810 & 41.3\phantom{0} & \phantom{0}17.6  & \phantom{0}38.0  & 19.0\phantom{00} & 6.18 & 169 \\
 MKW 4               & 0.0199 & \phantom{0}2.0\phantom{0}   & \phantom{0}36.3  & \phantom{0}33.9  & 14.4\phantom{00}  & 6.04 & 164 \\
 USGC S152       & 0.0153 & \phantom{0}0.95 & \phantom{0}14.3  & \phantom{0}13.5  & \phantom{0}4.42\phantom{0} & 5.36 & 119 \\
\hline
\end{tabular}
\label{tab:clusters}
\end{table*}
%%%%%%%%%%%%%%%%%%%%%%%%%%%%%%%%%%%%%%%%%%%%%%%%%%

%%%%%%%%%%%%%%%%%%%%%%%%%%%%%%%%%%%%%%%%%%%%%%%%%%
\begin{figure}
\includegraphics[angle=90,width=0.45\textwidth]{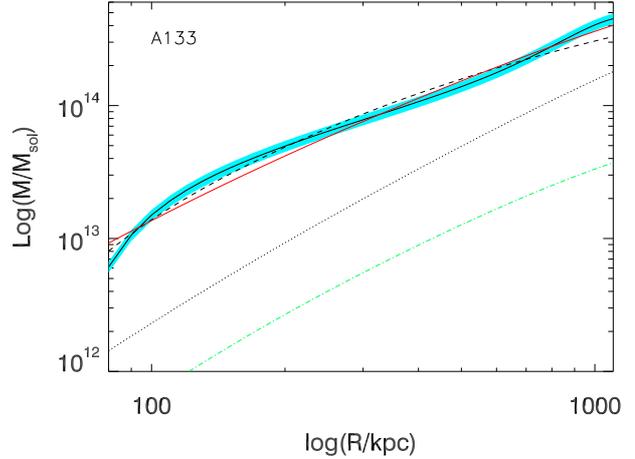}\vspace{5mm}\\
\caption{Plots of total mass within radius R (assuming spherical symmetry). The solid black line is the virial mass; The shaded region surrounding the virial mass plot represents the 1-$\sigma$ error, estimated at 10\% from the analysis of \citet{Vikhlinin:2006}; The dot-dashed green line is gas mass; The dotted black line is MOND (effective mass); The dashed black line is CDM; The solid red line is MDM with $\alpha = 100$.}
\label{fig:plots1}
\end{figure}

\begin{figure*}
\includegraphics[angle=90,width=0.45\textwidth]{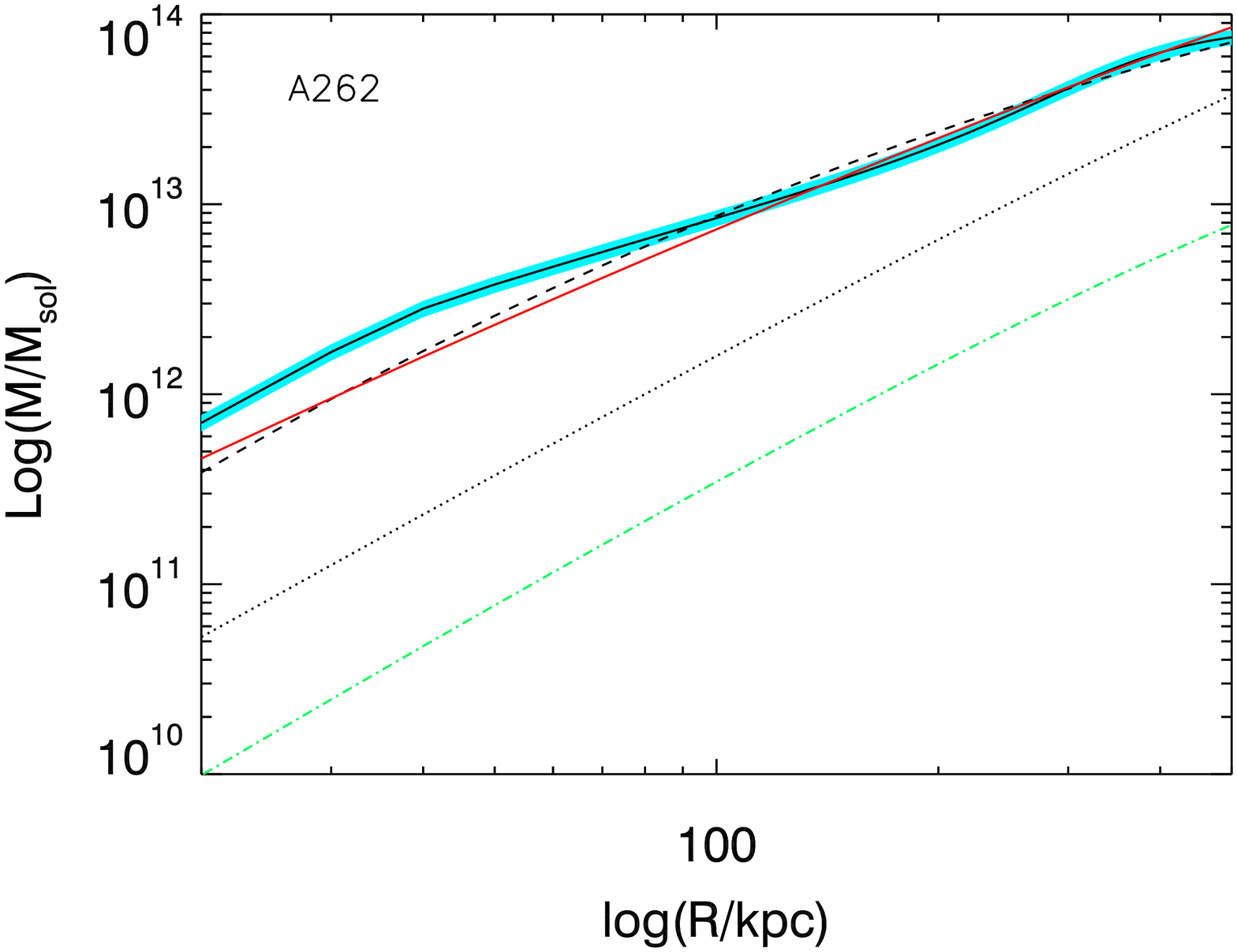}
\includegraphics[angle=90,width=0.45\textwidth]{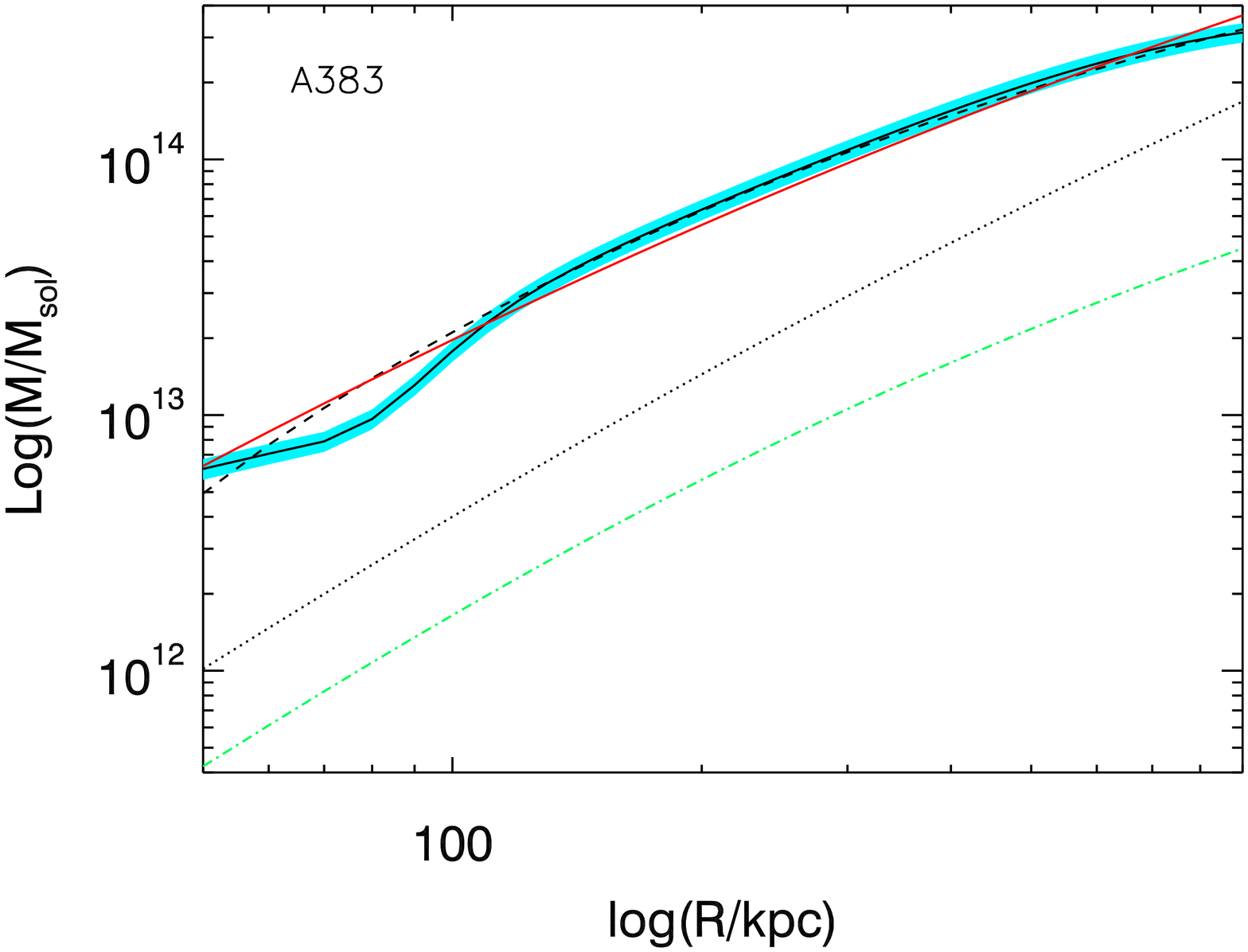}\vspace{5mm}\\
\includegraphics[angle=90,width=0.45\textwidth]{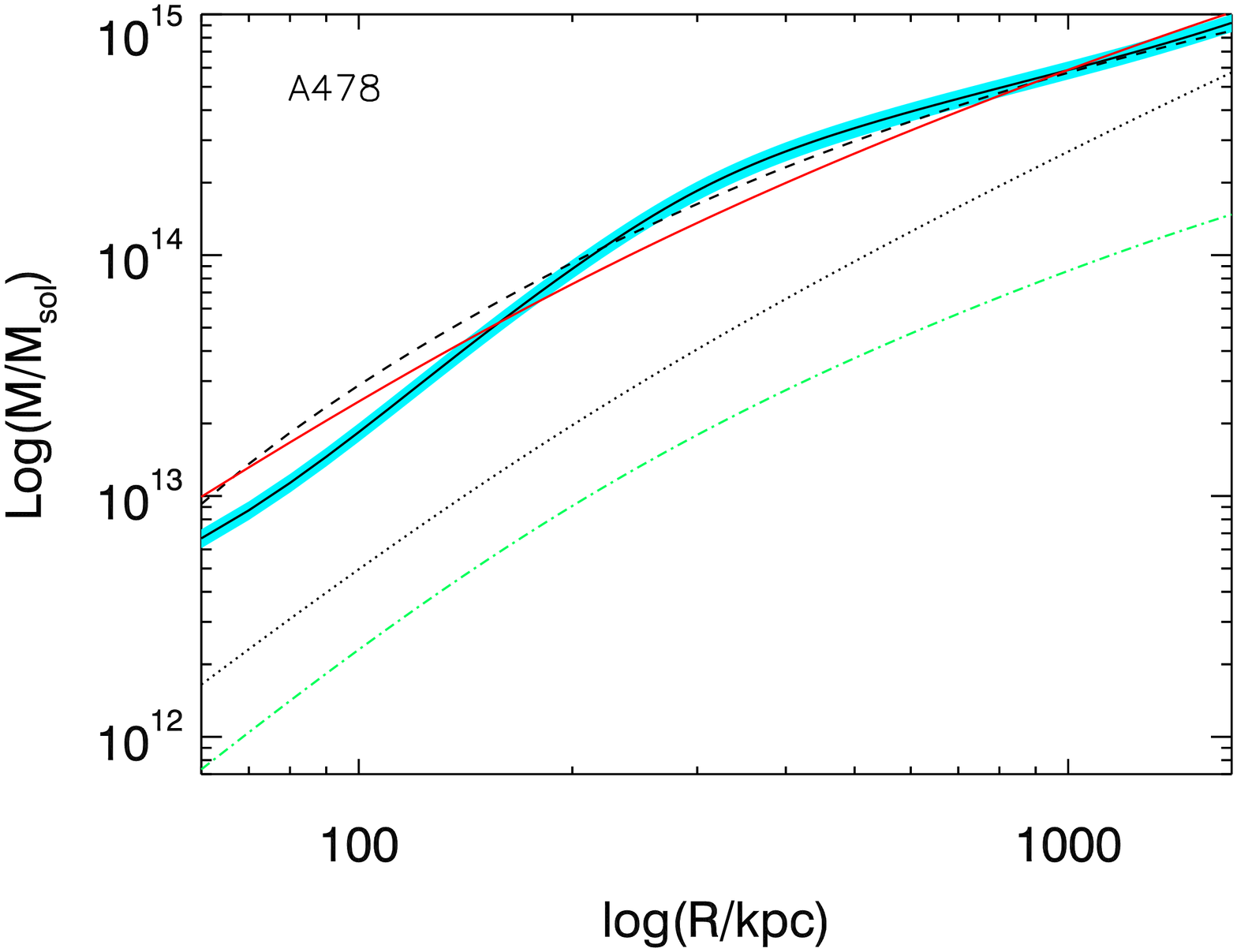}
\includegraphics[angle=90,width=0.45\textwidth]{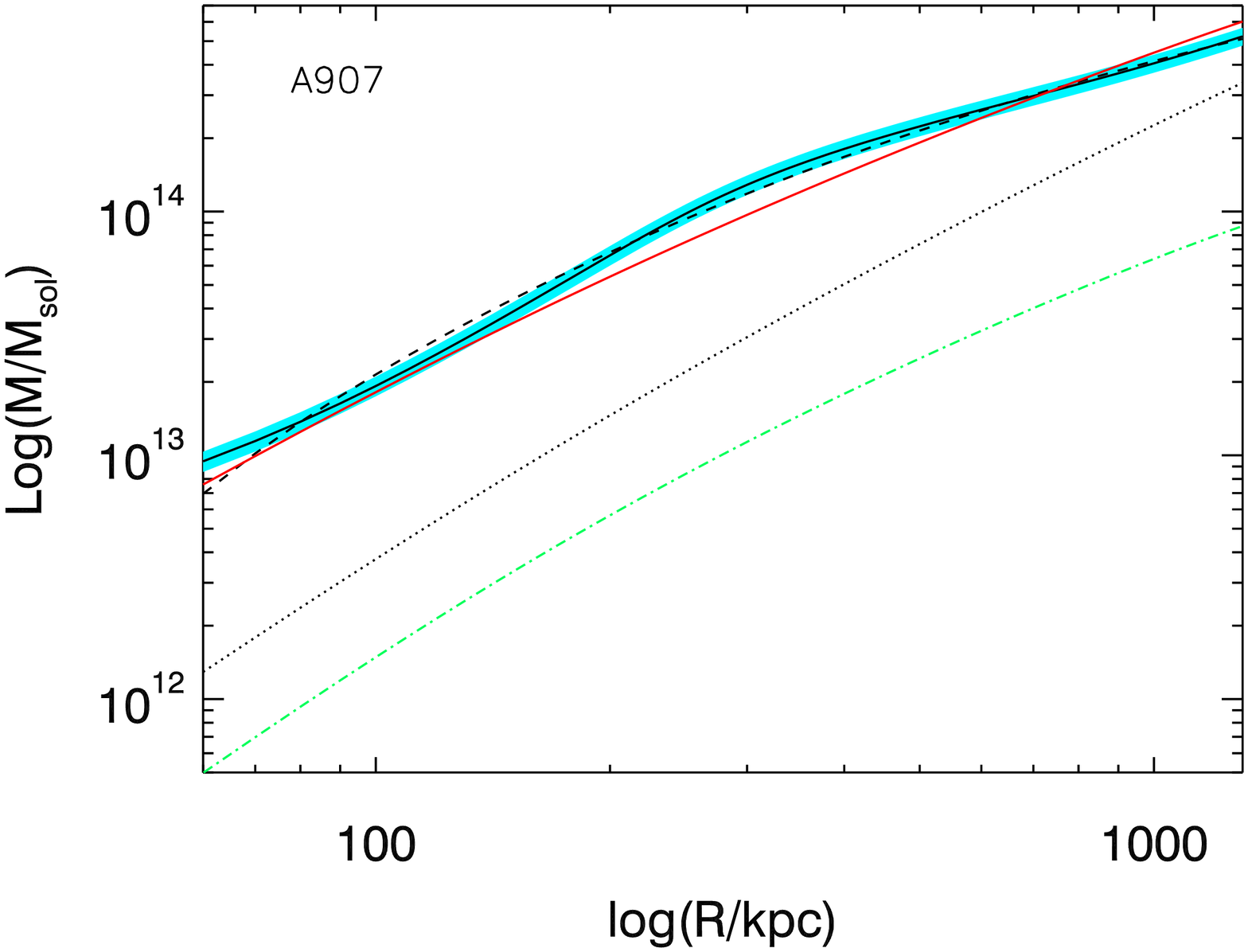}\vspace{5mm}\\
\contcaption{}
\end{figure*}

\begin{figure*}
%\begin{flushleft}
%\hspace{0.5cm}
\includegraphics[angle=90,width=0.45\textwidth]{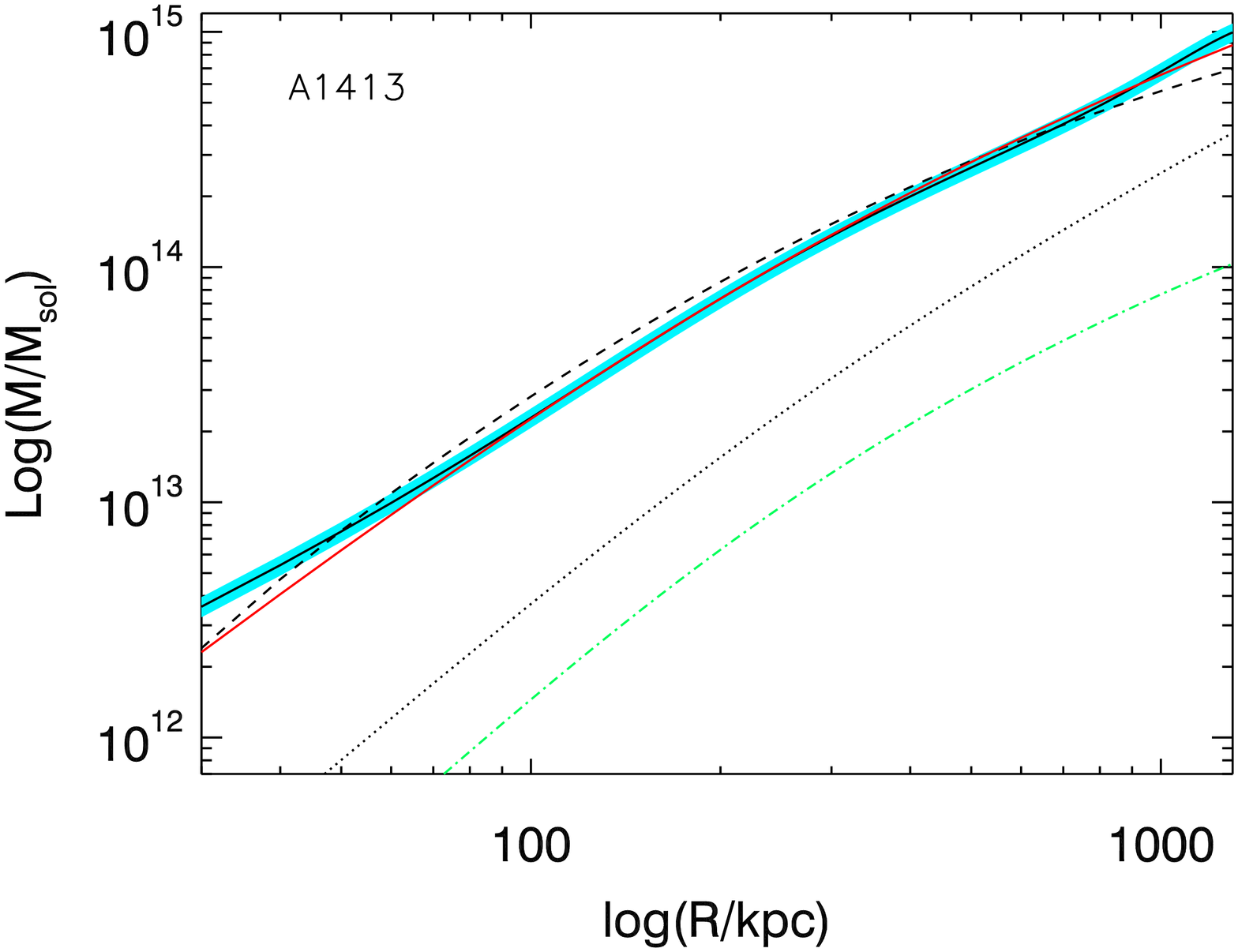}
\includegraphics[angle=90,width=0.45\textwidth]{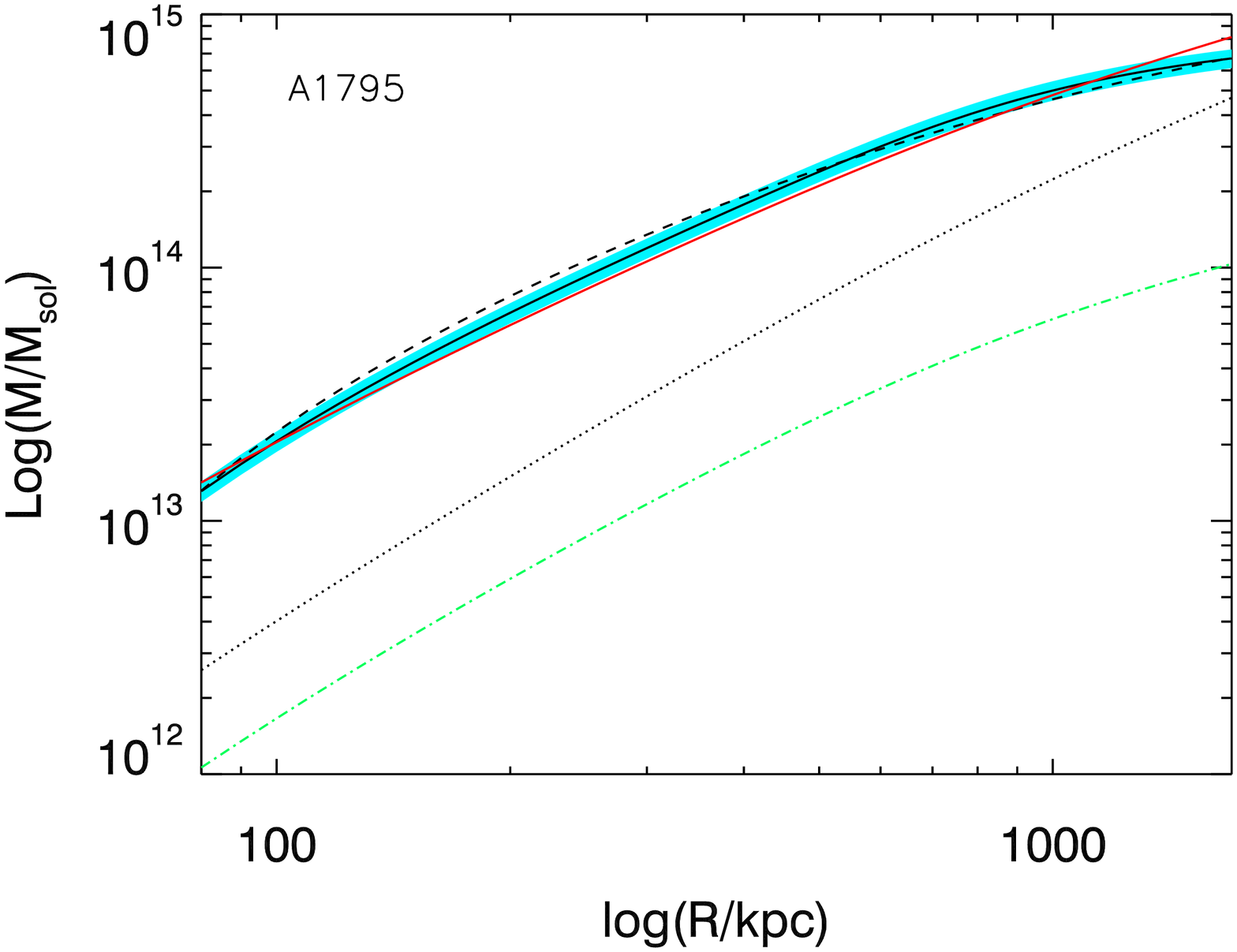}\vspace{5mm}\\
\includegraphics[angle=90,width=0.45\textwidth]{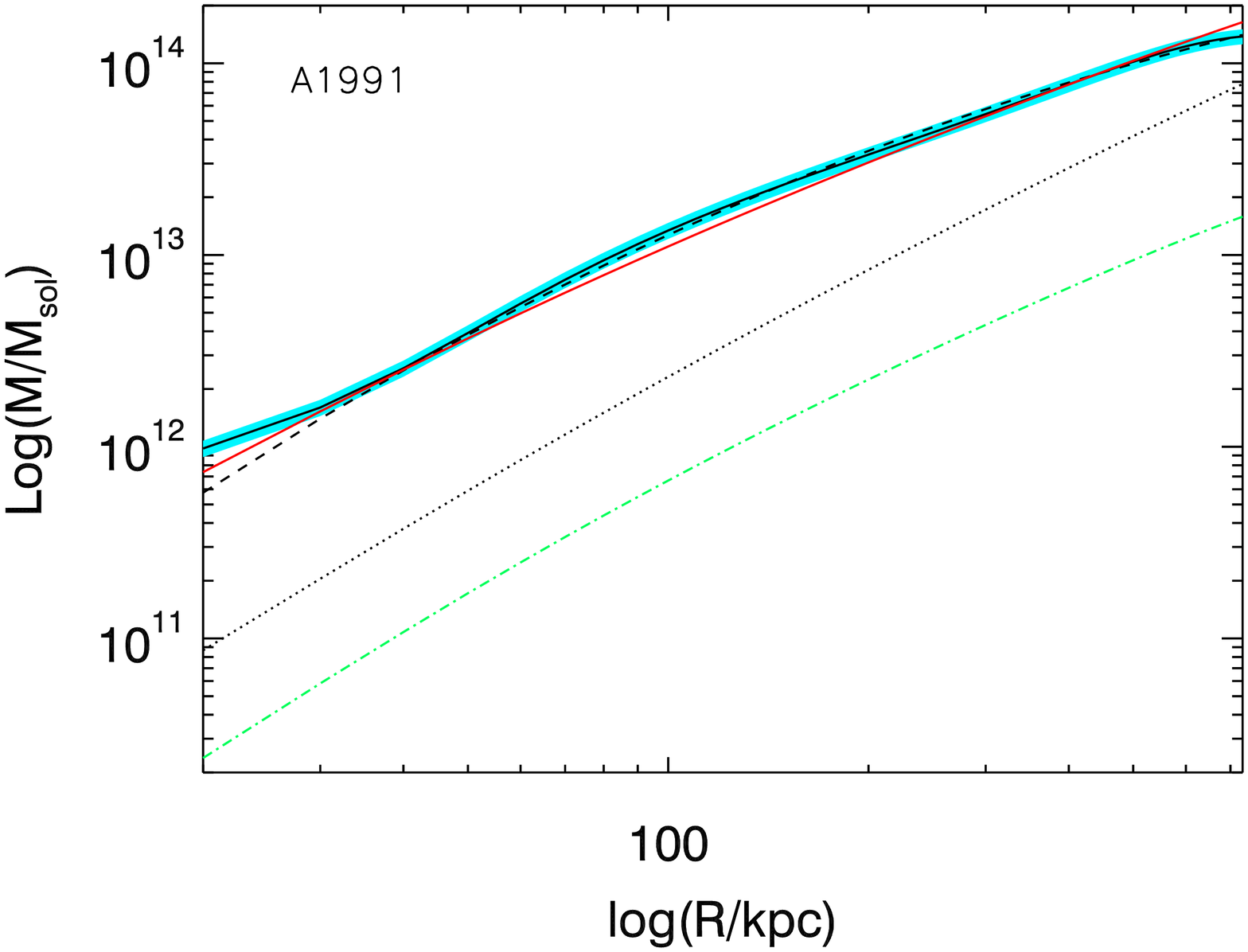}
\includegraphics[angle=90,width=0.45\textwidth]{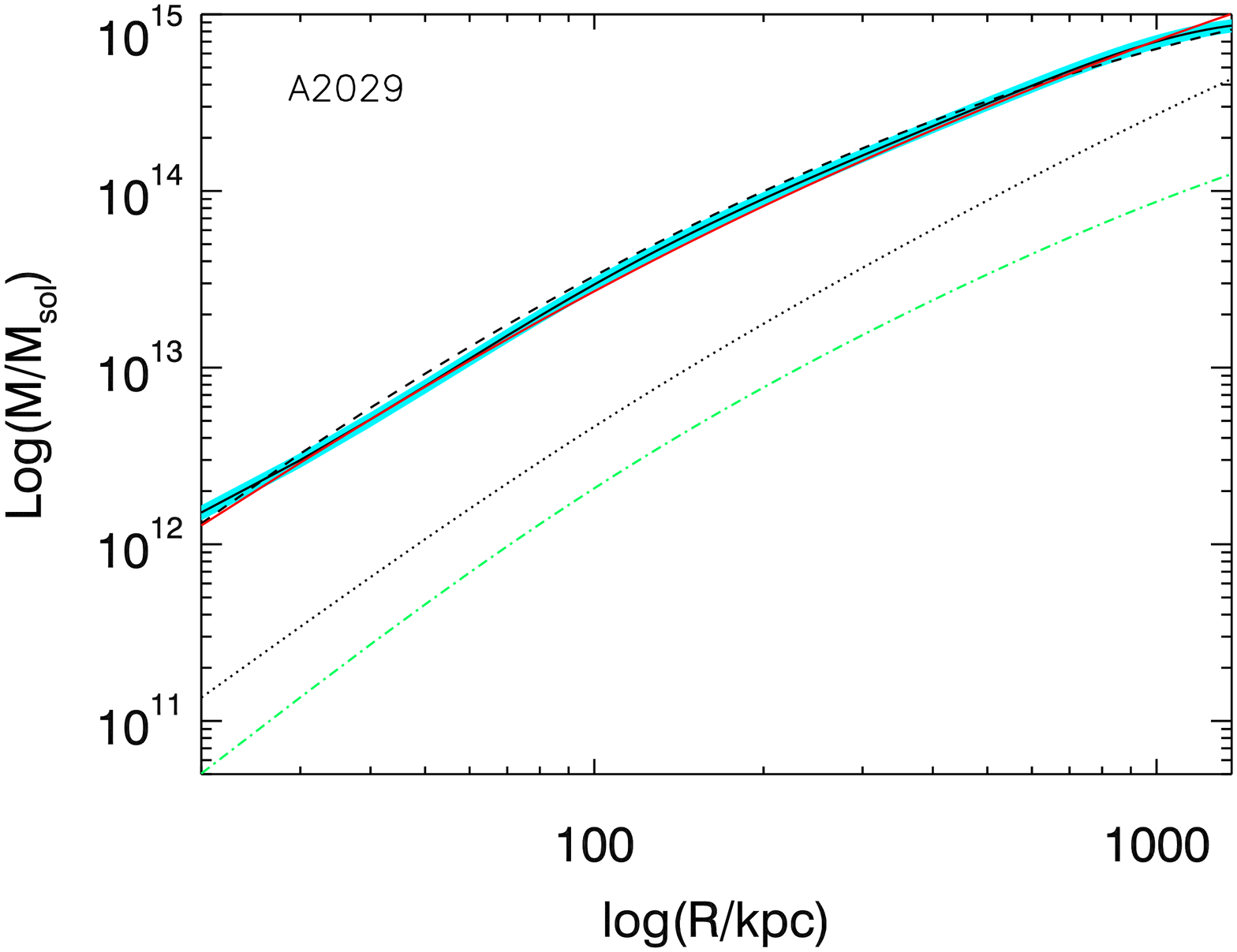}\vspace{5mm}\\
\includegraphics[angle=90,width=0.45\textwidth]{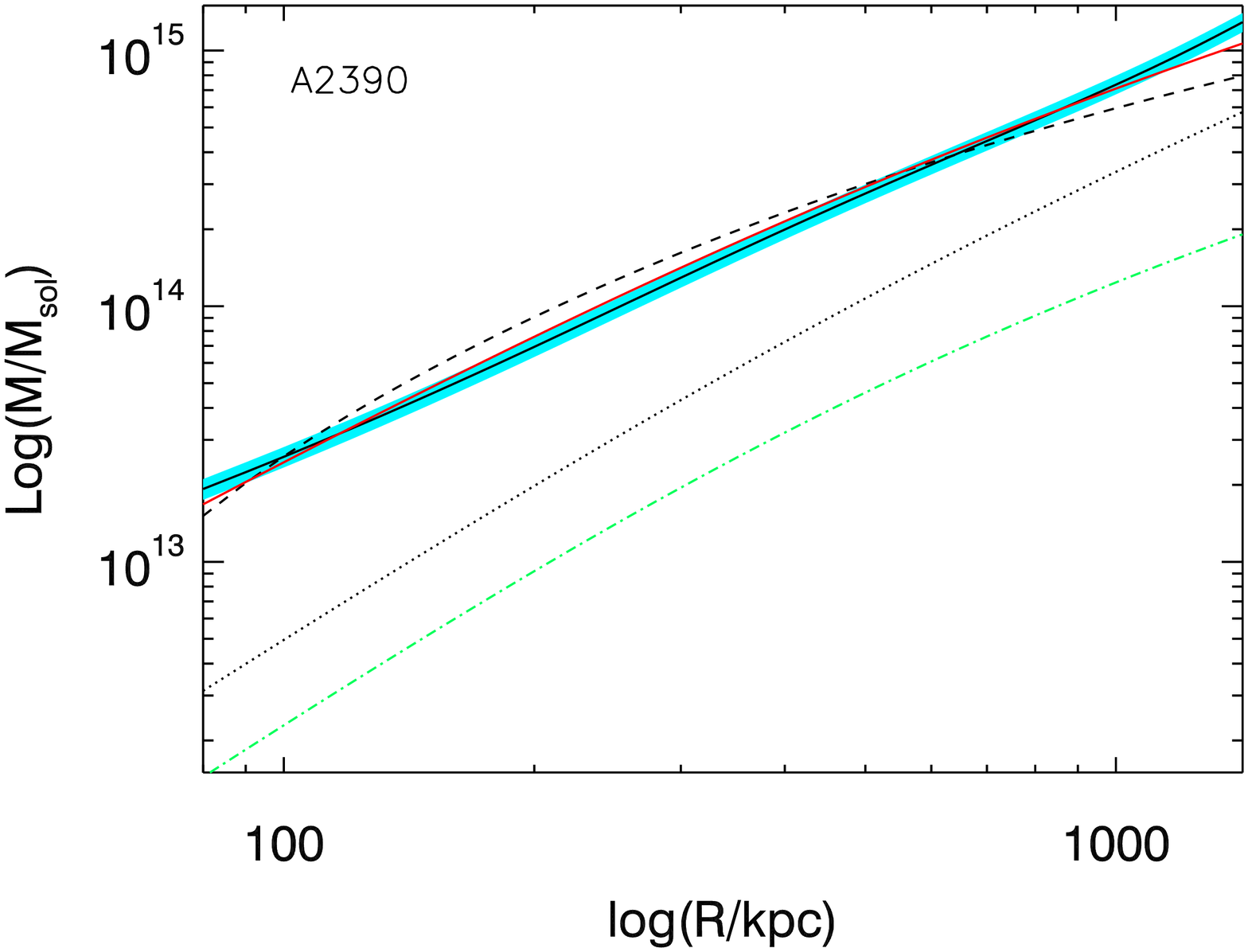}
\includegraphics[angle=90,width=0.45\textwidth]{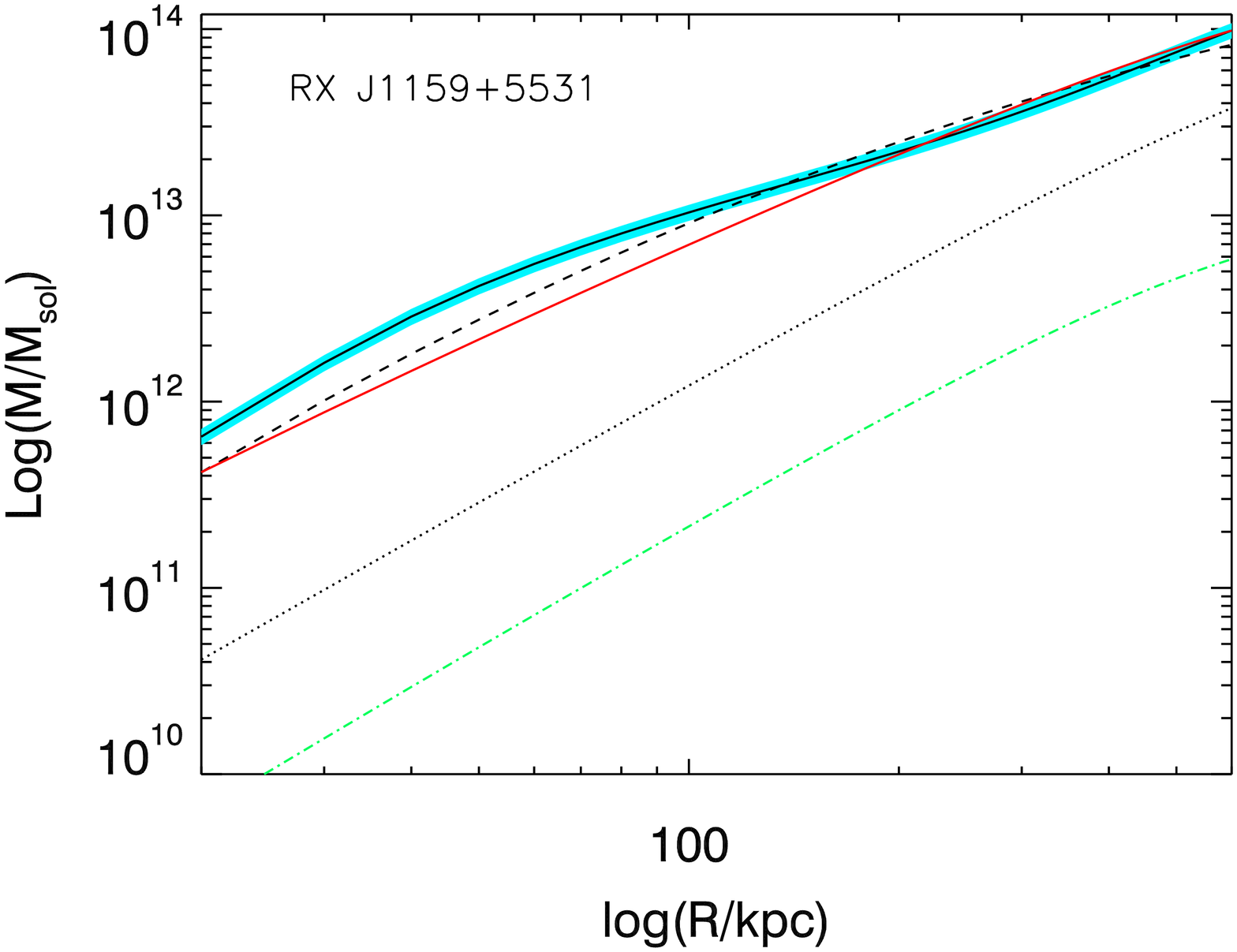}\vspace{5mm}\\
%\hspace{0.5cm}
%\end{flushleft}
\contcaption{}
\end{figure*}

\begin{figure*}
\includegraphics[angle=90,width=0.45\textwidth]{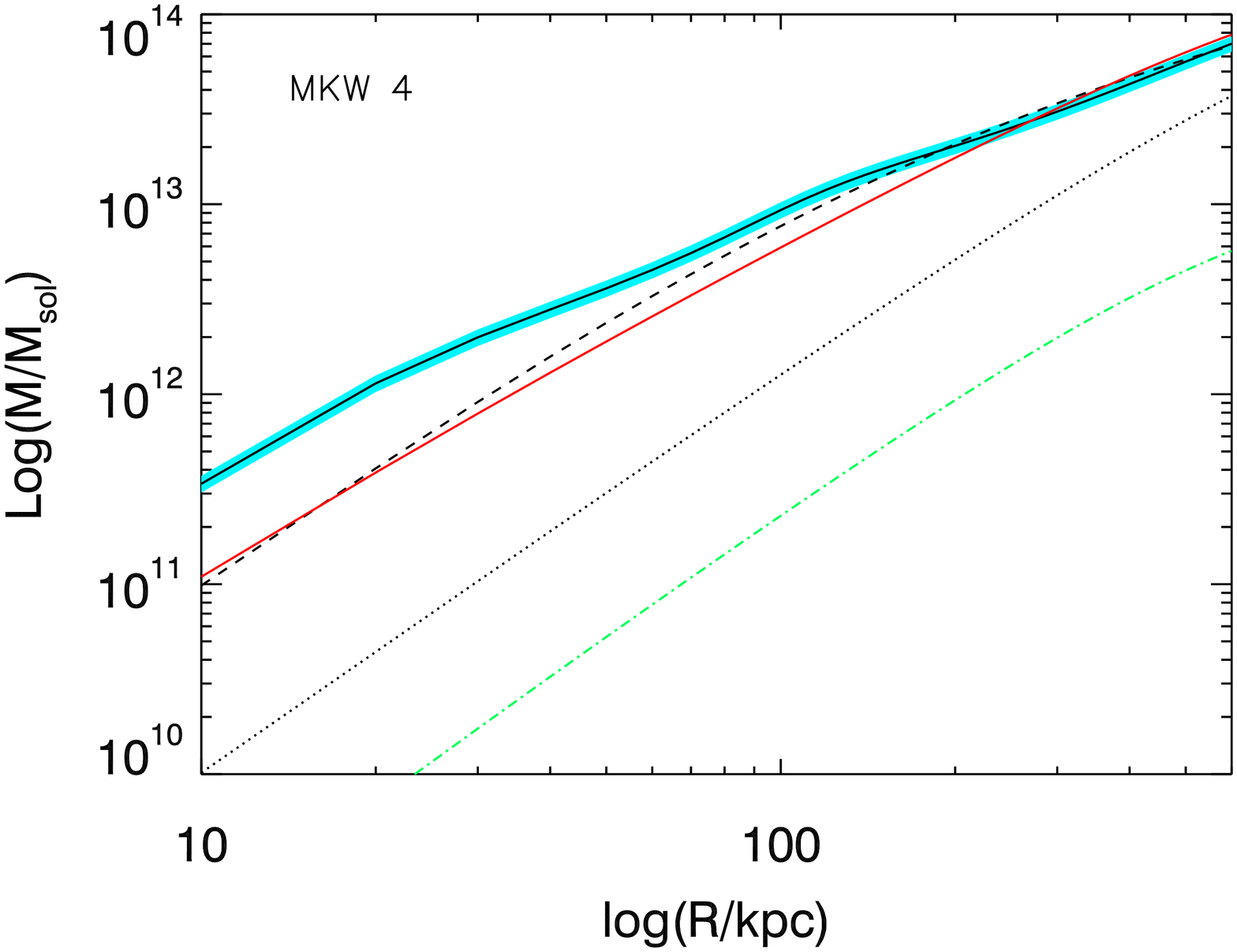}
\includegraphics[angle=90,width=0.45\textwidth]{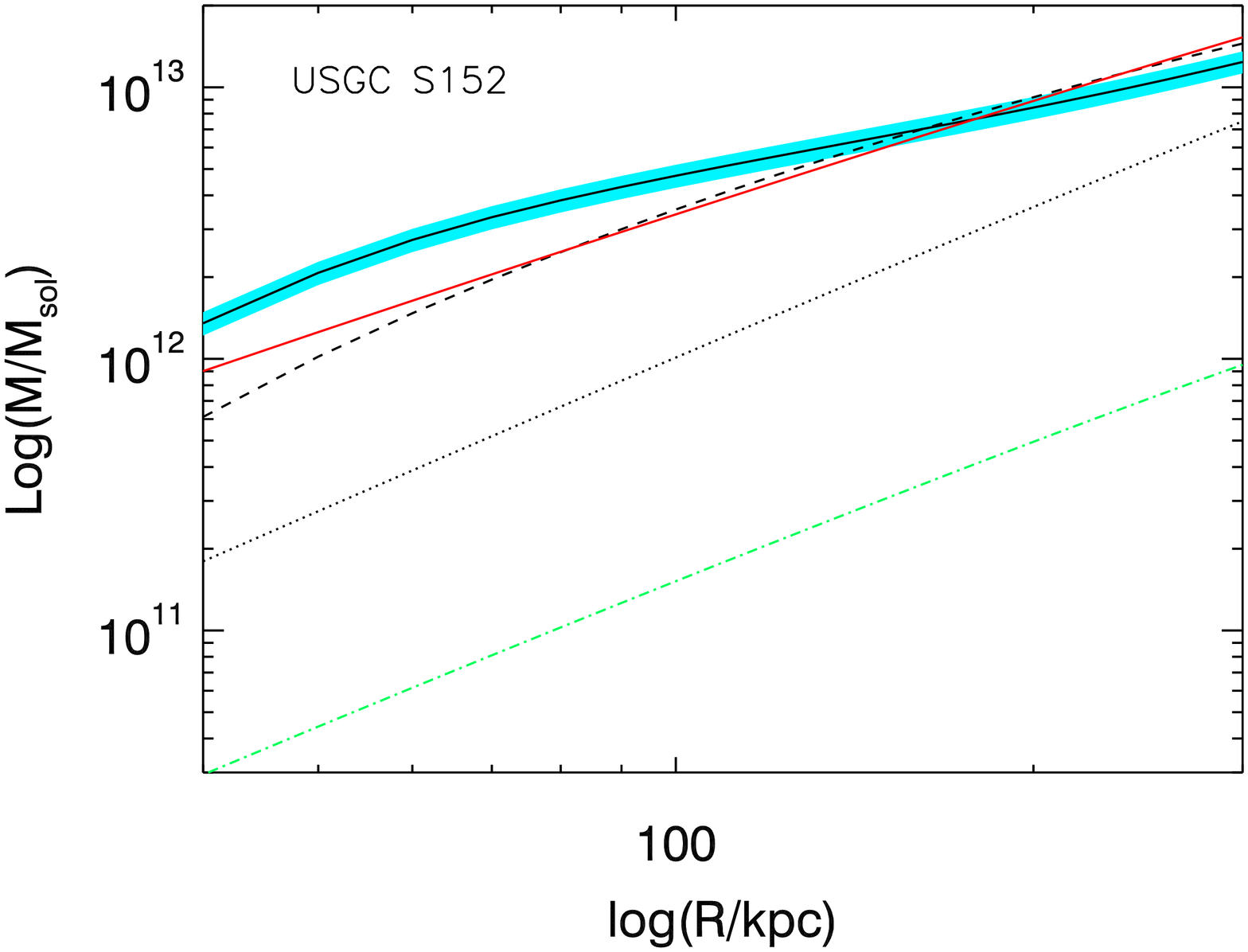}\vspace{5mm}\\
\contcaption{}
\end{figure*}
%%%%%%%%%%%%%%%%%%%%%%%%%%%%%%%%%%%%%%%%%%%%%%%%%%

%\clearpage
%\clearpage

%%%%%%%%%%%%%%%%%%%%%%%%%%%%%%%%%%%%%%%%%%%%%%%%%%
\section{Discussion}

In this paper we have presented further observational tests of Modified Dark Matter (MDM) by 
fitting the MDM mass profile to 13 low-redshift, relaxed X-ray-emitting clusters of galaxies.
We have shown that MDM performs as well as CDM, and considerably better than MOND (which only works well at the galactic scale), for all clusters in our sample. 
Moreover, using arguments from gravitational thermodynamics, we have justified, on theoretical 
grounds, our MDM mass profile.
Perhaps the most interesting aspect of our work is the dependence of the CDM mass profile on the
cosmological constant. We suggest that the CDM mass profiles can be reorganized in a form suggested
by the MDM mass profile. Such reorganization may constrain the nature of cold dark matter.

In \citet{Edmonds:2014}, we fit the MDM mass profile in Equation~(\ref{fMDM1}) to galactic rotation curves which differs from the one used for galaxy clusters in this paper by a function of radius (see Equation~\ref{clustermassp}). In the appendix, we show that the generalized mass profile in Equation~(\ref{clustermassp}) also works well for galactic rotation curves, albeit with a different constant scale value; $\alpha \sim 100$ for galaxy clusters and $\alpha \sim 1$ for galactic rotation curves. We do not yet know the underlying reason for the change in $\alpha$, though, as discussed in section~3, $\alpha$ may be related to boundary conditions  and can, in principle, scale with radius. Perhaps studying objects of similar scale, but under different physical conditions will be illuminating. The galaxy clusters in our sample were all low-redshift, virialized clusters. It would be interesting to constrain MDM mass profiles in clusters at higher redshifts, before they become virialized, assuming we can get reliable mass estimates in the non-virialized regions.

Along with the current work, we plan to study the constraints from gravitational lensing 
and colliding clusters such as the Bullet Cluster (1E 0657-558) and the Train Wreck Cluster (A520) on MDM. We will also test MDM at cosmic scales by studying the acoustic peaks in the CMB. 
In all of these follow-up studies, we will try to identify some distinctive observational signatures between MDM and CDM models.

Last but not least, it will be interesting and important to have a deeper understanding of the 
fundamental nature of the MDM quanta.  We would like to construct explicit models of
MDM quanta along the lines indicated by the logic of gravitational thermodynamics.
The deeper nature of such quanta should be found in the context of quantum gravity.
For example, such unusual non-local quanta could be found in recent
reformulations of quantum gravity and string theory \citep{FLM:2013,FLM:2014,FLM:2015}.
A more concrete theory for MDM quanta will allow us to test this idea 
at colliders, dark matter direct detection experiments and indirect detection experiments. For instance, we have speculated 
that the MDM quanta obey infinite statistics rather than the familiar Bose or Fermi statistics 
\citep{Ho:2012,Greenberg:1990,Shevchenko:2008}; This may lead to unusual particle phenomenology.

%%%%%%%%%%%%%%%%%%%%%%%%%%%%%%%%%%%%%%%%%%%%%%%%%%
\section*{Acknowledgments}

We are grateful for helpful discussions with J.~Beacom, C.~Frenk, S.~Horiuchi, J.~Khoury, P.~Mannheim and J.~Moffat. We are also grateful to the organizers of the Miami winter conference for providing an inspiring environment for work.
CMH was supported in part by the Office of the Vice-President for Research and Graduate Studies at Michigan State University. 
YJN was supported in part by the Bahnson Fund of UNC.
DM was supported by the U.S. Department of Energy under contract DE-FG02-13ER41917, Task A.

%%%%%%%%%%%%%%%%%%%%%%%%%%%%%%%%%%%%%%%%%%%%%%%%%%

%\newpage
\appendix
\section{Galactic Rotation Curves}

In \citet{Edmonds:2014}, we fit a sample of galactic rotation curves using the MDM mass profile
Equation~(\ref{fMDM1}).
%
%\begin{equation}
%M' = \frac{M}{\pi} \left( \dfrac{a_0}{a} \right)^2.
%\end{equation}
%
While this profile works well for galaxies, it does not fit the galaxy cluster data well. In this paper, we introduced a mass profile that works well for galaxy clusters (Equation \ref{clustermassp}), and we would like to see if the new mass profile can work at galactic scales as well as it does at cluster scales.

For the galaxy clusters in our sample, we found $\alpha \sim 100$ is required for the mass profile given in Equation~(\ref{clustermassp}). Since the fits to galactic rotation curves presented in \citet{Edmonds:2014} were fit so well with Equation~(\ref{fMDM1}), we expect $\alpha \sim 1$ with $r_\textrm{MDM} \gg r$ in the generalized mass profile. The mass-to-light ratio ($M/L$) is unknown, and is therefore determined by data fits. The solution space is not well-constrained if we allow $\alpha$ and $r_\textrm{MDM}$ to be fitting parameters in addition to $M\L$. Therefore, we fix the scale radius to be similar to those found for NFW halos; $r_\textrm{MDM} = 20$~kpc. It is interesting to note that this radius is near the observed turnover radius for the Milky Way \citep{Merrifield:1992,Battaner:2000}. However, the very extended rotation curve ($\sim 100$~kpc) observed in UGC~2885 first drops, but then rises again to remain flat beyond 50~kpc. If more data at very large radii becomes available, it may be possible to constrain the scale radius in dark matter mass profiles.

With the scale radus and the constant scale factor $\alpha$ fixed to 20~kpc and 0.5, respectively, we reproduce the plots from \citet{Edmonds:2014} using Equation \ref{clustermassp} in Figure~\ref{fig:grc}. In Table~\ref{tab:galaxies}, we provide M/L, the mass of the dark matter component, and the baryonic mass (inferred from M/L).

\begin{table*}
\caption{%
Sample of galaxies from \citet{Sanders:1996}. Asterisks denote galaxies with disturbed velocity fields \citep{Sanders:1996}. Low-Surface-Brightness galaxies (LSBs) are marked with an (L), and the rest are 
High-Surface-Brightness galaxies (HSBs). For each galaxy, the baryonic and MDM masses contained within the observed radii are given in columns 2 and 3, respectively. The mass-to-light ratios ($M/L$) determined by fitting the MDM mass profile is given in column 4. This table is similar to Table~1 in \citet{Edmonds:2014}, but with values corresponding to the generalized MDM mass profile developed in this paper using $r_\textrm{MDM} = 20$~kpc and $\alpha = 0.5$.
}
\begin{tabular}{lccc}
\hline
Name 
& $M_\mathrm{B}$
& $M_\mathrm{MDM}$
%& $M_\mathrm{CDM}$
& $M/L$ (MDM) \\
%& $R_s$ \\
%& $\alpha_\mathrm{CDM}$ 
& ($10^{10} M_\odot$) 
& ($10^{10} M_\odot$) 
%& ($10^{10} M_\odot$)
& ($M_\odot /L_\odot$) \\ 
%& (kpc) \\
\hline
 NGC 3726      & \phantom{0}2.94\phantom{00} & 15.8\phantom{00}           & 0.55   \\
 NGC 3769*     & \phantom{0}1.23\phantom{00} & 12.9\phantom{00}           & 0.44 \\
 NGC 3877      & \phantom{0}3.17\phantom{00} & \phantom{0}3.67\phantom{0} & 0.53   \\
 NGC 3893*     & \phantom{0}3.95\phantom{00} & \phantom{0}9.19\phantom{0} & 0.73   \\
 NGC 3917 (L)  & \phantom{0}1.50\phantom{00} & \phantom{0}4.04\phantom{0} & 0.73   \\
 NGC 3949      & \phantom{0}1.46\phantom{00} & \phantom{0}1.43\phantom{0} & 0.45  \\
 NGC 3953      & \phantom{0}8.55\phantom{00} & \phantom{0}8.17\phantom{0} & 0.74   \\
 NGC 3972      & \phantom{0}0.944\phantom{0} & \phantom{0}1.63\phantom{0} & 0.63  \\
 NGC 3992      & 15.0\phantom{000}           & 33.5\phantom{00}           & 1.9\phantom{0} \\
 NGC 4010 (L)  & \phantom{0}0.909\phantom{0} & \phantom{0}2.06\phantom{0} & 0.48   \\
 NGC 4013      & \phantom{0}4.44\phantom{00} & 17.6\phantom{00}           & 0.76  \\
 NGC 4051*     & \phantom{0}2.65\phantom{00} & \phantom{0}3.82\phantom{0} & 0.56   \\
 NGC 4085      & \phantom{0}0.832\phantom{0} & \phantom{0}0.982           & 0.50   \\
 NGC 4088*     & \phantom{0}3.72\phantom{00} & \phantom{0}9.73\phantom{0} & 0.43  \\
 NGC 4100      & \phantom{0}4.14\phantom{00} & 11.0\phantom{00}           & 0.95  \\
 NGC 4138      & \phantom{0}2.55\phantom{00} & \phantom{0}6.83\phantom{0} & 0.80   \\
 NGC 4157      & \phantom{0}5.04\phantom{00} & 17.3\phantom{00}           & 0.67   \\
 NGC 4183 (L)  & \phantom{0}0.946\phantom{0} & \phantom{0}5.28\phantom{0} & 0.59   \\
 NGC 4217      & \phantom{0}3.99\phantom{00} & \phantom{0}6.97\phantom{0} & 0.60   \\
 NGC 4389*     & \phantom{0}0.586\phantom{0} & \phantom{0}1.69\phantom{0} & 0.37   \\
 UGC 6399 (L)  & \phantom{0}0.257\phantom{0} & \phantom{0}0.832           & 0.60   \\
 UGC 6446 (L)  & \phantom{0}0.513\phantom{0} & \phantom{0}2.65\phantom{0} & 0.53  \\
 UGC 6667 (L)  & \phantom{0}0.235\phantom{0} & \phantom{0}0.801           & 0.53  \\
 UGC 6818* (L) & \phantom{0}0.0997           & \phantom{0}0.484           & 0.30   \\
 UGC 6917 (L)  & \phantom{0}0.720\phantom{0} & \phantom{0}1.87\phantom{0} & 0.83  \\
 UGC 6923 (L)  & \phantom{0}0.172\phantom{0} & \phantom{0}0.415           & 0.42  \\
 UGC 6930 (L)  & \phantom{0}0.771\phantom{0} & \phantom{0}3.52\phantom{0} & 0.70   \\
 UGC 6973*     & \phantom{0}1.61\phantom{00} & \phantom{0}1.72\phantom{0} & 0.48   \\
 UGC 6983 (L)  & \phantom{0}0.869\phantom{0} & \phantom{0}3.45\phantom{0} & 1.1\phantom{0}  \\
 UGC 7089 (L)  & \phantom{0}0.182\phantom{0} & \phantom{0}0.910           & 0.30   \\
\hline
\end{tabular}
\label{tab:galaxies}
\end{table*}

%%%%%%%%%%%%%%%%%%%%%%%%%%%%%%%%%%%%%%%%%%%%%%%%%%%%%%%%%%%%%%%%%%%%%%%%%%%%%%%%%%%%%%%%%%%%%%%%%%%%%%%%%%%%%%%%%%%%%%%%%%%
\begin{figure*}
  \includegraphics[angle=90,width=0.45\textwidth]{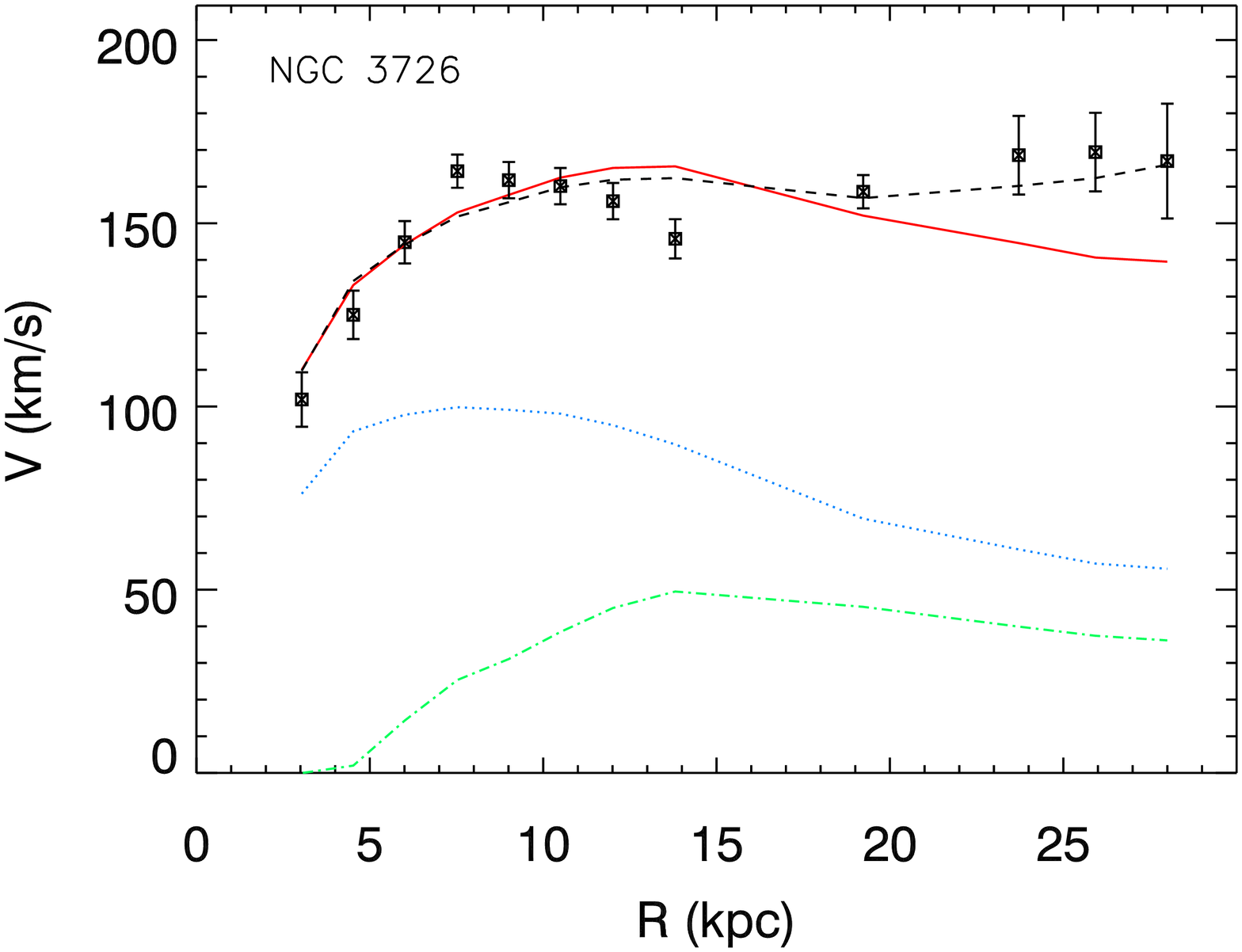}%\hspace{5mm}
  \includegraphics[angle=90,width=0.45\textwidth]{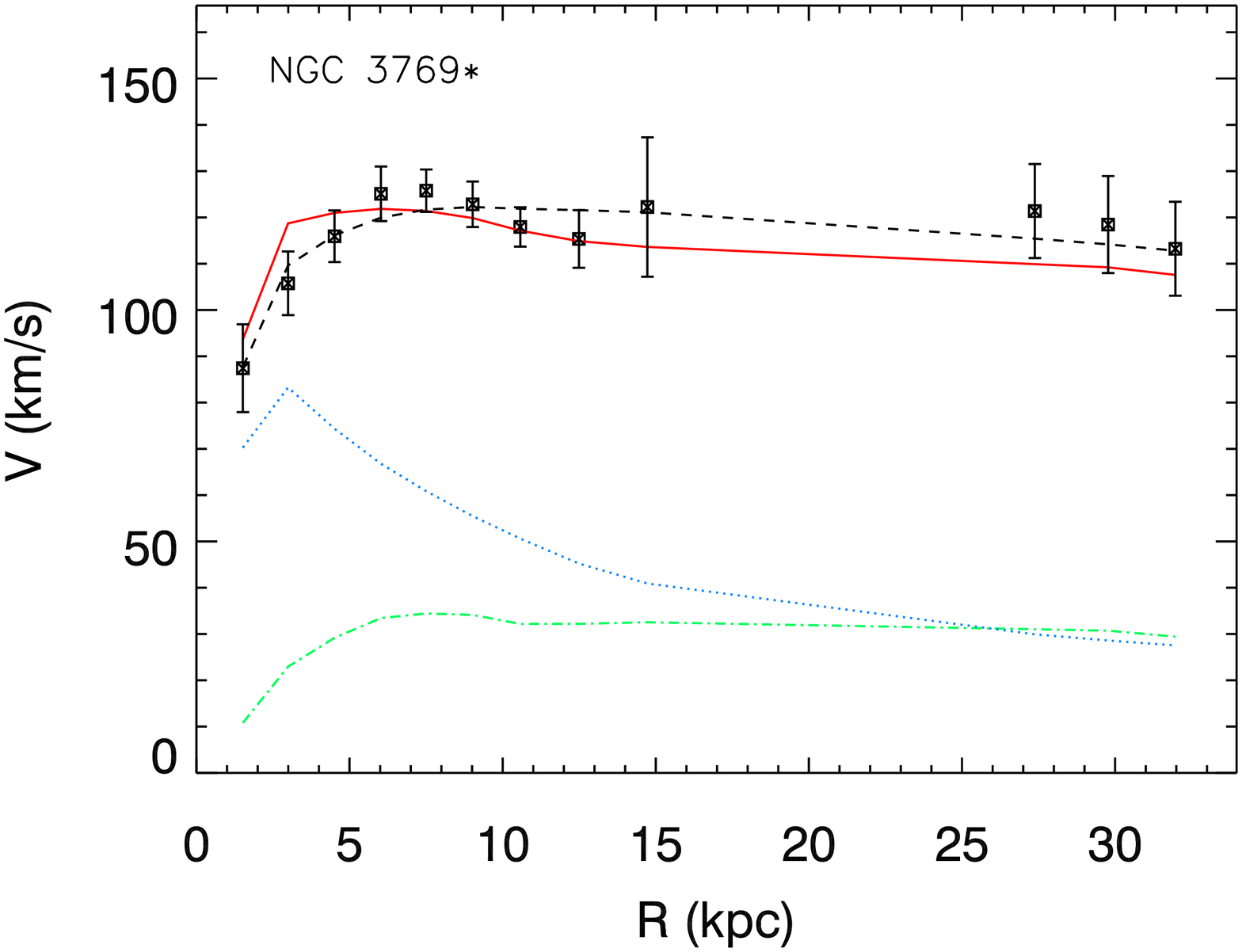}\vspace{5mm}\\
  \includegraphics[angle=90,width=0.45\textwidth]{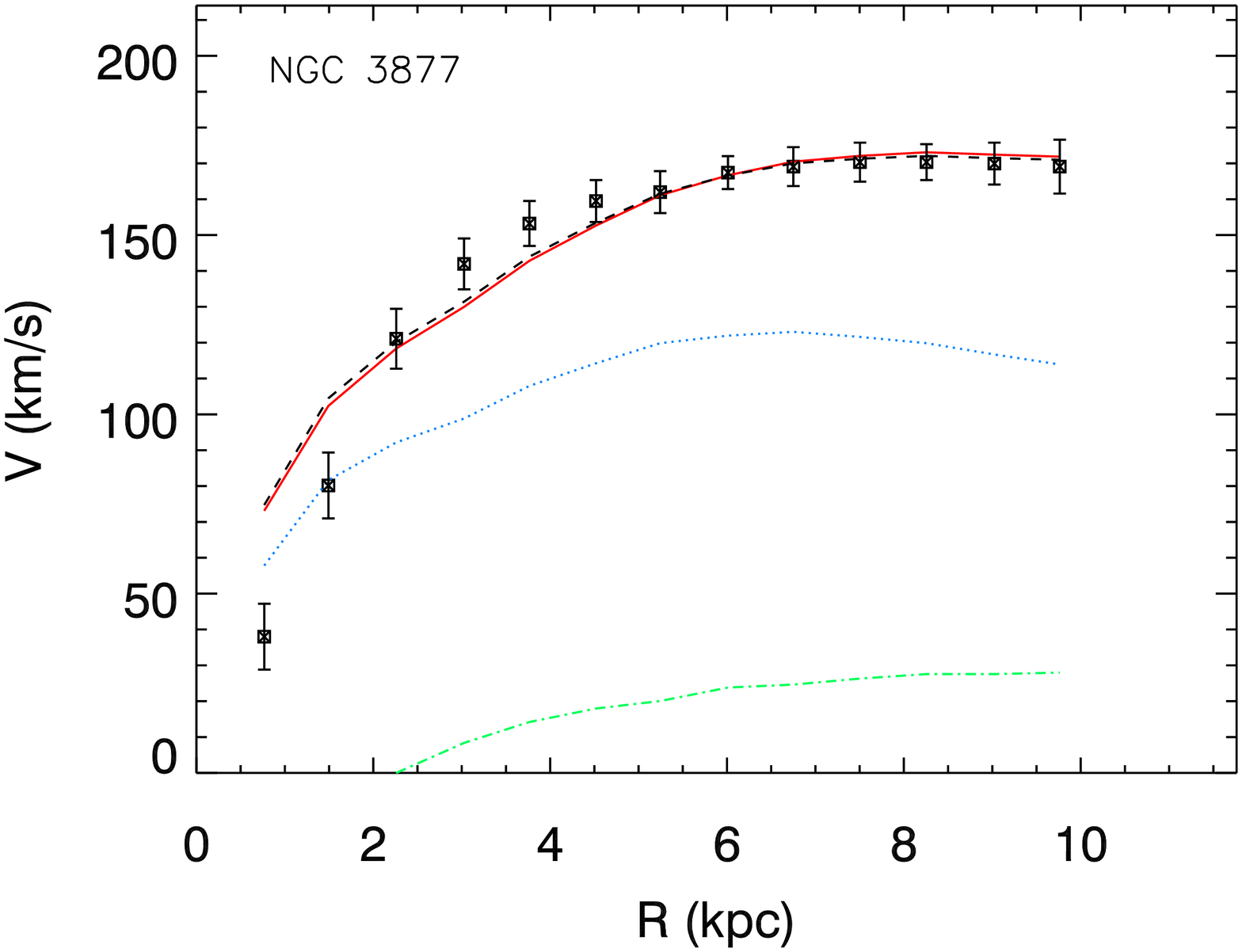}%\hspace{5mm}
  \includegraphics[angle=90,width=0.45\textwidth]{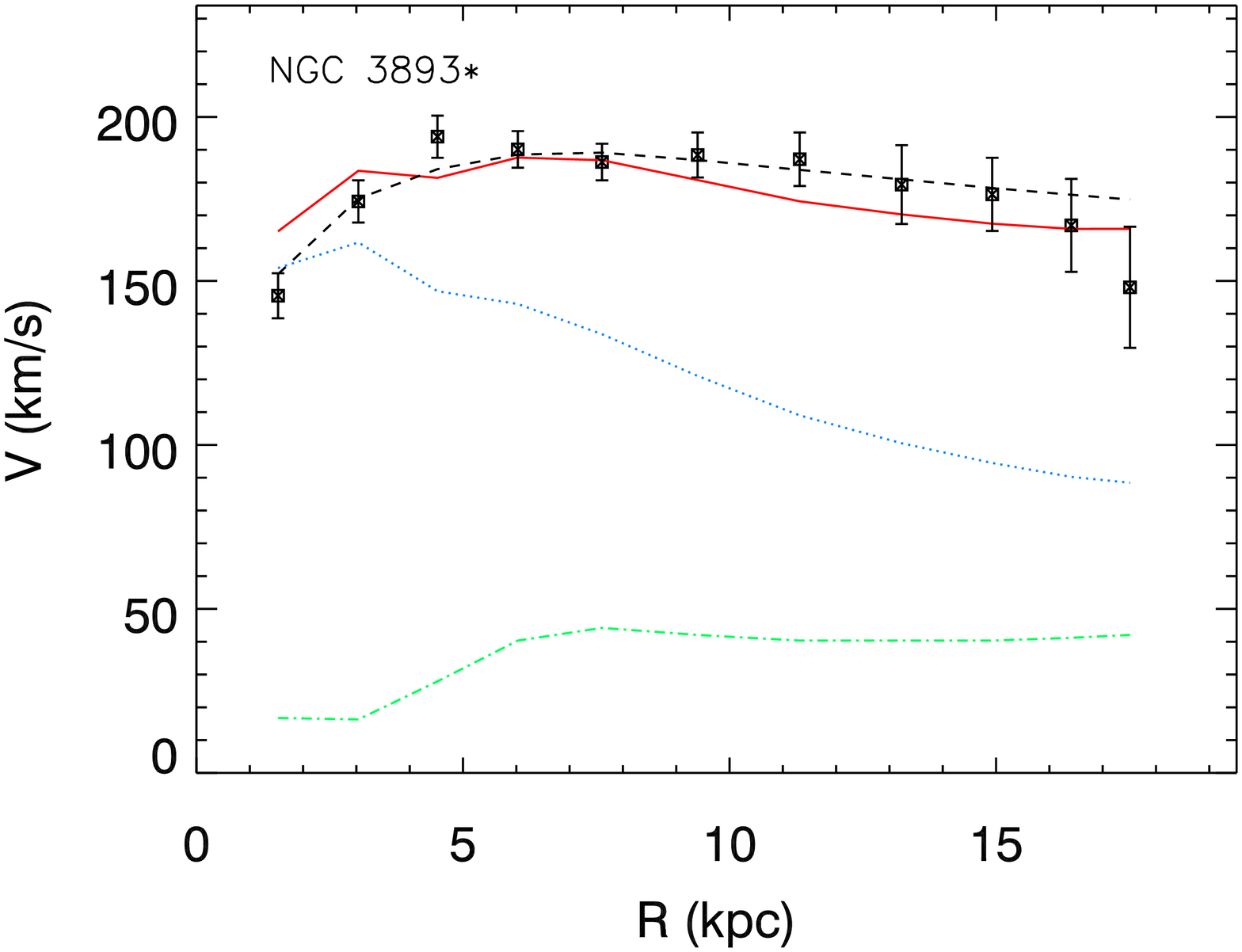}\vspace{5mm}\\
  \includegraphics[angle=90,width=0.45\textwidth]{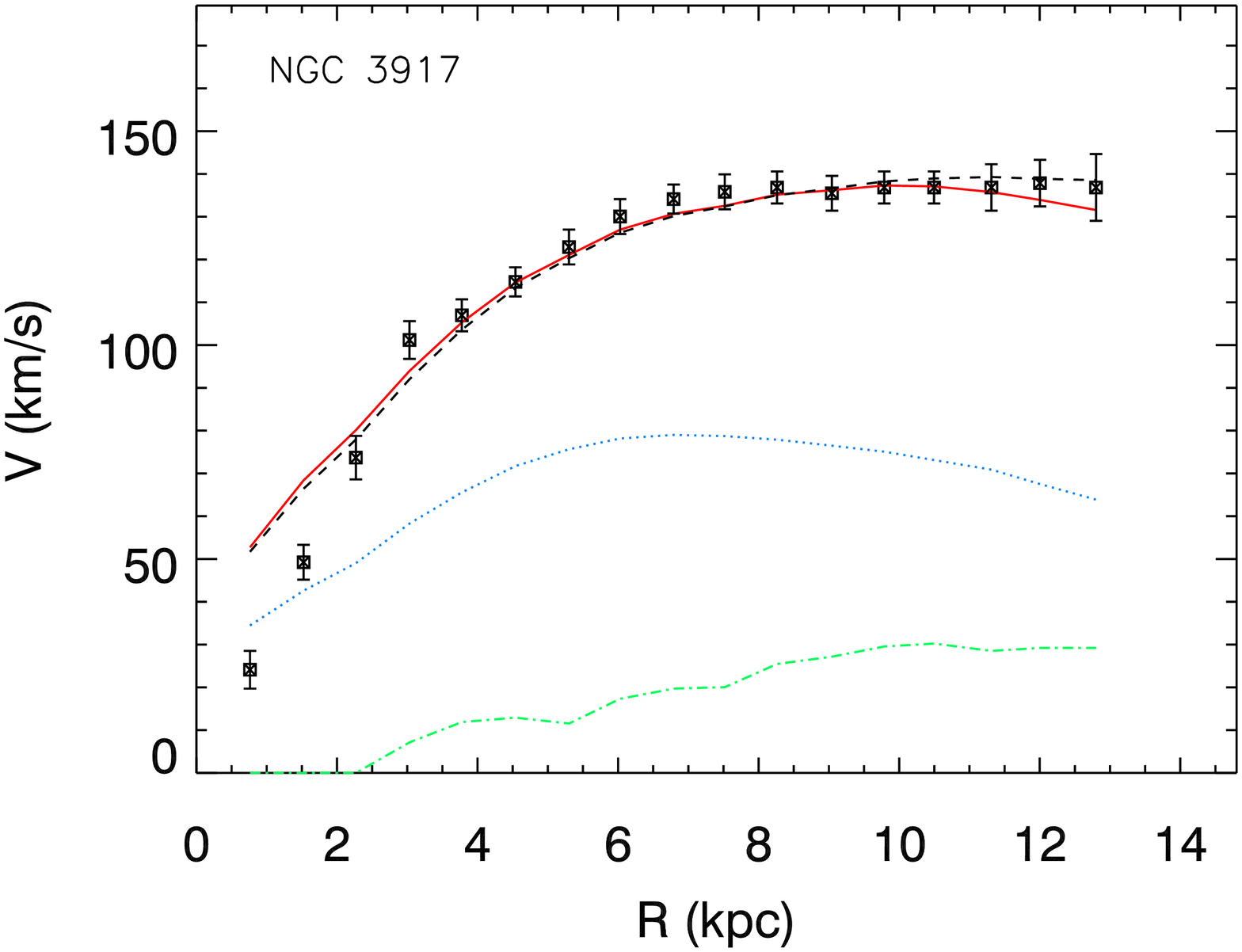}%\hspace{5mm}
  \includegraphics[angle=90,width=0.45\textwidth]{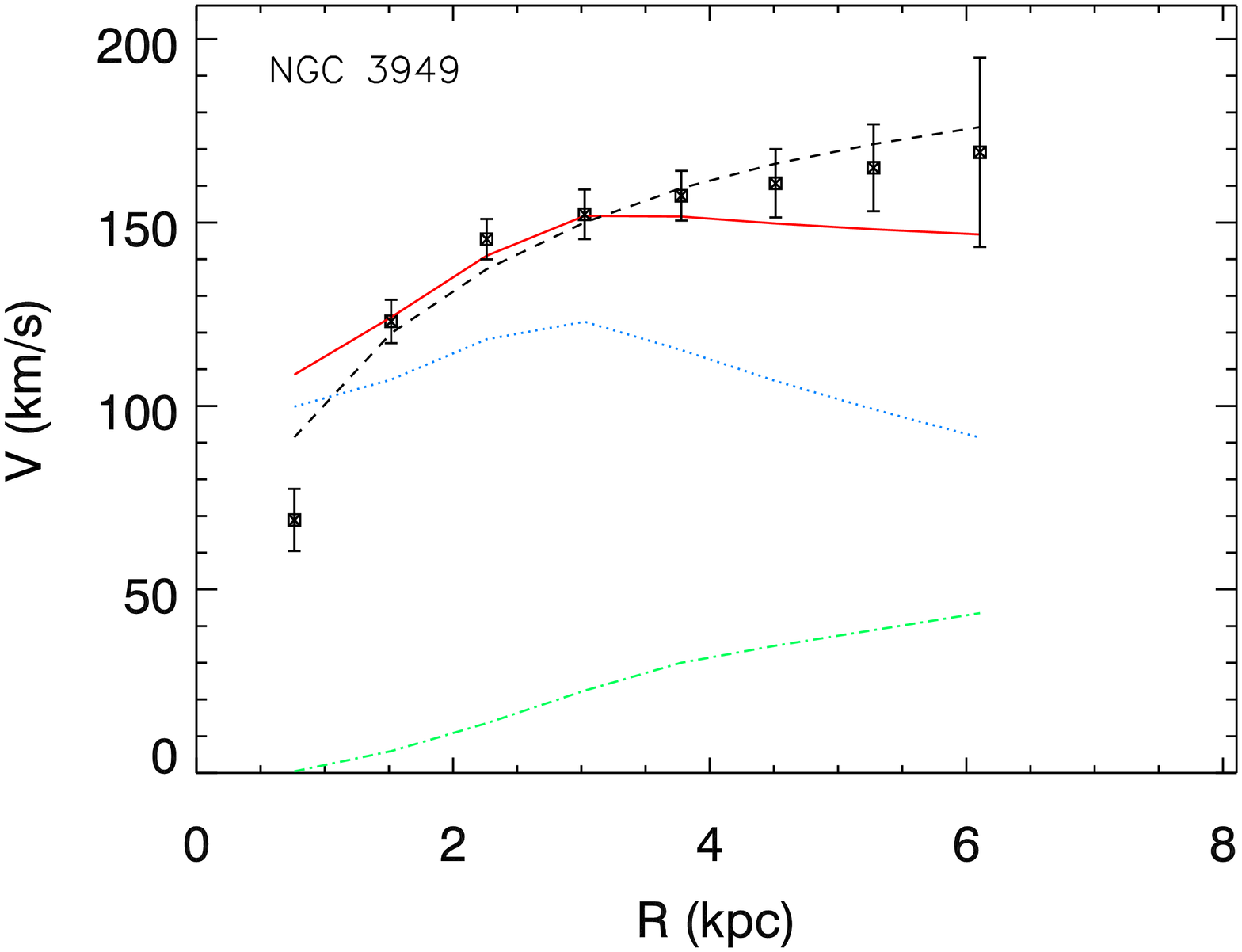}\vspace{5mm}\\
  \caption{Galactic rotation curves. The observed rotation curve is depicted by points with error bars. The solid red and dashed black lines are the MDM and CDM rotation curves, respectively. Newtonian curves for the stellar and gas components of the baryonic matter are depicted by dotted blue and dot-dashed green lines, respectively. The mass of the stellar component is derived from the $M/L$ ratio determined from MDM fits to the rotation curve. These figures are similar to those presented in \citet{Edmonds:2014}, but using the generalized mass profile presented in this paper using $r_\textrm{MDM} = 20$~kpc and $\alpha = 0.5$.}
  \label{fig:grc}
\end{figure*}

\begin{figure*}
  \includegraphics[angle=90,width=0.45\textwidth]{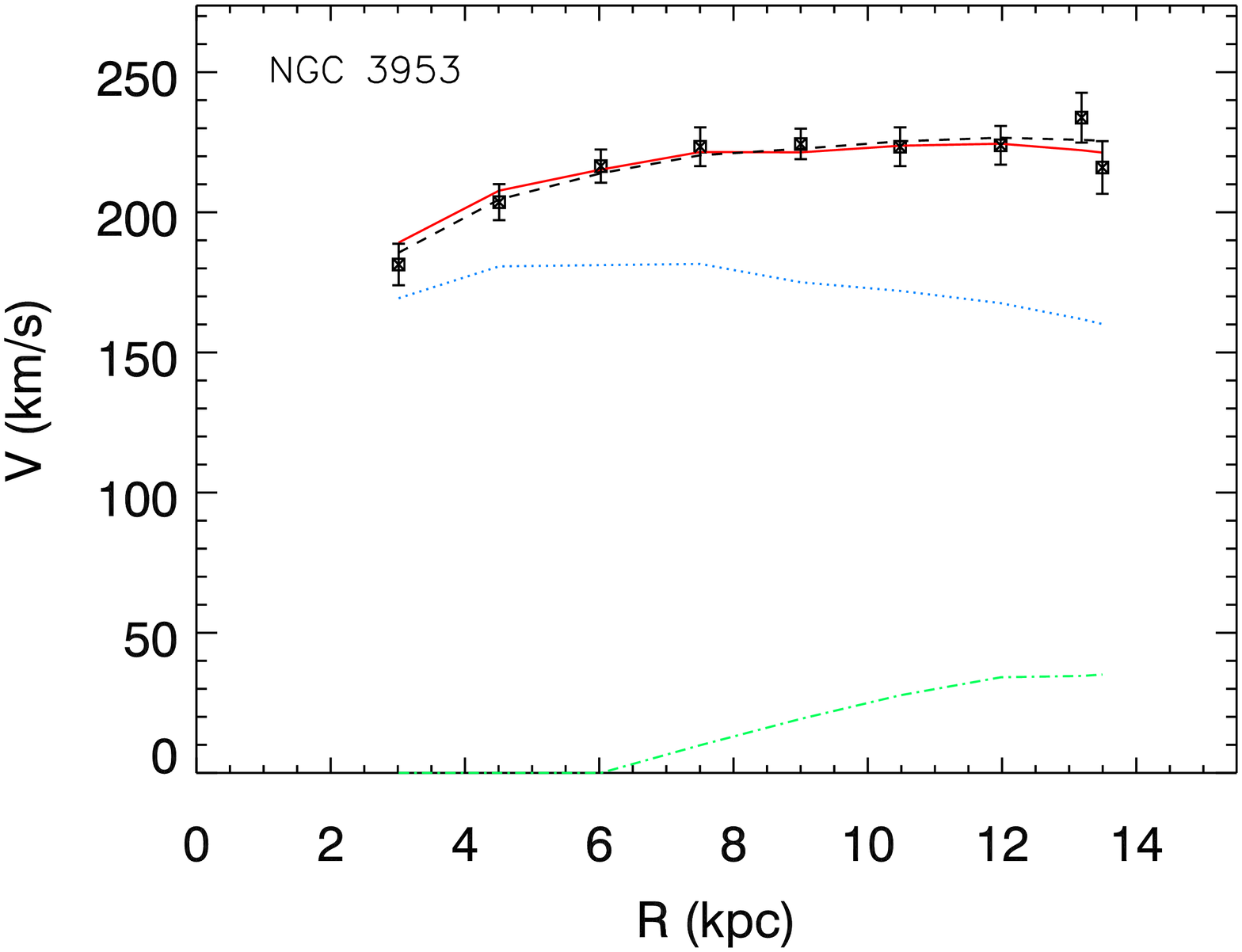}%\hspace{5mm}
  \includegraphics[angle=90,width=0.45\textwidth]{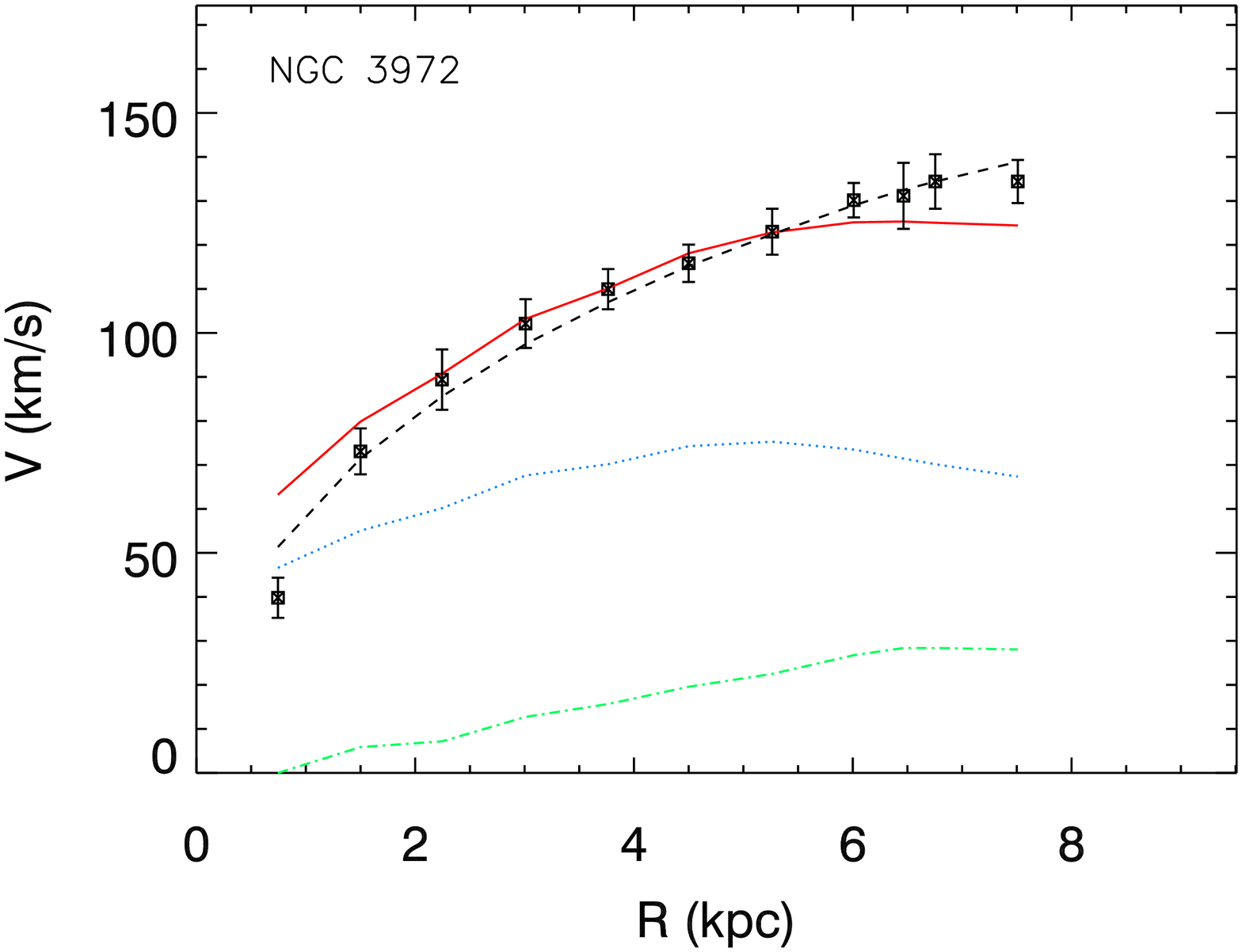}\vspace{5mm}\\
  \includegraphics[angle=90,width=0.45\textwidth]{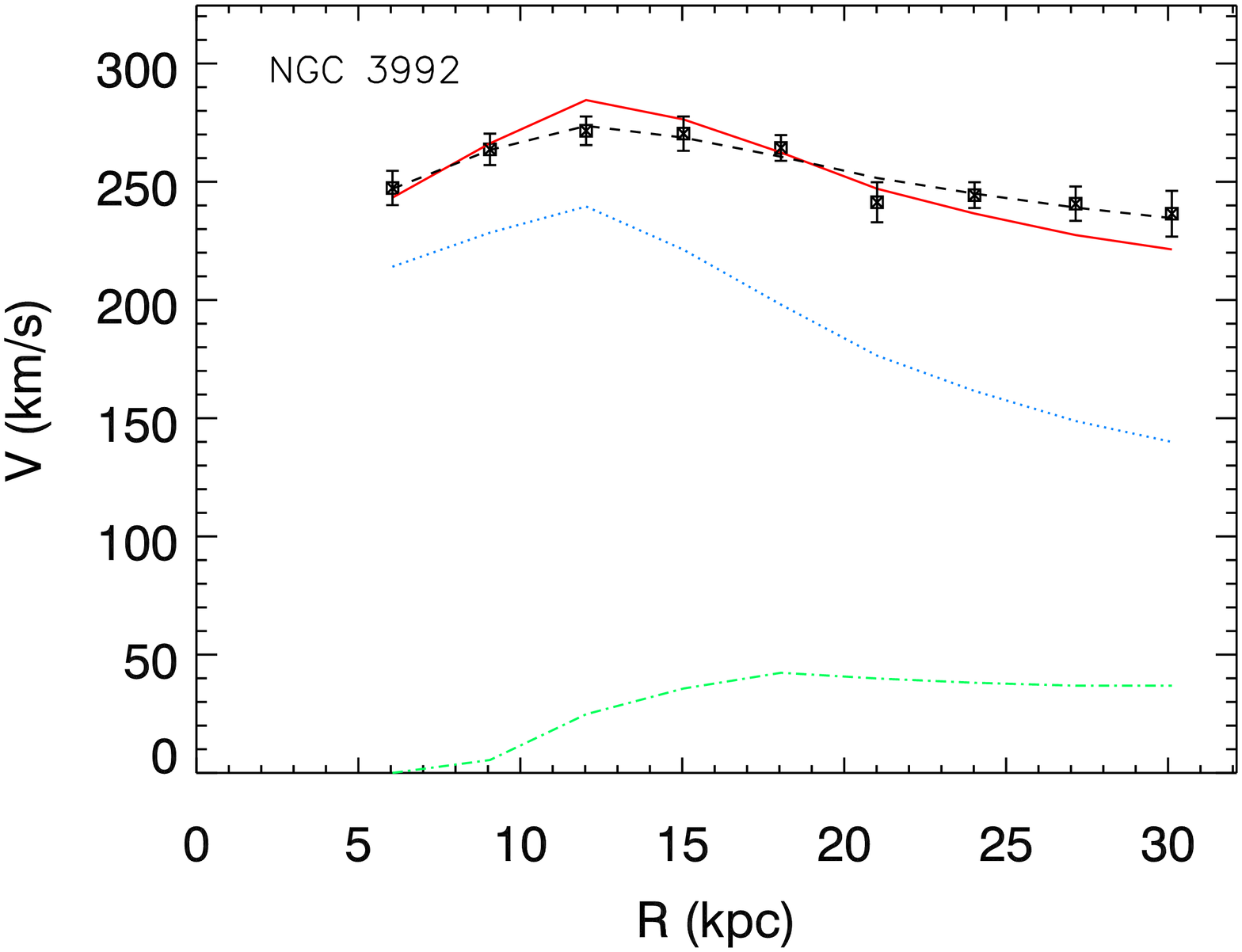}%\hspace{5mm}
  \includegraphics[angle=90,width=0.45\textwidth]{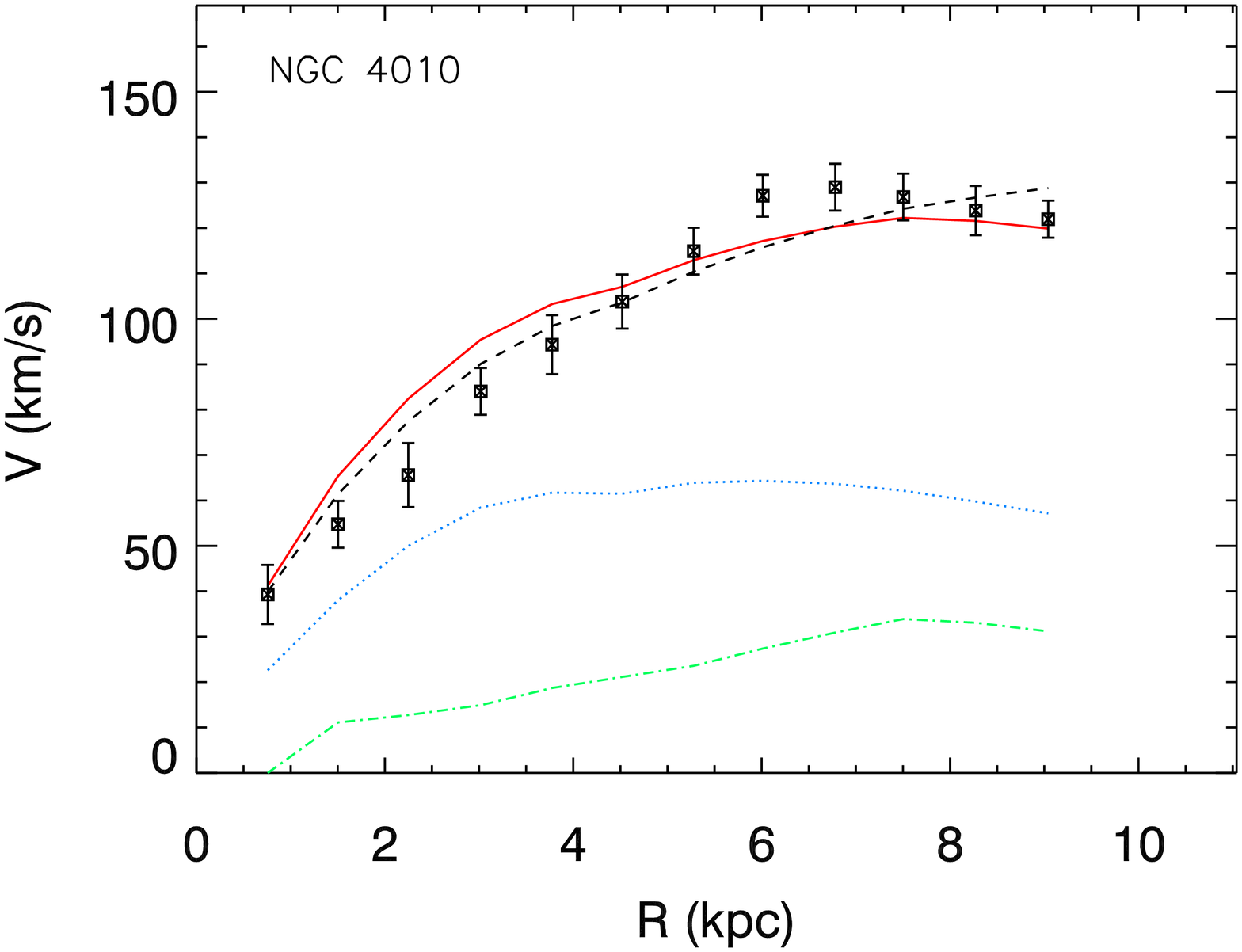}\vspace{5mm}\\
  \includegraphics[angle=90,width=0.45\textwidth]{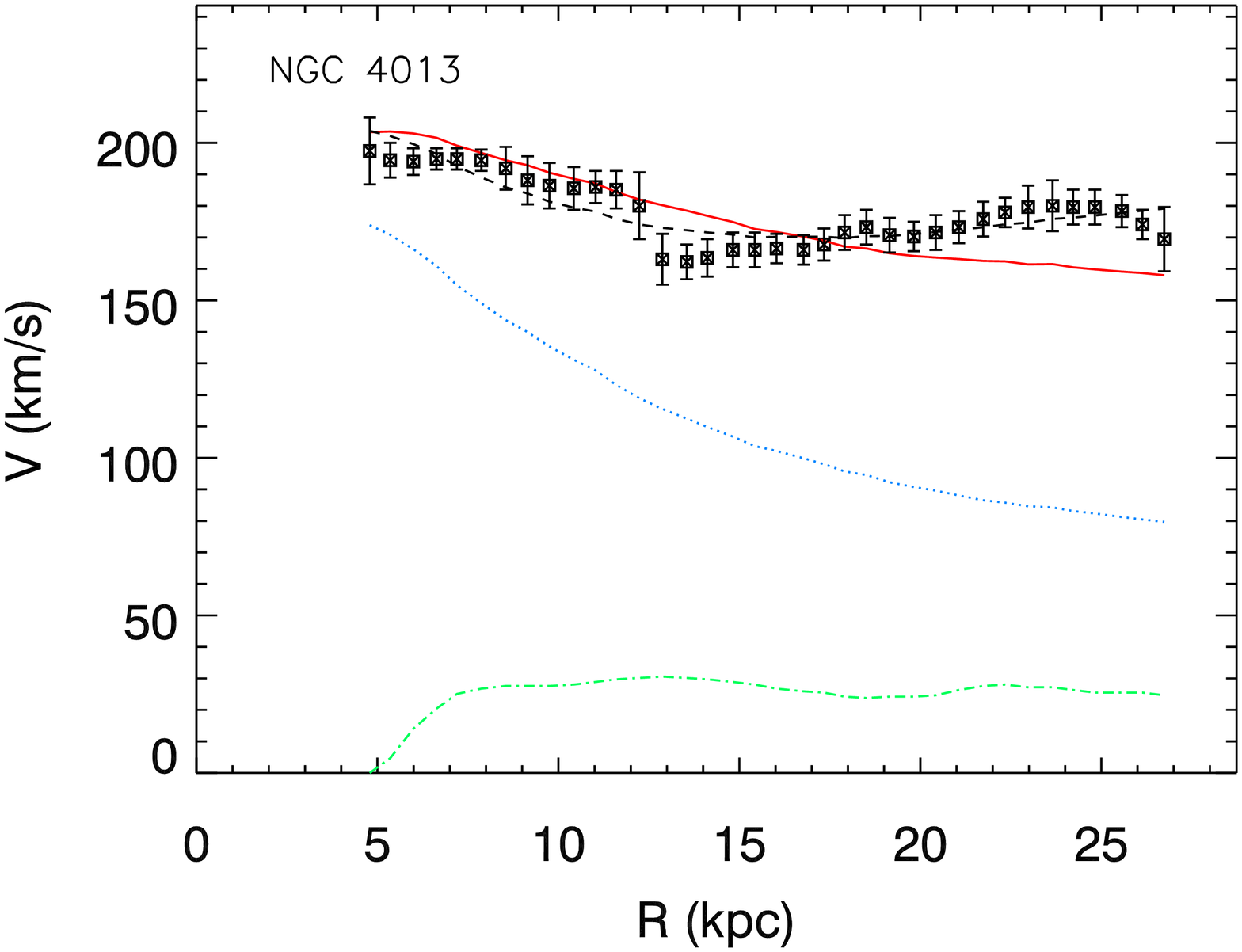}%\hspace{5mm}
  \includegraphics[angle=90,width=0.45\textwidth]{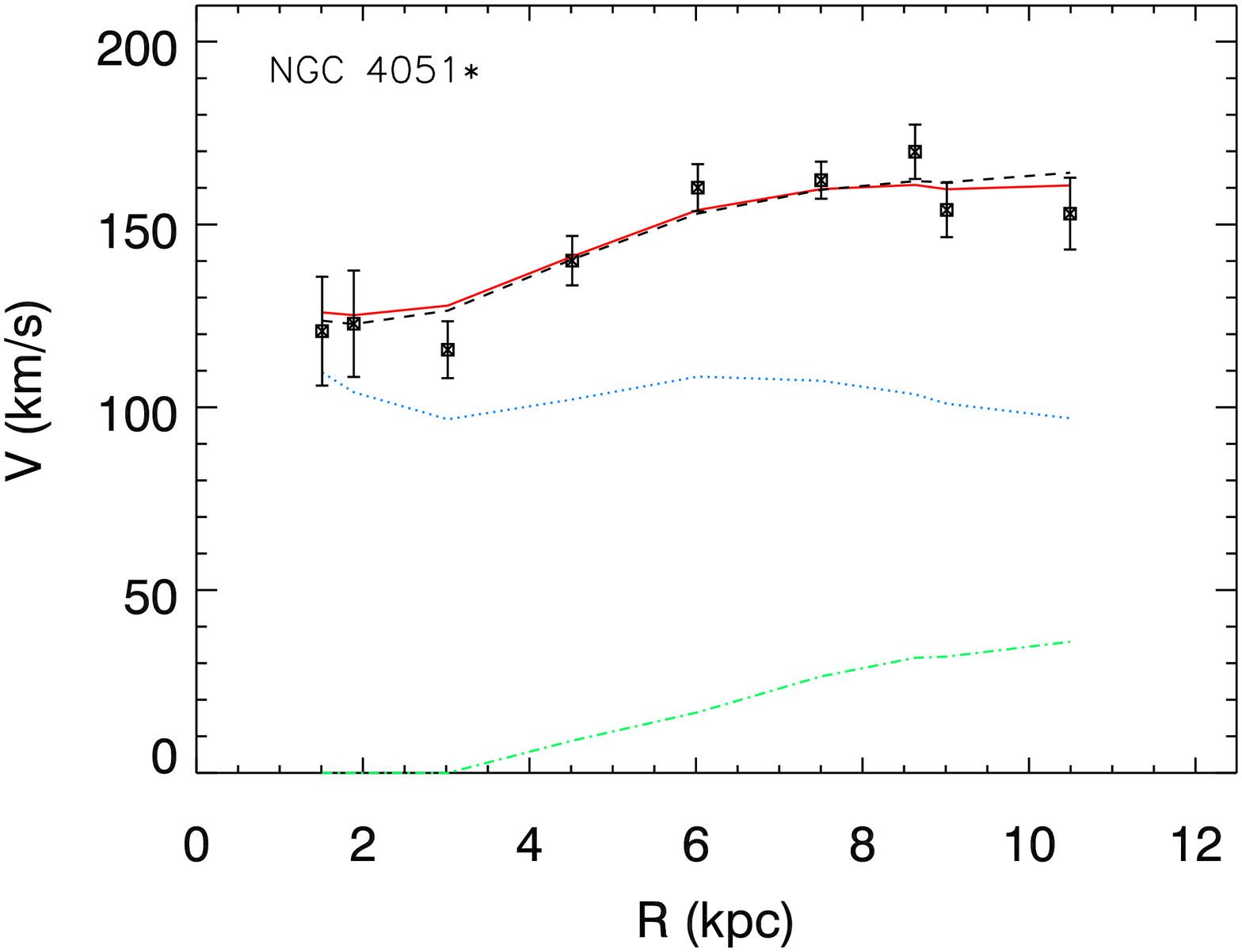}\vspace{5mm}\\
  \contcaption{}
\end{figure*}

\begin{figure*}
  \includegraphics[angle=90,width=0.45\textwidth]{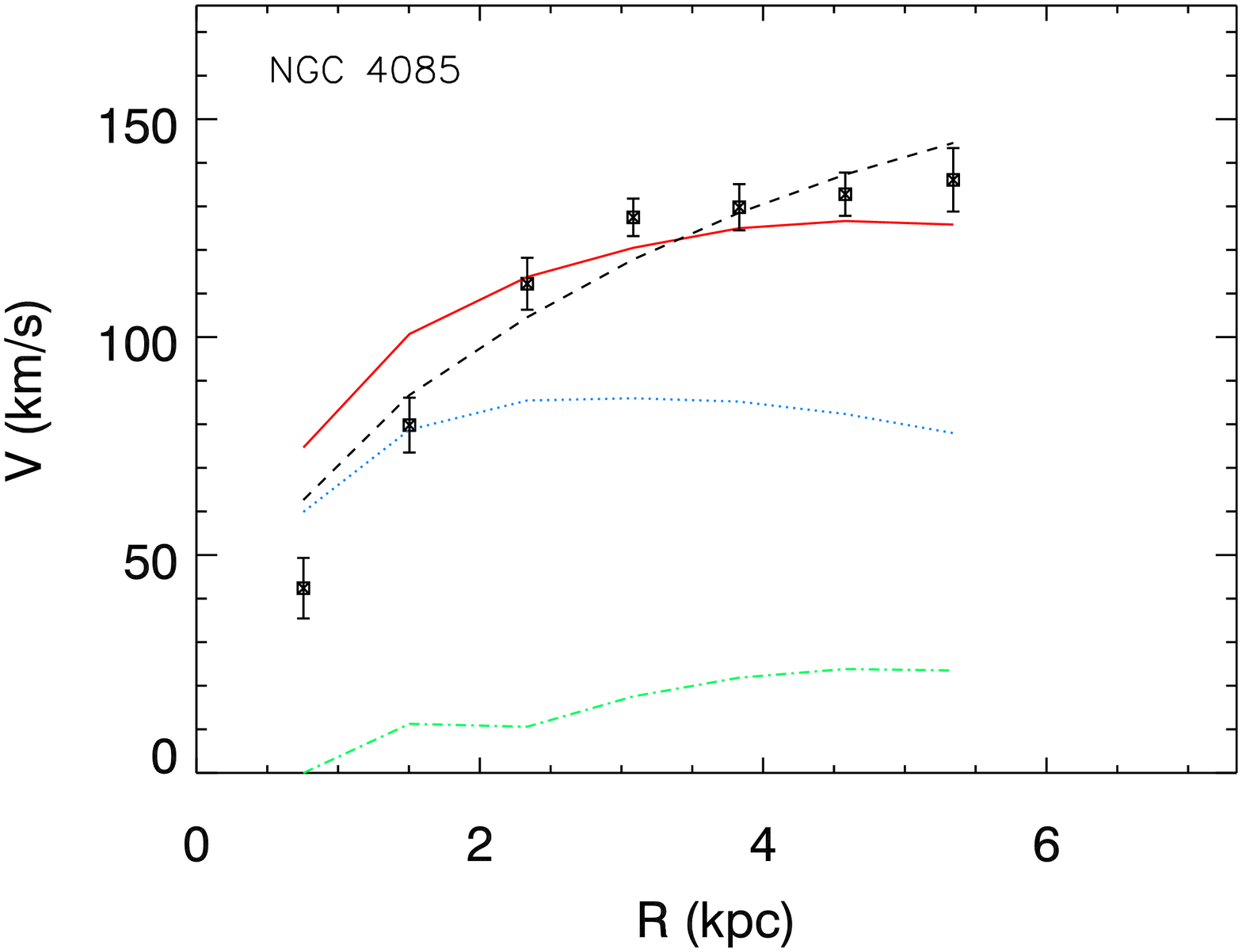}%\hspace{5mm}
  \includegraphics[angle=90,width=0.45\textwidth]{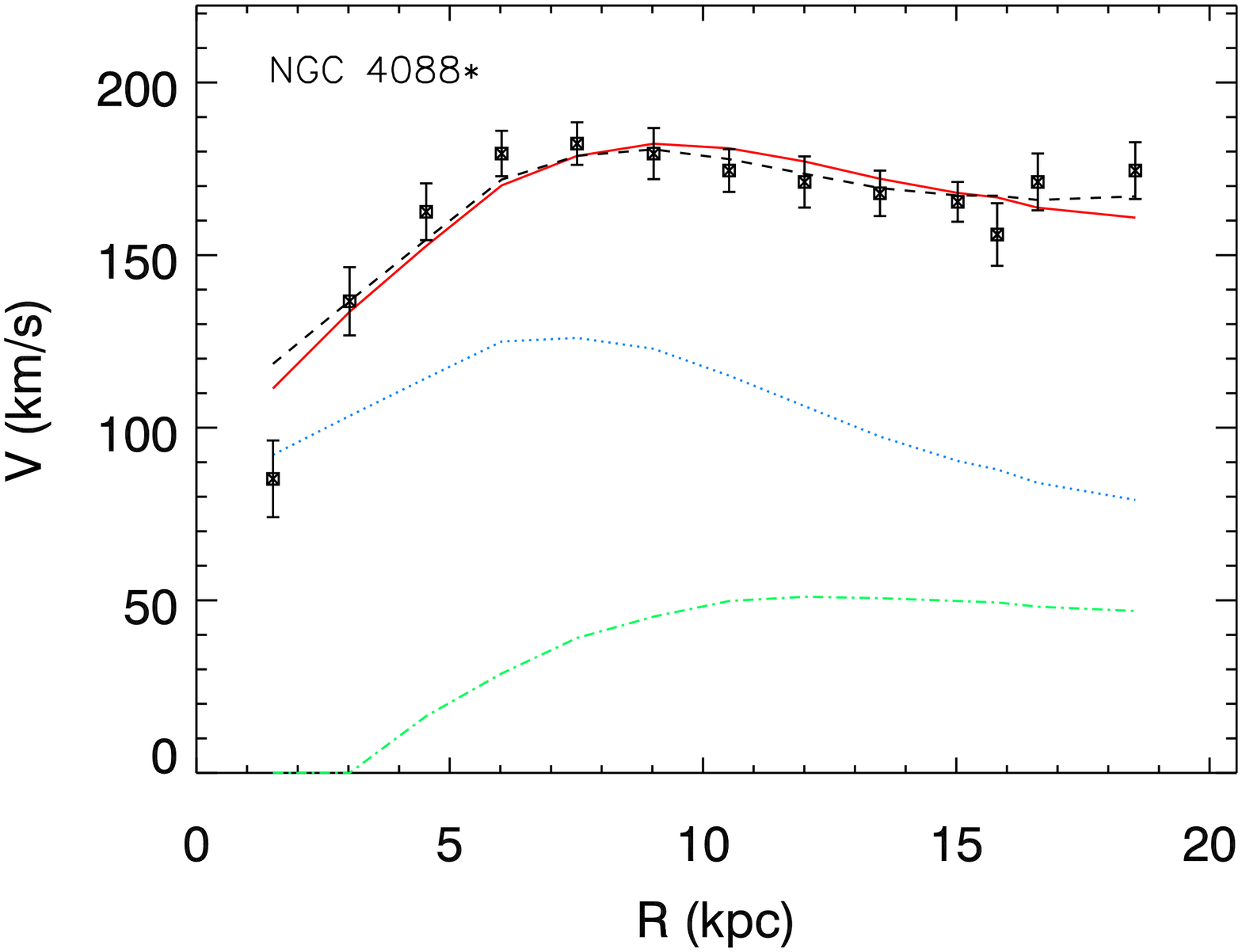}\vspace{5mm}\\
  \includegraphics[angle=90,width=0.45\textwidth]{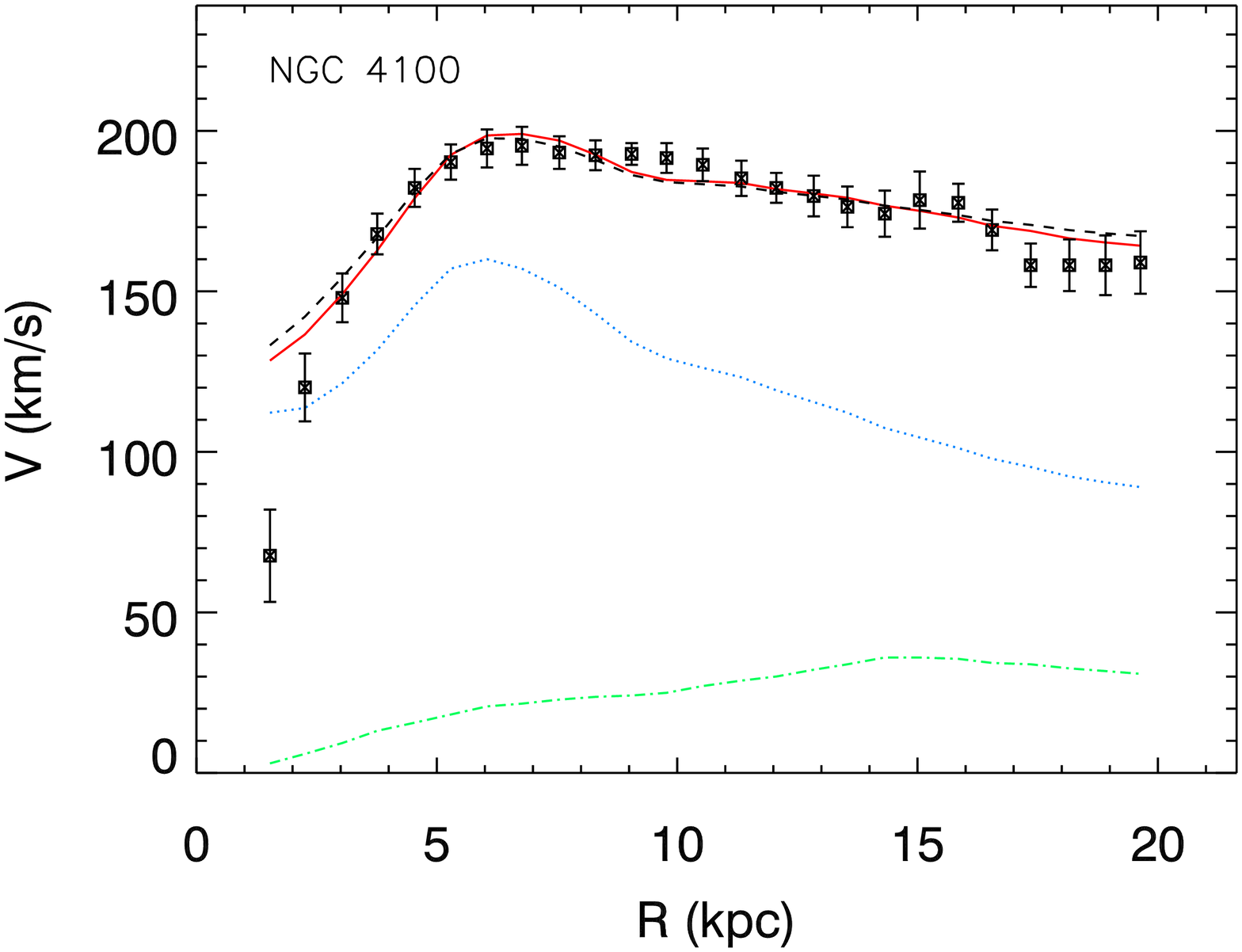}%\hspace{5mm}
  \includegraphics[angle=90,width=0.45\textwidth]{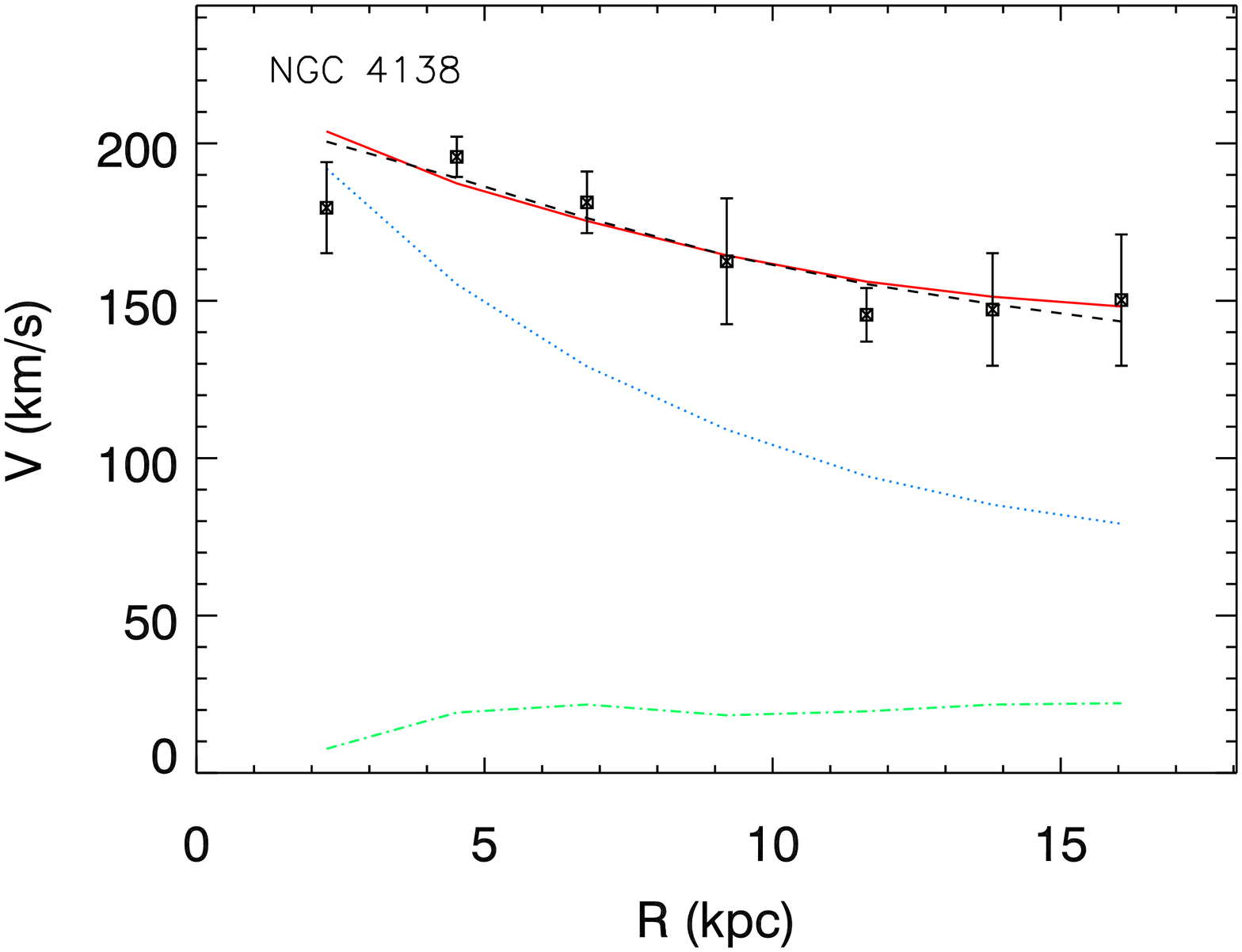}\vspace{5mm}\\
  \includegraphics[angle=90,width=0.45\textwidth]{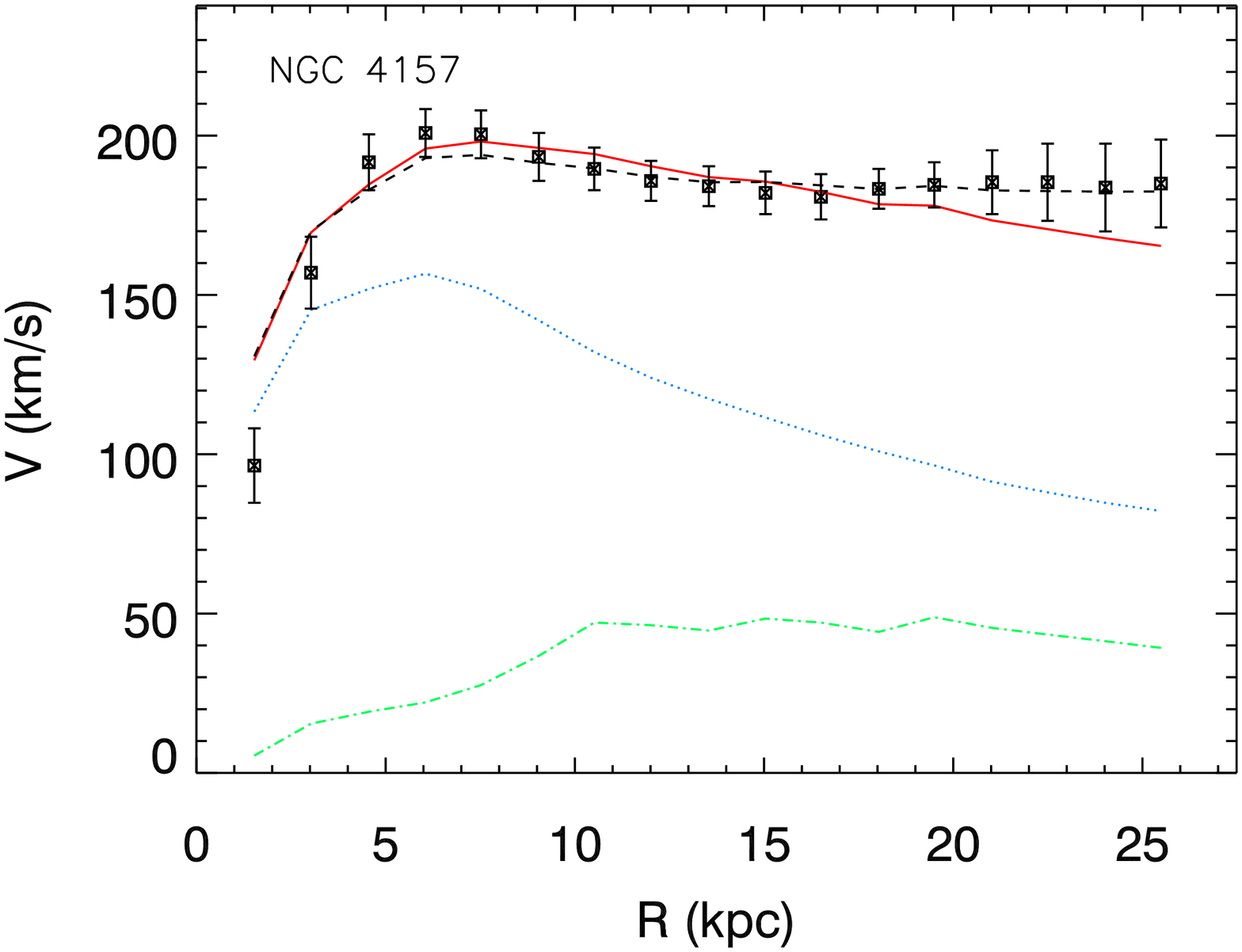}%\hspace{5mm}
  \includegraphics[angle=90,width=0.45\textwidth]{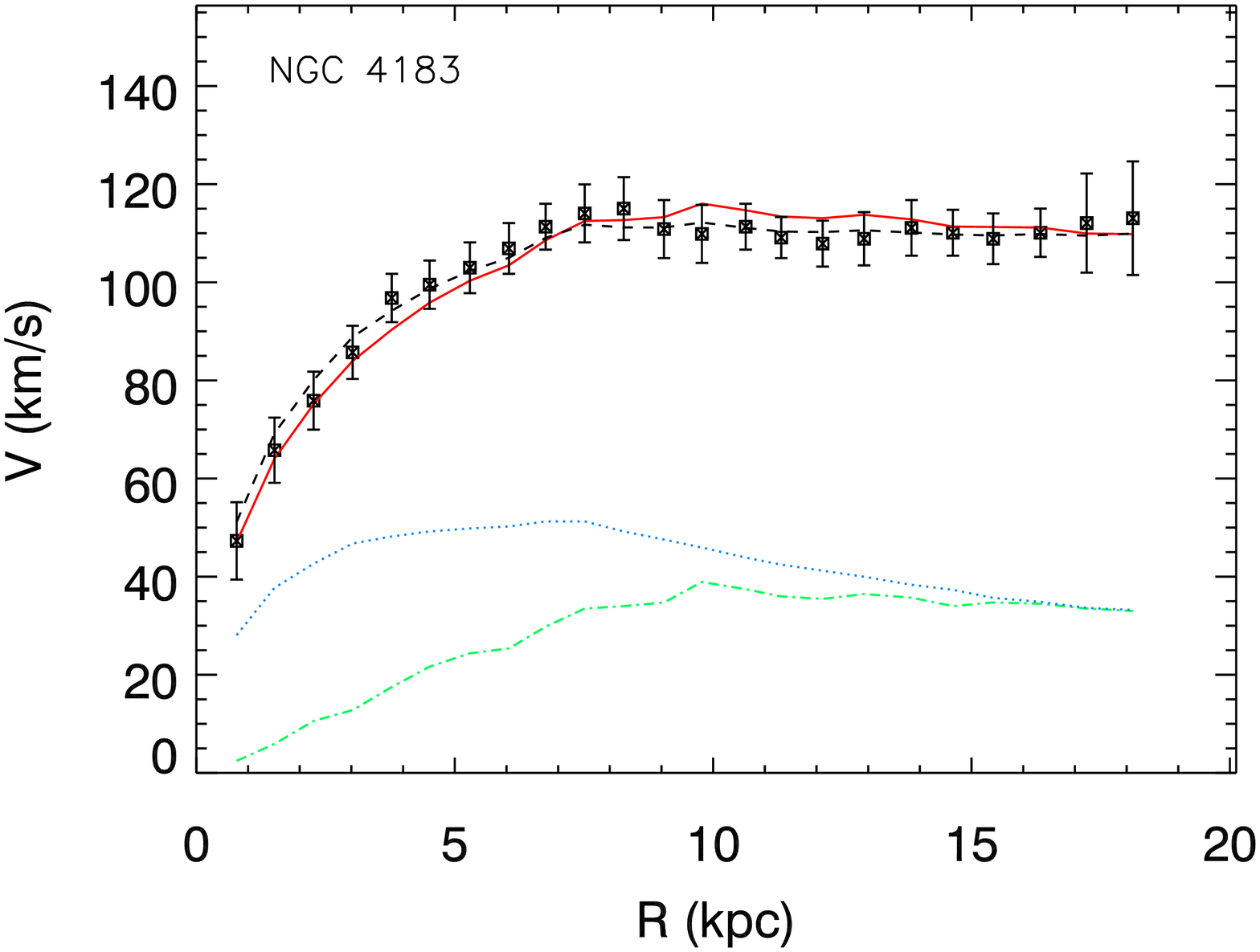}\vspace{5mm}\\
  \contcaption{}
\end{figure*}

\begin{figure*}
  \includegraphics[angle=90,width=0.45\textwidth]{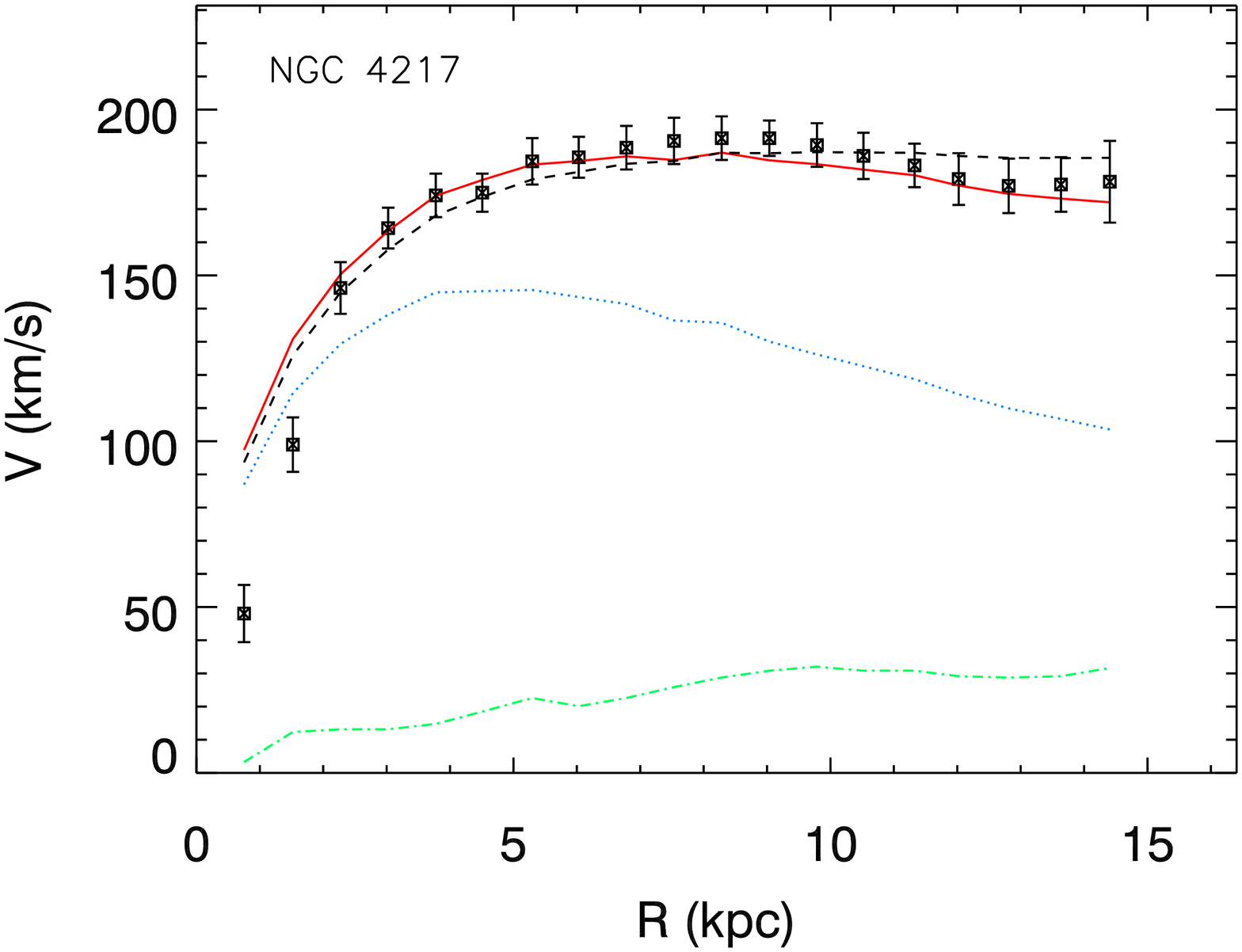}%\hspace{5mm}
  \includegraphics[angle=90,width=0.45\textwidth]{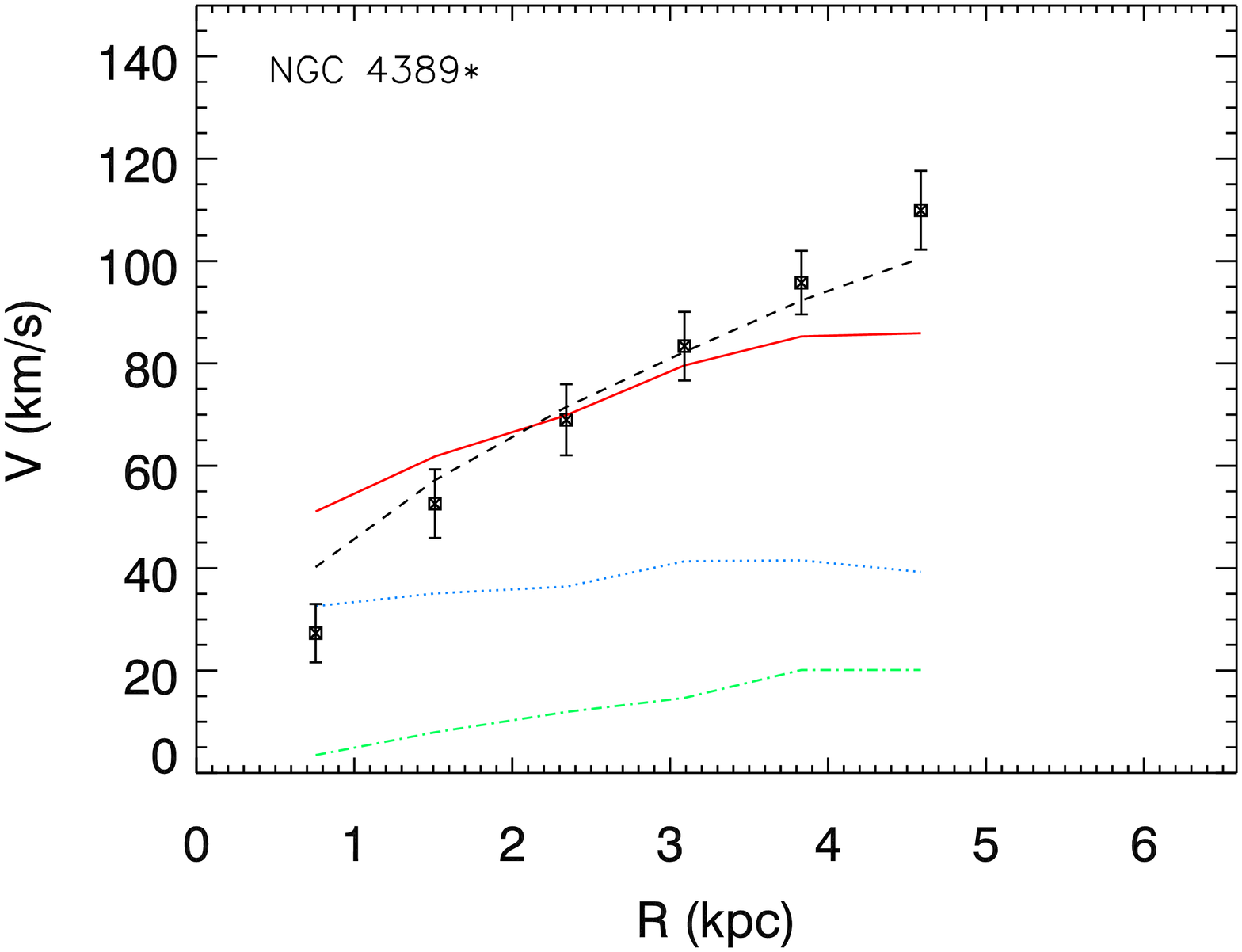}\vspace{5mm}\\
  \includegraphics[angle=90,width=0.45\textwidth]{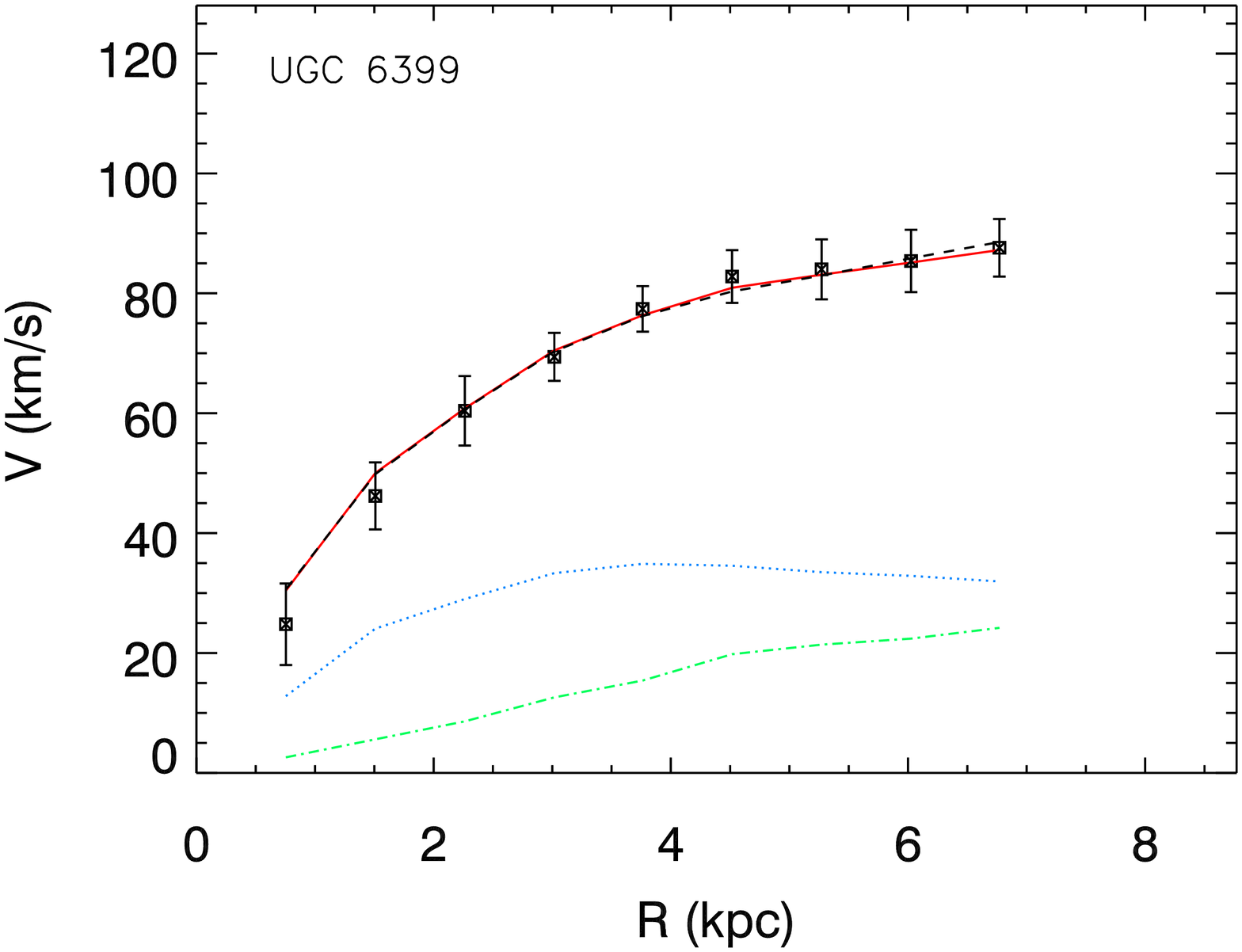}%\hspace{5mm}
  \includegraphics[angle=90,width=0.45\textwidth]{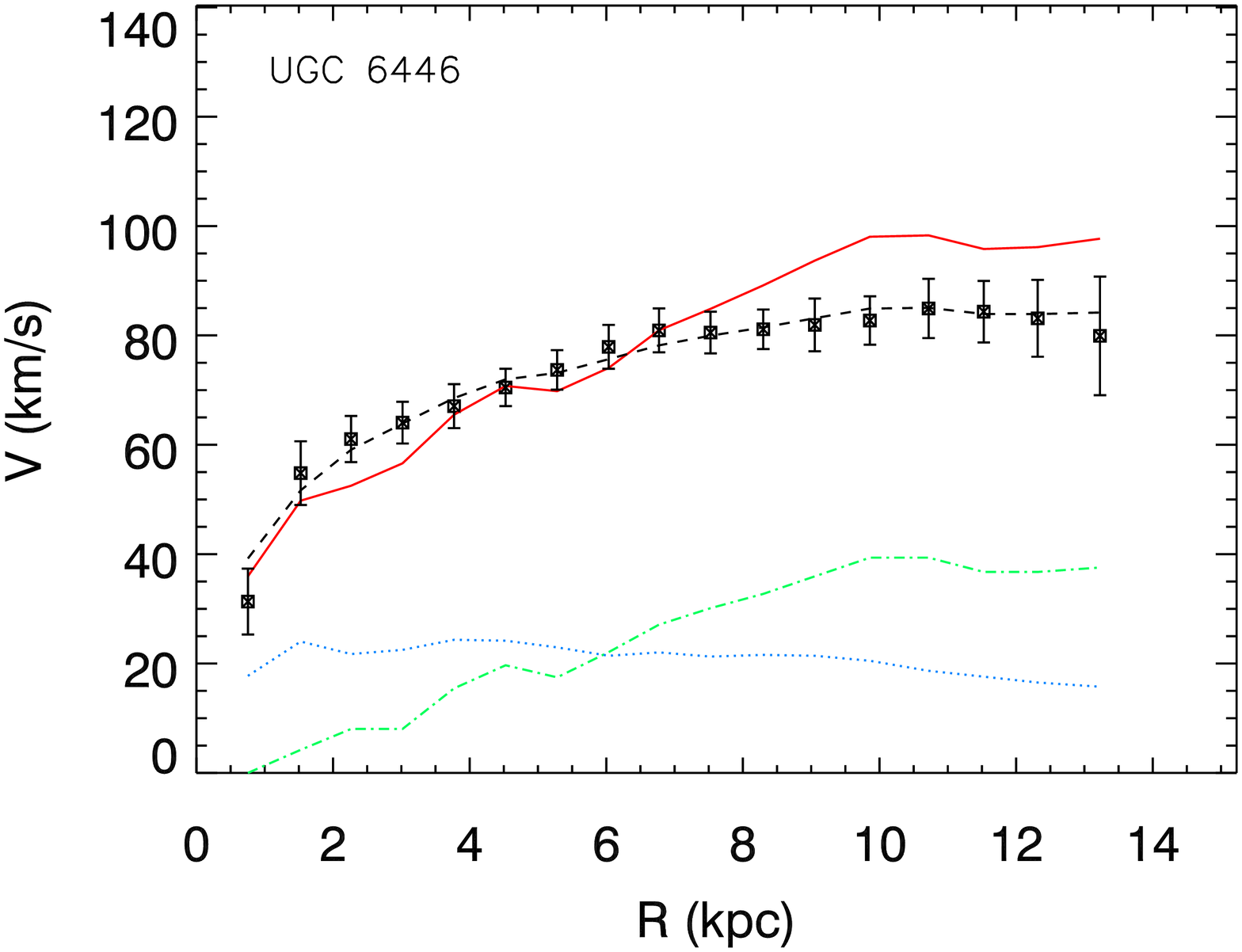}\vspace{5mm}\\
  \includegraphics[angle=90,width=0.45\textwidth]{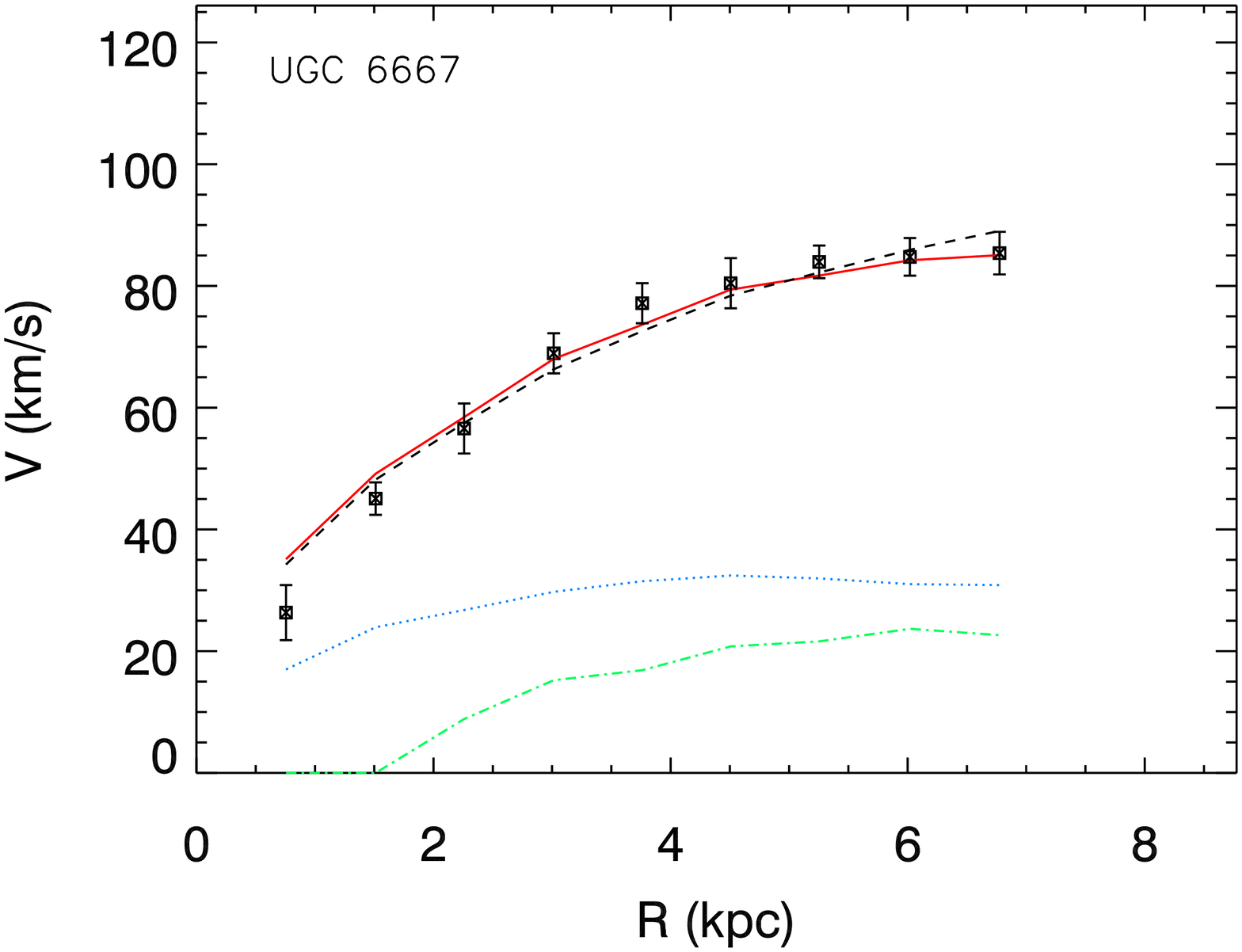}%\hspace{5mm}
  \includegraphics[angle=90,width=0.45\textwidth]{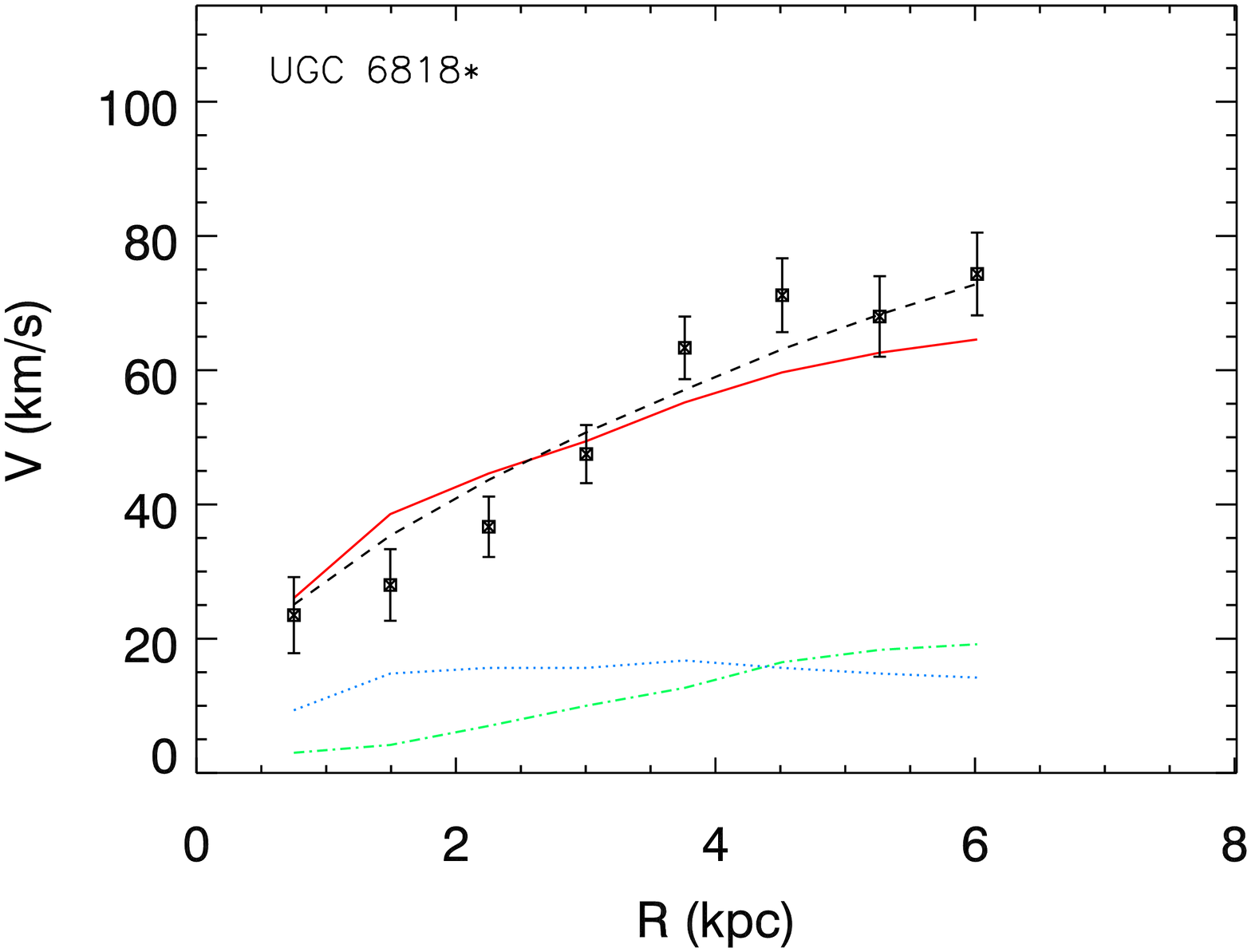}\vspace{5mm}\\
  \contcaption{}
\end{figure*}

\begin{figure*}
  \includegraphics[angle=90,width=0.45\textwidth]{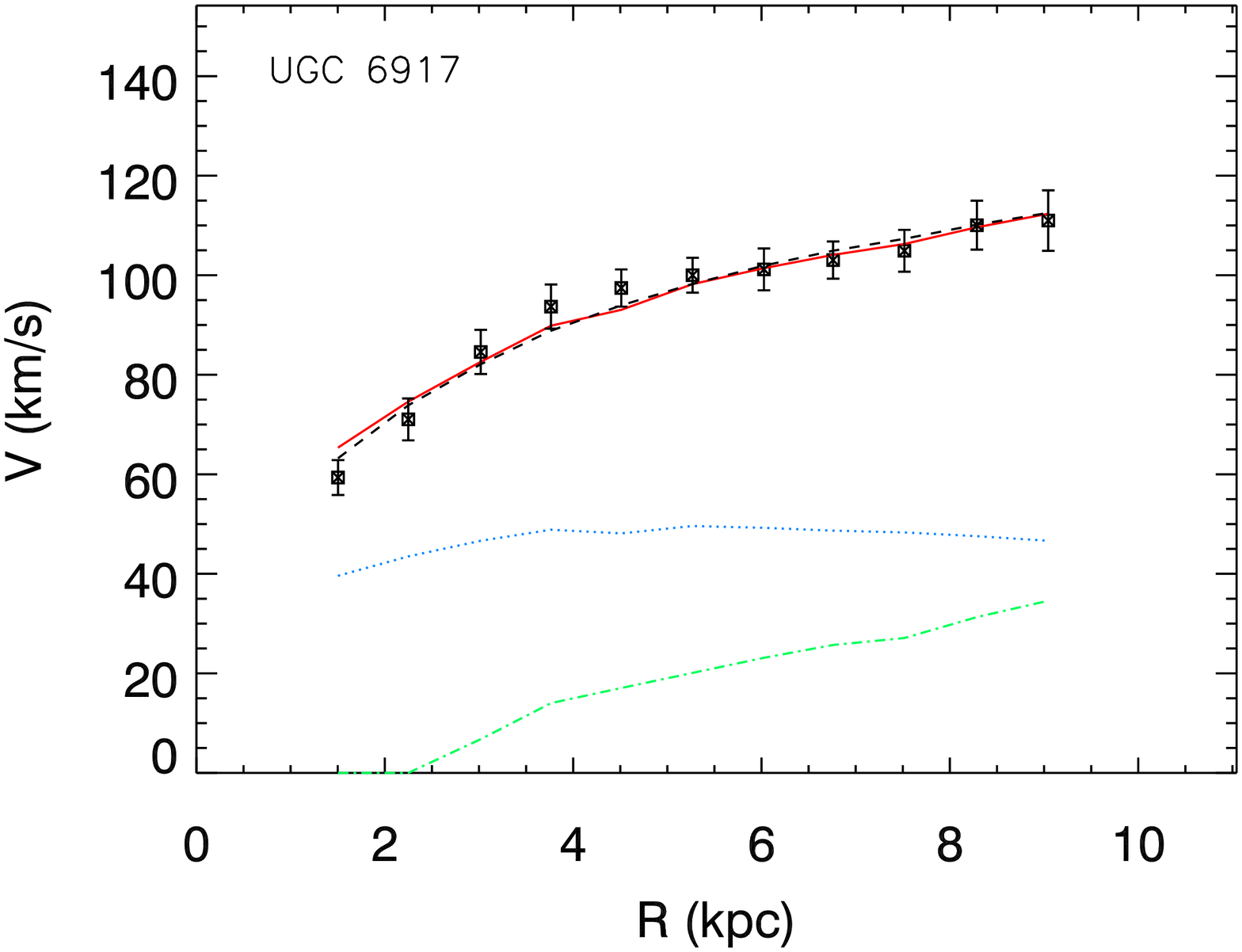}%\hspace{5mm}
  \includegraphics[angle=90,width=0.45\textwidth]{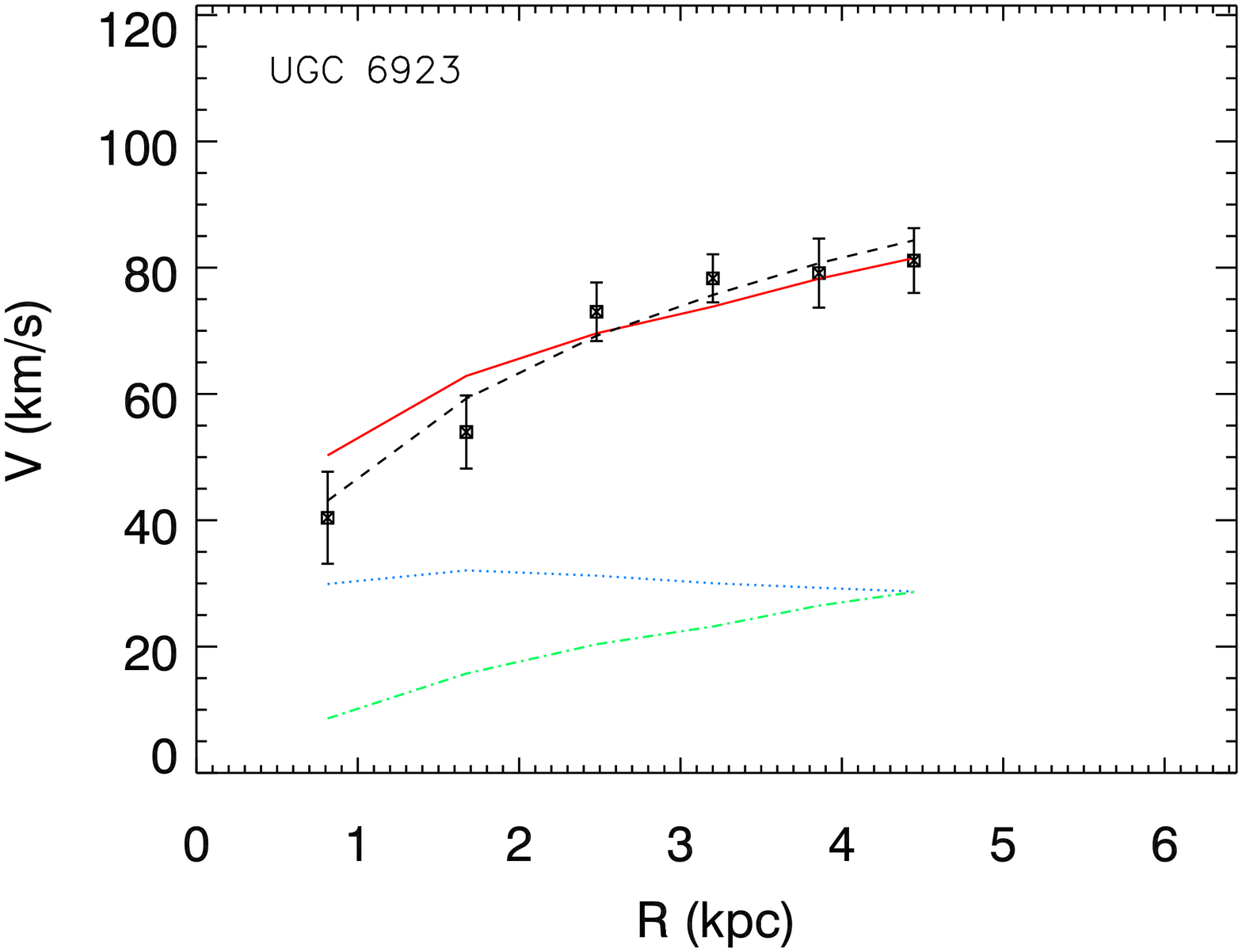}\vspace{5mm}\\
  \includegraphics[angle=90,width=0.45\textwidth]{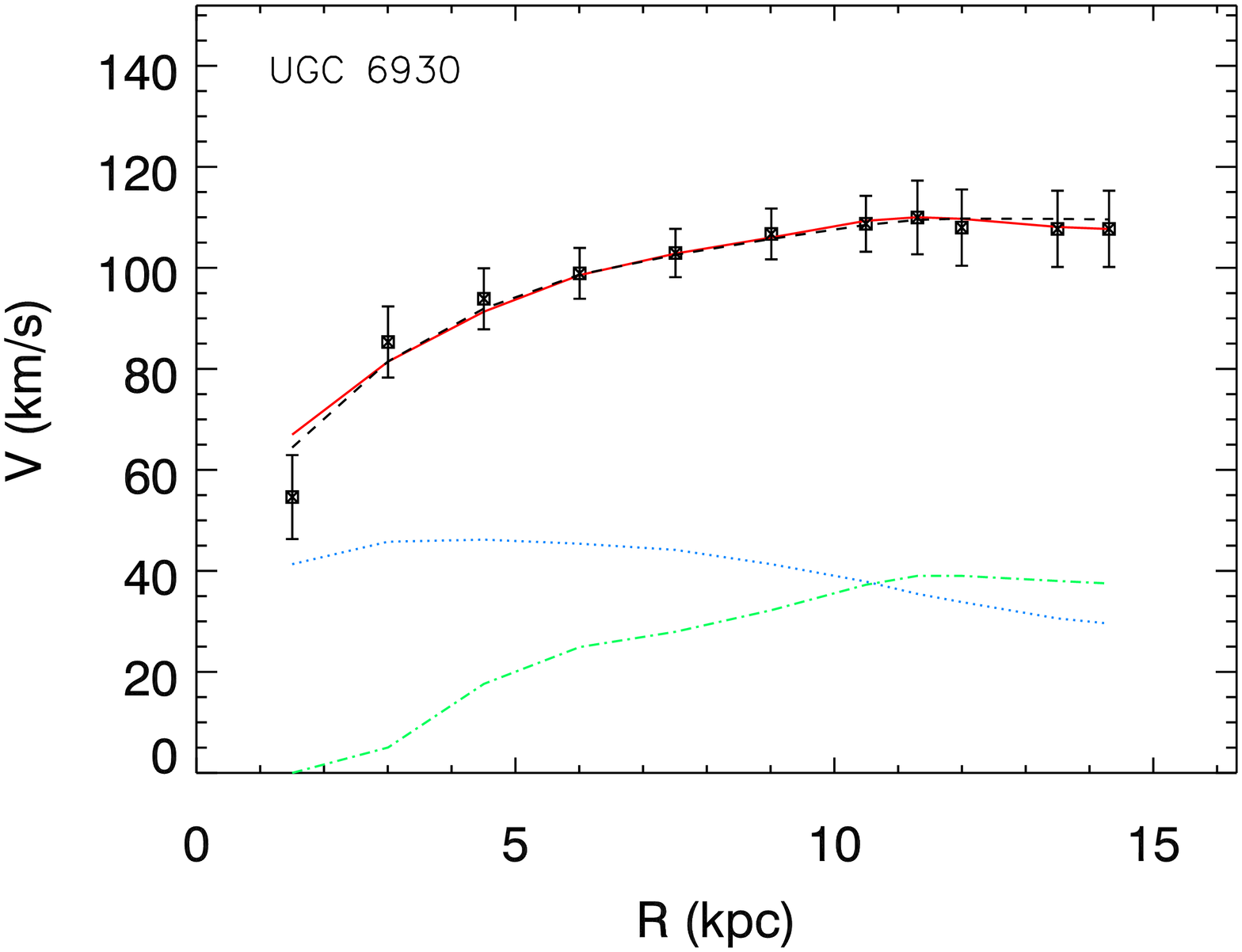}%\hspace{5mm}
  \includegraphics[angle=90,width=0.45\textwidth]{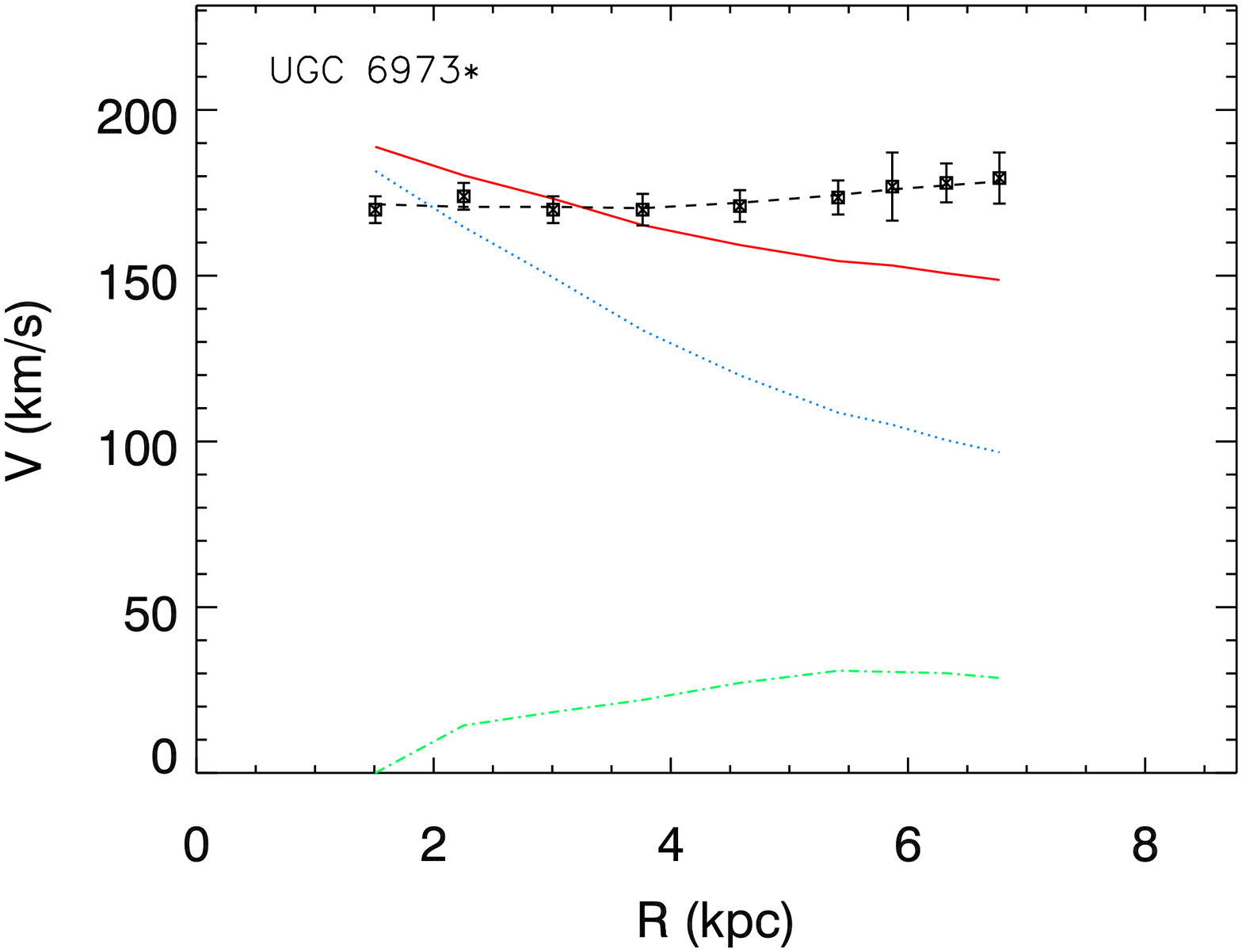}\vspace{5mm}\\
  \includegraphics[angle=90,width=0.45\textwidth]{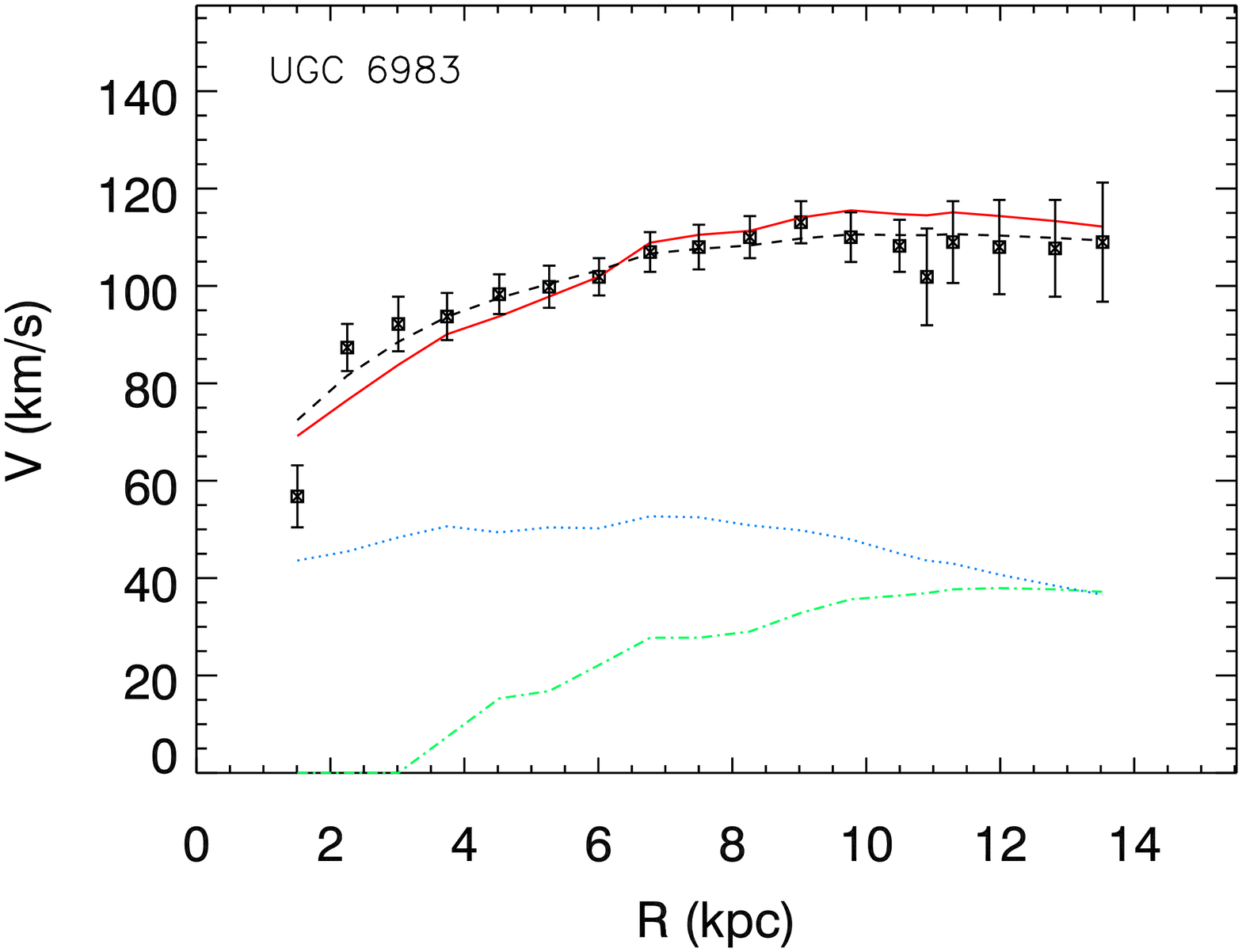}%\hspace{5mm}
  \includegraphics[angle=90,width=0.45\textwidth]{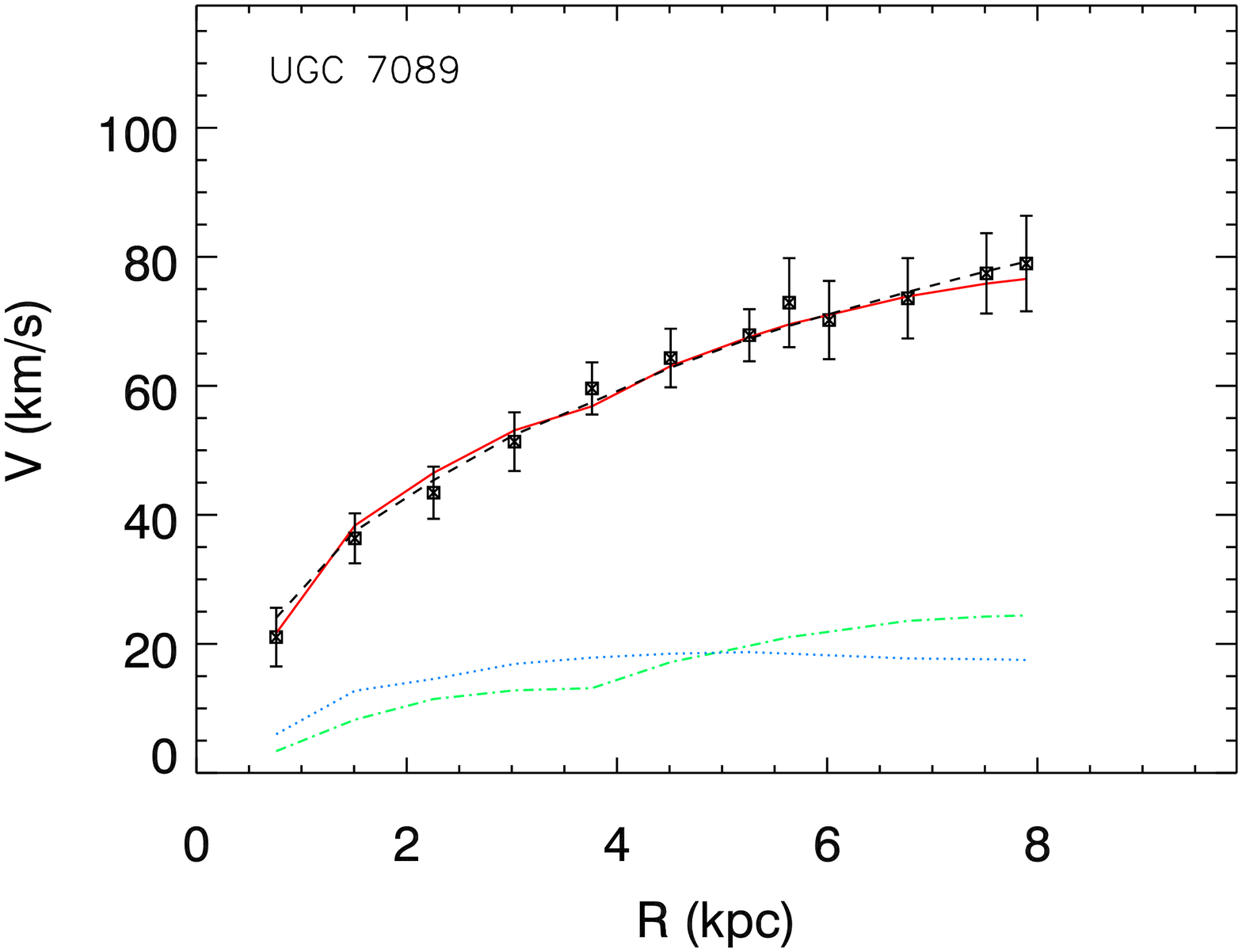}\vspace{5mm}\\
  \contcaption{}
\end{figure*}
%%%%%%%%%%%%%%%%%%%%%%%%%%%%%%%%%%%%%%%%%%%%%%%%

\clearpage
\clearpage

%%%%%%%%%%%%%%%%%%%%%%%%%%%%%%%%%%%%%%%%%%%%%%%%%%
%%%%%%%%%%%%%%%%%%%%%%%%%%%%%%%%%%%%%%%%%%%%%%%%%%
\label{lastpage}

\bsp	% typesetting comment

%%%%%%%%%%%%%%%%%%%%%%%%%%%%%%%%%%%%%%%%%%%%%%%%%%%%%%%%
%%%%%%%%%%%%%%%%%%%%%%%%%%%%%%%%%%%%%%%%%%%%%%%%%%%%%%%%
%%%%%%%%%%%%%%%%%%%%%%%%%%%%%%%%%%%%%%%%%%%%%%%%%%%%%%%%
%%%%%%%%%%%%%%%%%%%%%%%%%%%%%%%%%%%%%%%%%%%%%%%%%%%%%%%%
%%%%%%%%%%%%%%%%%%%%%%%%%%%%%%%%%%%%%%%%%%%%%%%%%%%%%%%%
\end{document}